\documentclass[a4paper,12pt]{article}
\usepackage{bm}
\usepackage{amsmath}
\usepackage{amssymb}
\usepackage[dvips]{graphicx}
\usepackage[dvips]{psfrag}

\makeatletter
\@addtoreset{equation}{section}
\renewcommand{\theequation}{\thesection.\@arabic\c@equation}
\makeatother
\makeatletter
\renewcommand\appendix{\par
  \setcounter{section}{0}%
  \setcounter{subsection}{0}%
  \gdef\thesection{\@Alph\c@section }
  \renewcommand{\theequation}
  {\Alph{section}.\arabic{equation}}
}
\makeatother

\def\nn{\nonumber}
\def\lb{\label}
\def\ci{\cite}
\newcommand{\Ref}[1]{(\ref{#1})}
\def\a{\alpha}
\def\b{\beta}
\def\g{\gamma}
\def\d{\delta}
\def\D{\Delta}
\def\G{\Gamma}
\def\e{\epsilon}
\def\o{\omega}
\def\Om{\Omega}
\def\th{\theta}
\def\s{\sigma}
\def\lam{\lambda}
\def\vp{\varphi}

\newcommand{\rd}{\mathrm{d}}
\newcommand{\p}{\partial}
\newcommand{\N}{\nabla}

\def\bra{\langle}
\def\ket{\rangle}
\def\l{\left}
\def\r{\right}
\def\f{\frac}
\def\tu{\tilde u}

\def\bk{\bm k}
\def\bl{\bm l}

\def\tr{\mathrm{Tr}}

\def\MM{\mathcal {M}}
\def\MN{\mathcal N}
\def\MO{\mathcal {O}}

\def\wt{\widetilde}
\begin{document}

\begin{titlepage}

\vspace*{-15mm}   
\baselineskip 10pt   
\begin{flushright}   
\begin{tabular}{r} 
\end{tabular}   
\end{flushright}   
\baselineskip 24pt   
\vglue 10mm   

\begin{center}
{\Large\bf Black Hole as a Quantum Field Configuration}
\vspace{8mm}   
\baselineskip 18pt   

\renewcommand{\thefootnote}{\fnsymbol{footnote}}

Hikaru~Kawai$^a$\footnote[2]{hkawai@gauge.scphys.kyoto-u.ac.jp} and 
Yuki~Yokokura$^b$\footnote[4]{yuki.yokokura@riken.jp}

\renewcommand{\thefootnote}{\arabic{footnote}}

\vspace{5mm}   

{\it  
 $^a$ Department of Physics, Kyoto University, 
 Kitashirakawa, Kyoto 606-8502, Japan \\
$^b$ iTHEMS, RIKEN, Wako, Saitama 351-0198, Japan}

\vspace{10mm}   

\end{center}

\begin{abstract}
We describe 4D evaporating black holes as quantum field configurations 
by solving the semi-classical Einstein equation 
$G_{\mu\nu}=8\pi G \langle \psi|T_{\mu\nu}|\psi \rangle$ 
and quantum matter fields in a self-consistent manner. 
As the matter fields we consider $N$ massless free scalar fields ($N$ is large). 
We find a spherically symmetric self-consistent solution of the metric $g_{\mu\nu}$ and the state $|\psi\rangle$. 
Here, $g_{\mu\nu}$ is locally $AdS_2\times S^2$ geometry, 
and $|\psi\rangle$ provides $\langle \psi|T_{\mu\nu}|\psi \rangle=\langle0|T_{\mu\nu}|0 \rangle+T_{\mu\nu}^{(\psi)}$, 
where $|0\rangle$ is the ground state of the matter fields in the metric 
and $T_{\mu\nu}^{(\psi)}$ consists of the excitation of s-waves that describe the collapsing matter and 
Hawking radiation with the ingoing negative energy flow. 
This object is supported by a large tangential pressure $\langle0|T^\theta{}_\theta|0 \rangle$ due to 
the vacuum fluctuation of the bound modes with large angular momenta $l\gg1$. 
This describes the interior of the black hole when the back reaction of the evaporation is taken into account. 
In this picture, the black hole is a compact object with a surface (instead of horizon) that looks like 
a conventional black hole from the outside and eventually evaporates without a singularity. 
If we count the number of configurations $\{|\psi\rangle\}$ that satisfy the self-consistent equation, 
we reproduce the area law of the entropy. 
This tells that the information is carried by the s-waves inside the black hole.  
$|\psi\rangle$ also describes the process that 
the negative ingoing energy flow created with Hawking radiation 
is superposed on the collapsing matter to decrease the total energy while the total energy density remains positive. 
Finally, as a special case, we consider conformal matter fields and show that 
the interior metric is determined by the matter content of the theory, 
which leads to a new constraint to the matter contents for the black hole to evaporate. 
\end{abstract}

\baselineskip 18pt   

\end{titlepage}

\newpage

\tableofcontents

\section{Introduction}\lb{s:intro}
In quantum theory, black holes evaporate \ci{Hawking}. 
This property may change the definition of black holes from the classical one. 
It should be determined by quantum dynamics of matter and spacetime. 
By collapse of a star, an object should be formed and evaporate in a finite time. 
The Penrose diagram of the space-time should have the same topology structure as the Minkowski space-time, 
where there is no event horizon or singularity but may be a trapping horizon. 
Such an object should be the black hole. 
This view is an accepted consensus in the context of quantum theory \ci{Frolov1}-\ci{Ho5}
\footnote{Recently, astrophysical phenomena related to the quantum properties of black holes 
have been actively studied \ci{Cardoso0}-\ci{Oshita2}.}. 
In this paper, we consider this problem in field theory and find a picture of the black hole. 

First, we briefly describe our basic idea. 
(In section \ref{s:idea}, we will provide a more detailed picture.) 
Suppose that we throw a test spherical shell or particle into an evaporating spherically symmetric black hole. 
If the black hole evaporates completely in a finite time without any singular phenomenon, 
the particle should come back after evaporation. 
Let's take a closer look at this process.  
As the particle comes close to the Schwarzschild radius of the black hole, 
it becomes ultra-relativistic and behaves like a massless particle \ci{Landau_C}. 
Then, if we take into account 
the time-dependence of the metric due to the back reaction of evaporation, 
the particle does \textit{not} enter the \textit{time-dependent} Schwarzschild radius 
but moves along an ingoing null line just outside it \ci{KMY}, 
which will be explicitly shown below \Ref{r_t}. 
After the black hole evaporates, the particle returns to the outside.  

Here, one might think that this view is strange. 
In a conventional intuition, a collapsing matter should enter soon the \textit{slowly} decreasing Schwarzschild radius 
because a typical time scale of evaporation $\Delta t\sim a^3/l_p^2$ is much larger than that of collapse $\Delta \tau \sim a$. 
Here, $a\equiv2GM$ is the Schwarzschild radius of the black hole with mass $M$, $l_p\equiv \sqrt{\hbar G}$ is the Planck length, 
$t$ is the time coordinate at infinity, and $\tau$ is that of a comoving observer along the collapsing matter. 
However, it doesn't make sense to compare these two time scales, which are measured by different clocks.
In the above argument, we have considered the both time evolutions of the spacetime and particle in a common time.  
We will see (around \Ref{r_t2}) that 
the back reaction of the evaporation plays a non-negligible role 
in determining the motion of the particle at a Planck length distance from the Schwarzschild radius.

Now, let us consider a process in which a spherical matter collapses to form a black hole. 
We can focus on the motion of each of the spherical layers that compose the matter because of the spherical symmetry. 
As a layer approaches the Schwarzschild radius that corresponds to the energy of itself and the matter inside it, 
it moves at the speed of light.  
At the same time, the time-dependent spacetime (\textit{without} a horizon structure) causes particle creation \ci{BD,Parker}
(which is so-called \textit{pre-Hawking radiation} \cite{Barcelo1,Barcelo2,KMY,KY2}), 
and the energy begins to decrease\footnote{Note here that it is essential for particle creation that the spacetime is time-dependent 
but \textit{not} that the spacetime has a horizon structure \ci{BD,Parker}.}. 
Indeed, we will show in section \ref{s:idea} that this pre-Hawking radiation 
has the same magnitude as the usual Hawking radiation. 
Then, applying the above result about the motion of the particle, 
we can see that the layer keeps falling just outside the time-dependent Schwarzschild radius.  
As this occurs for all the layers, the entire of the collapsing matter just shrinks to form a compact object, 
which is filled with the matter and radiation. 
Here, it should be noted that a strong tangential pressure occurs inside to stabilize the object against the gravitational force \ci{KMY}. 
The pressure is consistent with 4D Weyl anomaly and 
so strong that the interior is anisotropic locally (that is, the interior is not a fluid) 
and the dominant energy condition breaks down \ci{KY1,KY2,KY3}. 
Therefore, this object doesn't contradict Buchdahl's limit \ci{Bbound}.
(We will explain the origin of the pressure later.)

\begin{figure}[h]
\begin{center}
\includegraphics*[scale=0.2]{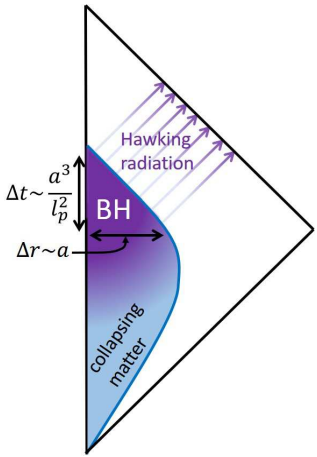}
\caption{The Penrose diagram of the black hole in quantum theory. 
The matter becomes ultra-relativistic in the final stage of the collapse, 
and particle creation occurs inside it. 
Then, a dense object is formed with the size $\D r \sim a$ and evaporates 
in the time $\D t\sim \f{a^3}{l_p^2}$.}
\label{f:Penrose}
\end{center}
\end{figure}

The object has (instead of horizon) a null \textit{surface} just outside the Schwarzschild radius 
and looks like the classical black hole from the outside, whose spatial size is $\D r \sim a$. 
Here, the surface is the boundary between the interior dense region and the exterior dilute region. 
Eventually, it evaporates in a time $\Delta t\sim a^3/l_p^2$.\footnote{\lb{foot:smallBH}
A small black hole with $a=\MO(l_p)$ should be described by some bound state in string theory 
and it may decay in a finite time. 
Therefore, we postulate that 
the remaining small part disappears in a finite time which is much smaller 
than $\D t\sim\MO(\f{a^3}{l_p^2})$.} 
This object should be the black hole in quantum theory \ci{KMY,KY1,KY2,KY3}. 
As we will see, there is no singularity.  
(No trans-Planckian quantity appears if the theory has many degrees of freedom of matter fields.) 
Therefore, the Penrose diagram is given by Fig.\ref{f:Penrose}, 
and the spacetime region of $\D r \sim a$ and $\Delta t\sim a^3/l_p^2$ corresponds to the black hole. 
In this paper, we show that this story can be realized in field theory. 

\subsection{Strategy and result}\lb{s:stra}
The above idea has been checked partially 
by a simple model \ci{KMY}, a phenomenological discussion \ci{KY1} and the use of conformal matter fields \ci{KY3}. 
It also holds for charged black holes and slowly rotating black holes \ci{KY2}. 
Furthermore, by a thermodynamical discussion, 
the entropy density inside the object is evaluated and integrated 
over the volume to reproduce the area law \ci{KY2}. 
Therefore, it seems that this picture is plausible 
and works universally for various black holes. 

However, there still remain several questions about this picture. 
What is the self-consistent state $|\psi\ket$? 
What configurations do the matter fields take inside?
How is the large tangential pressure produced inside?
Can we reproduce the entropy area law by counting microscopic states of fields?
How does the energy of the collapsing matter decrease? 
These are crucial for understanding what the black hole is and how the information of the matter comes back after evaporation. 

In this paper, to answer these questions, 
we analyze time evolution of a 4D spherical collapsing matter 
by solving the semi-classical Einstein equation 
\begin{equation}\lb{Einstein}
G_{\mu\nu}=8\pi G \bra \psi|T_{\mu\nu}|\psi\ket
\end{equation}
in a self-consistent manner, and we find the metric $g_{\mu\nu}$ and state $|\psi\ket$ 
which represent the interior of the black hole. 
Here, we treat gravity as a classical metric $g_{\mu\nu}$ 
while we describe the matter as $N$ massless free quantum scalar fields. 
$\bra\psi|T_{\mu\nu}|\psi\ket$ is the renormalized expectation value of 
the energy-momentum tensor operator in $g_{\mu\nu}$ 
that contains the contribution from both the collapsing matter and the Hawking radiation. 

\begin{figure}[h]
 \begin{center}
 \includegraphics*[scale=0.25]{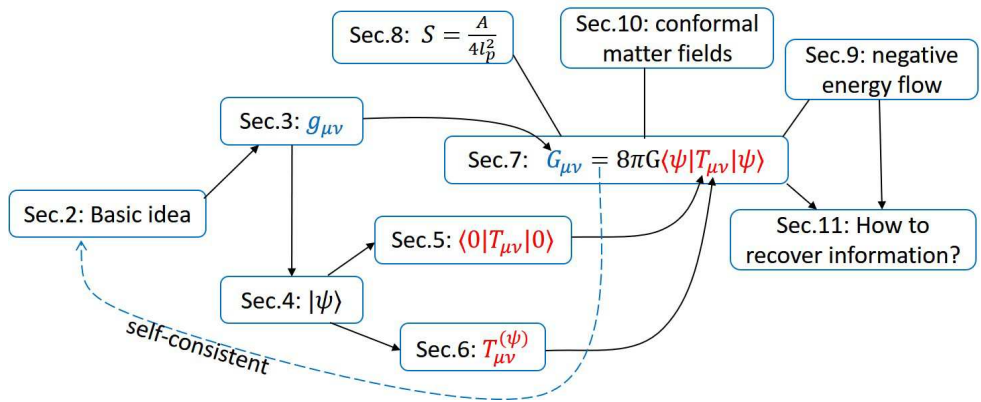}
 \caption{Flow chart of the composition of this paper.}
 \label{f:chart}
 \end{center}
 \end{figure}

We explain our self-consistent strategy and the results. 
The flow chart of Fig.\ref{f:chart} represents the composition of this paper. 
In section \ref{s:idea}, 
we first explain our basic idea in a more concrete manner by using a simple model. 
We also show how the pre-Hawking radiation occurs in the time-evolution of the system. 
In section \ref{s:metric}, 
we employ and generalize the model 
and construct a candidate metric $g_{\mu\nu}$. 
In particular, the interior metric is shown to be static as a result of dynamics. It can be expressed as 
\begin{equation}\lb{AB}
ds^2=-\f{e^{A(r)}}{B(r)}dt^2+B(r)dr^2+r^2d\Omega^2.
\end{equation}
We write down two functions $A(r), B(r)$ in terms of two phenomenological functions: 
one is the intensity of Hawking radiation $\s$ and 
the other is a function $\eta$ that provides the ratio between the radial pressure and energy density. 
We also show that this metric is locally $AdS_2\times S^2$ geometry. 

We are interested in the black holes most likely to be formed in gravitational collapse. 
As shown in section \ref{s:excite}, 
the statistical fluctuation of the mass is evaluated as $\D M \sim m_p$, 
where $m_p\equiv \sqrt{\f{\hbar}{G}}$ is the Planck mass. 
Therefore, from a macroscopic perspective, all black holes with mass $\in [M,M+m_p]$ are the same. 
We consider the set of states $\{|\psi\ket\}$ 
that represent the interior of such statistically identified black holes.

In section \ref{s:field}, 
we examine the potential energy of the partial waves of the scalar fields in the interior metric. 
Modes with angular momenta $l\geq1$ are trapped inside, 
and they emerge in the collapsing process even if they don't exist at the beginning.  
We show that 
if such bound modes are excited, the energy increases by more than $\MO\l(\f{m_p}{\sqrt{N}}\r)$, 
which means that the number of excited bound modes is at most order of $\MO(\sqrt{N})$ in the set $\{|\psi\ket\}$. 
Therefore, those modes can be regarded as the ground state because $\MO(\sqrt{N})$ 
is negligible compared to the number of total modes $\MO(\f{a^2}{l_p^2})$ (which is shown in section \ref{s:excite}). 
On the other hand, s-wave modes can enter and exit the black hole and represent the collapsing matter and Hawking radiation.
Thus, the state $|\psi\ket$ provides
\begin{equation}\lb{TTT}
\langle \psi|T_{\mu\nu}|\psi \rangle=\langle0|T_{\mu\nu}|0 \rangle+T_{\mu\nu}^{(\psi)},
\end{equation}
where $|0 \rangle$ is the ground state in the interior metric, and 
$T_{\mu\nu}^{(\psi)}$ is the contribution from the excitations of the s-waves. 

In section \ref{s:EMT}, we evaluate $\langle0|T_{\mu\nu}|0 \rangle$. 
We first solve the equation of motion of the scalar fields in the interior metric.  
We calculate the regularized energy-momentum tensor $\bra 0|T_{\mu\nu}|0\ket_{reg}$ 
in the dimensional regularization. 
Then, we renormalize the divergences and obtain the finite expectation value, $\bra 0|T_{\mu\nu}|0\ket_{ren}^\prime$. 
This contains contributions from the finite renormalization terms $(\a_0,\b_0)$, 
which correspond to the renormalized coupling constants of $R^2$ and $R_{\mu\nu}R^{\mu\nu}$ in the action, respectively.  

In section \ref{s:excite}, 
we combine Bekenstein's discussion of black-hole entropy \ci{Bekenstein} and our picture of the interior of the black hole 
to infer the form of $T_{\mu\nu}^{(\psi)}$, which is fixed by a parameter $\wt a_{\psi}$. 

In section \ref{s:sol}, we solve \Ref{Einstein} by using the ingredients obtained so far:
$g_{\mu\nu}$, $\bra 0|T_{\mu\nu}|0\ket_{ren}^\prime$, and $T_{\mu\nu}^{(\psi)}$. 
We determine the self-consistent values of $(\s,\eta,\wt a_{\psi})$ 
for a certain class of the finite renormalization terms $(\a_0,\b_0)$. 
We then check the various consistency. 
Especially, we see that there is no singularity,  
and that the vacuum fluctuation of the bound modes with $l\gg1$ creates 
the large tangential pressure $\langle0|T^\theta{}_\theta|0 \rangle_{ren}^\prime$. 

In section \ref{s:entropy}, 
we consider the stationary black hole which has grown up adiabatically in the heat bath. 
We count the number of the states $\{|\psi\ket\}$ of the s-waves inside the black hole to evaluate the entropy, 
reproducing the area law. 
This implies that the information is carried by the s-waves. 

In section \ref{s:energy}, 
to understand the mechanism by which the energy of the collapsing matter decreases, 
we assume a s-wave model for simplicity to describe the outermost region of the black hole 
and study the time evolution of quantum fields. 
We see that $|\psi\rangle$ describes the process that 
the negative ingoing energy flow created with Hawking radiation 
is superposed on the collapsing matter to decrease the total energy while the total energy density remains positive. 

Thus, three independently conserved energy-momenta appear in this solution: 
that of the bound modes of the vacuum $\langle0|T_{\mu\nu}|0 \rangle$, 
that of the collapsing matter, and that of the pair of the Hawking radiation and negative energy flow, 
where the last two contribute to $T_{\mu\nu}^{(\psi)}$. 
These three form the self-consistent configuration of the black hole. 

In section \ref{s:conformal}, 
as a special case, we consider conformal matter fields and show that 
the parameters $(\s,\eta)$ are determined by the matter content of the theory. 
Interestingly, the consistency of $\eta$ provides a condition to the matter content. 
For example, the Standard Model with right-handed neutrino satisfies the constraint but a model without it doesn't. 
Therefore, this can be regarded as a new constraint (like the weak-gravity conjecture \cite{Swamp1,Swamp2}) 
which is required in order for the black hole to evaporate. 

In section \ref{s:concl}, we conclude and discuss future directions. 
Especially, we discuss how the information comes back after evaporation 
if there are interactions between the collapsing matter, Hawking radiation and negative energy flow. 

In Appendixes, we give the derivation of various key equations. 
In particular, we explain the difference between our pre-Hawking radiation and the usual Hawking radiation 
in Appendix \ref{A:pre}. 

\section{Basic idea}\lb{s:idea}
We explain our basic idea of the black hole more precisely \ci{KMY,KY1,KY2,KY3}, 
which makes the motivation of this paper clearer. 
The discussion is composed of 3 steps. 
In step1, we examine the motion of a thin shell (with an infinitely small mass) near an evaporating black hole 
and anticipate what will be formed as a result of the time evolution of a spherical collapsing matter. 
In step2, we study the time evolution of the pre-Hawking radiation induced by the shell, 
including the effect of a small finite mass of the shell and the radiation. 
In step3, we construct a simple model (multi-shell model) to realize the prediction given in step1, 
and show that the pre-Hawking radiation has the same magnitude as that of the usual Hawking radiation. 
We also discuss a surface pressure on the shell induced by the pre-Hawking radiation. 

\subsection{Step1: Motion of a shell near the evaporating black hole}
\begin{figure}[h]
 \begin{center}
 \includegraphics*[scale=0.23]{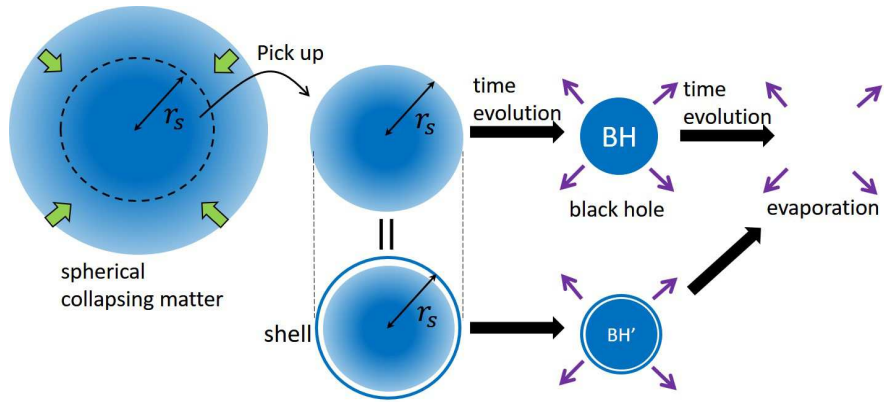}
 \caption{Time evolution of a part of a spherical collapsing matter.}
 \label{f:collapse}
 \end{center}
 \end{figure}

\lb{s:step1} 
Imagine that a spherical collapsing matter with a continuous distribution starts 
to collapse (see the left of Fig.\ref{f:collapse}). 
We pick up a part of it with a radius $r_s$. 
When the radius comes close to the Schwarzschild radius of the mass inside it, 
the entire of the part behaves lightlike. 
Then, we can discuss the time evolution of the part 
without considering the outside because the outside matter 
does not come in and has no influence to the inside due to the spherical symmetry (see the center of the upper in Fig.\ref{f:collapse}). 
We suppose that the part becomes a black hole and evaporates eventually (see the right of the upper in Fig.\ref{f:collapse}). 
Now, we consider the outermost part of the collapsing matter 
as a spherical thin shell (which is located at $r=r_s$) with an infinitely small energy (see the center of the lower in Fig.\ref{f:collapse}). 
Focusing on the motion of the shell, 
it will approach the evaporating black hole consisting of the rest (see the right of the lower in Fig.\ref{f:collapse}).
(Note again that the rest part is not affected by the shell and becomes the evaporating black hole.) 
As we show in the following, 
the black hole evaporates to disappear \textit{before} the shell catches up with ``the horizon", 
and thus the shell will never enter ``the horizon". 

Because of the spherical symmetry,
the gravitational field that the shell feels is determined by 
the energy of itself and the matter inside it, no matter what is outside the shell. 
Therefore, the metric which determines the motion of the shell near the black hole is given 
approximately by the outgoing Vaidya metric \ci{Vaidya} 
\begin{equation}\lb{SchV}
ds^2=-\l(1-\f{a(u)}{r}\r)du^2 -2 du dr + r^2 d\Om^2,  
\end{equation}
where $u$ is the null coordinate that represents an outgoing radial null geodesic 
as $u=$const. 
$M(u)\equiv \f{a(u)}{2G}$ is the energy inside the shell at time $u$ 
(including the energy of the shell itself)\footnote{\lb{foot:ADM}Note that 
the mass $M(u)$ may be different from the usual Bondi energy 
even if we neglect the matter outside the part and regard it as an isolated black hole. 
Outside the shell, there exists dilute radiation with energy density $\sim T_H^4\sim \f{\hbar}{a^4}$, where $T_H\sim \f{\hbar}{a}$. 
Then, the outside region may have energy $\sim \f{\hbar}{a^4} \times \f{4\pi a^3}{3}\sim \f{\hbar}{a}$, 
which makes the difference.
In the following, however, we use the name ``Bondi energy" and ``ADM energy" to represent the mass we are considering.}. 
The Einstein tensor has only $G_{uu}=-\f{\dot a}{r^2}$ and satisfies $G^\mu{}_\mu=0$, 
and therefore this metric can represent the outgoing null energy flow with total flux $J=4\pi r^2 \bra T_{uu}\ket$\footnote{Later, 
we will give a more proper definition of the energy flux. See \Ref{Jdef}.}
\footnote{\lb{foot:outside}One might think that the metric \Ref{SchV} is not proper to describe the region near the evaporating black hole 
because there is not an outgoing energy flow but an ingoing negative energy flow near the horizon in 2D models \ci{BD,DFU,BH_model,Honega}. 
As shown in section \ref{s:energy}, such a negative energy flow has energy density $\bra -T^t{}_t \ket\sim \MO \l(\f{1}{G a^2}\r)<0$.
We will include this contribution into the interior of the black hole. 
Therefore, by using \Ref{SchV}, we describe the exterior of the black hole approximately 
in that we neglect a dilute radiation around it with $\bra -T^t{}_t \ket \ll \MO\l(\f{1}{G a^2}\r)$.}. 
Thus, we assume that $a(u)$ decreases according to the Stefan-Boltzmann law of Hawking temperature $T_H=\f{\hbar}{4\pi a(u)}$:
\begin{equation}\lb{da}
\f{da(u)}{du}=-\f{\s}{a(u)^2}.
\end{equation}
Here, $\s$ is the intensity of the Hawking radiation, 
which is determined by dynamics of the theory. 
In general, it takes the form of $\s=kNl_p^2$, 
where $k$ is an $O(1)$ constant. 

Suppose that the shell consists of many particles. 
If a particle of them comes close to $a(u)$, the motion is governed by 
the equation for an ingoing radial null geodesic,
\begin{equation}\lb{r_t}
\f{dr_s(u)}{du}=-\f{r_s(u)-a(u)}{2r_s(u)},
\end{equation}
no matter what mass and angular momentum the particle has\footnote{See Appendix I in \ci{KY2} for the precise derivation}.
Here, $r_s(u)$ is the radial coordinate of the particle (or the shell). 

\begin{figure}[h]
 \begin{center}
 \includegraphics*[scale=0.22]{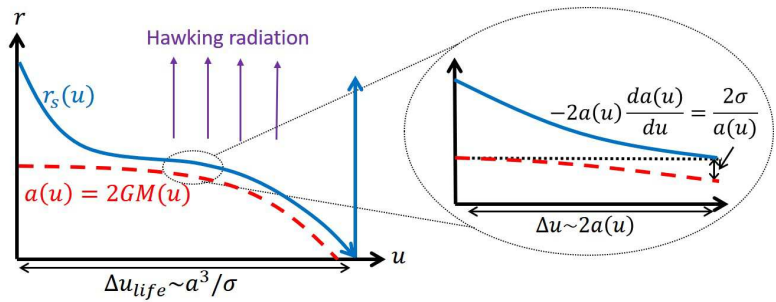}
 \caption{Motion of a particle (or a spherical shell) near the evaporating black hole. 
 $r_s(u)$ is the position of the particle (or the shell) and $a(u)$ is the Schwarzschild radius decreasing as \Ref{da}. 
  Left: The black hole evaporates before the particle reaches it, 
 and the particle comes back after evaporation. 
 Right: The particle cannot catch up with the Schwarzschild radius $a(u)$ due to the back reaction of the evaporation.}
 \label{particle}
 \end{center}
 \end{figure}

At this point, we can see a general property of \Ref{r_t}: 
Once a particle starts from a position outside $r=a(u)$, 
the particle comes close to $r=a(u)$ but does not pass it. 
Therefore, if $a(u)$ decreases to zero in a finite time, 
the particle will reach the center $(r=0)$ in a finite time and return to $r\to \infty$. 
See the left of Fig.\ref{particle}. 
Note that the radius $a(u)$ becomes zero in a finite time ($\D u\sim \f{a^3}{\s}$), 
and the time coordinate $u$ describes the outside spacetime region $(r>a(u))$ globally.

Here, we stress that we have no coordinate singularity in this analysis. 
First, we note that the Vaidya metric \Ref{SchV} with \Ref{da} is 
the metric around the trajectory $r_s(u)$ of the particles that compose the outermost shell of 
the collapsing matter (remember the center of the lower in Fig.\ref{f:collapse}). 
It seems that the metric has coordinate singularity at $r=a(u)$. 
However, because $a(u)$ is assumed to become zero in a finite time, 
the particle moves along \Ref{r_t} and stays always outside $r=a(u)$. 
In particular, when $a(u)$ becomes zero, $r_s(u)$ is still positive. 
After that, $r_s(u)$ propagates in the flat space and reaches $r=0$ in a finite time\footnote{If $a$ were constant, 
the particle would keep approaching $r=a$ for an infinite amount of time $\D u=\infty$, which means the coordinate singularity of $u$ at $r=a$. }. 
Actually, as we will see in \Ref{dl}, 
$r_s(u)$ is always apart from $r=a(u)$ at least by the proper length $\sim \sqrt{N}l_p$, 
which is physically long if $N$ is large (see the discussion below \Ref{largeN}).
Of course, at the final stage of the evaporation, which exceeds the semi-classical approximation, 
a curvature singularity may occur.  
However, the black hole at that moment has only a few of Planck mass 
and should be considered as a stringy object (see also footnote \ref{foot:smallBH}). 
(We can show that even if we used \Ref{SchV} to describe such a final stage, 
$r_s(u)$ and $a(u)$ would never become zero at the same time. See Appendix \ref{A:r_a}.) 
Thus, the particle keeps moving outside $r=a(u)$ without coordinate singularity. 
Finally, we emphasize that 
when we consider a particle that starts to fall before the shell we have focused, 
the metric around the trajectory of the particle is not the Vaidya metric with $a(u)$. 
Rather, it is given by another Vaidya metric with the Schwarzschild radius of the energy of the matter inside 
the shell to which the particle belongs. 
See the following discussion and subsection \ref{s:step3}.

Let us examine more specifically where $r_s(u)$ will approach 
when $a(u)$ evolves according to \Ref{da}.
We are interested in the difference $\Delta r(u)\equiv r_s(u)-a(u)$, 
which is much smaller than $a(u)$, $\Delta r(u)\ll a(u)$. 
Then, $r_s(u)$ in the denominator of \Ref{r_t} can be replaced with $a(u)$ approximately, 
and \Ref{r_t} becomes 
\begin{equation}\lb{r_t2}
\f{d\Delta r(u)}{du}\approx-\f{\Delta r(u)}{2a(u)}-\f{da(u)}{du}.
\end{equation}
The first term in the r.h.s. is negative, which is the effect of collapse, 
and the second one is positive due to \Ref{da}, which is the effect of evaporation. 
The second term is negligible when $\Delta r(u)\sim a(u)$, 
but it becomes comparable to the first term when the particle is so close to $a(u)$ that 
$\Delta r(u)\sim \f{l_p^2}{a(u)}$. 
Then, the both terms are balanced so that the r.h.s of \Ref{r_t2} vanishes, 
and we have $\Delta r(u)=-2a(u)\f{da(u)}{du}$. 
This means that any particle moves asymptotically as 
\begin{equation}\lb{r_t3}
r_s(u)\to a(u)-2a(u)\f{da(u)}{du}=a(u)+\f{2\s}{a(u)},
\end{equation}
and so does the shell. 
By solving \Ref{r_t2} explicitly, 
we can check that this approach occurs exponentially 
in the time scale $\D u\sim 2a$ (see Appendix \ref{A:particle}). 

This behavior can be understood as follows. See the right of Fig.\ref{particle}. 
The particle approaches the radius $a(u)$ in the time $\D u\sim 2a$. 
During this time, the radius $a(u)$ itself is slowly shrinking as \Ref{da}. 
Therefore, $r_s(u)$ cannot catch up with $a(u)$ completely and is always apart from $a(u)$ by $-2a\f{d a}{du}=\f{2\s}{a}$. 

Thus, considering the time evolution of both the particle and spacetime together, 
we have reached the conclusion that any particle never enter ``the horizon". 
Therefore, the shell (consisting of the particles) will move as \Ref{r_t3} just outside ``the horizon".

\begin{figure}[h]
 \begin{center}
 \includegraphics*[scale=0.25]{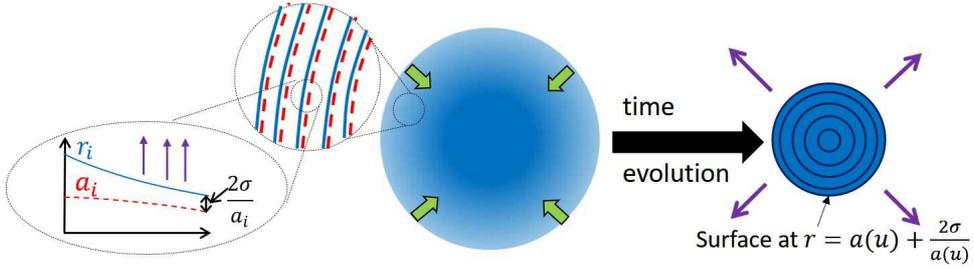}
 \caption{Left: The collapsing matter can be considered as consisting of many shells. 
 Right: It evolves to a dense object with a surface at $r=a(u)+\f{2\s}{a(u)}$. }
 \label{f:idea}
 \end{center}
 \end{figure}

Because the above argument holds for any radius $r_s$ (recall the center of the upper in Fig.\ref{f:collapse}), 
we can imagine that the entire matter consists of many spherical thin shells. 
See Fig.\ref{f:idea}. 
That is, when we focus on any part of it with radius $r_i$ and mass $\f{a_i}{2G}$, 
the shell (particle) at $r_i$ moves asymptotically as $r_i\to a_i +\f{2\s}{a_i}$ 
if the inner part evaporates as $\f{da_i}{du_i}=-\f{\s}{a_i^2}$ (see the left of Fig.\ref{f:idea}). 
Here, $u_i$ is the local time just outside the shell. 
If this happens in the whole part of the matter, it implies that 
many shells pile up and form a dense object with the total mass $M$ 
(see the right of Fig.\ref{f:idea}.) 
The object has, instead of a horizon, a \textit{surface} as 
the boundary between the high-density interior and the low-density exterior at 
\begin{equation}\lb{R}
r=a(u)+\f{2\s}{a(u)}\equiv R(a(u)),
\end{equation}
where the total radius $a(u)$ decreases as \Ref{da}. 
This object looks like a conventional black hole from the outside because $\f{2\s}{a}\ll a$, 
while it is not vacuum and has an internal structure. 
Eventually, it evaporates in $\D u \sim \f{a^3}{\s}$, and the Penrose diagram should be represented as in Fig.\ref{f:Penrose}. 
This should be the black hole in quantum theory. 
In the following 2 steps, we will gradually describe this picture precisely.

Before going to the next step, one might wonder here if the above idea can be realized in field theory or not. 
To see this point simply, 
we examine the distance $\D r=\f{2\s}{a}$ more because it is the typical length scale in this picture. 
The proper length $\D l$ is evaluated as
\footnote{For the general spherically-symmetric metric in the $(u,r)$ coordinate,
the proper radial length is given by $\D l = \sqrt{g_{rr}-\f{(g_{ur})^2}{g_{uu}}}\D r$ \ci{Landau_C}.}
\begin{equation}\lb{dl}
\Delta l= \sqrt{\f{R(a)}{R(a)-a}}\f{2\s}{a}\approx \sqrt{2\s},
\end{equation}
where \Ref{SchV} and \Ref{R} have been used. 
We here assume that $N$ is large but finite, for example, $\MO(1000)$:  
\begin{equation}\lb{largeN}
\s\sim N l_p^2\gg l_p^2.
\end{equation}
Then, $\Delta l$ is larger than $l_p$, 
and $\Delta r=\f{2\s}{a}$ is sufficiently long in order for the effective local field theory to be valid. 
In particular, the position of the surface \Ref{R} is meaningful physically. 
A way to show the validity of the field theory more precisely 
is to construct a concrete solution by solving \Ref{Einstein} and confirm its self-consistency. 
From section \ref{s:metric}, we will do it. 
\subsection{Step2: Pre-Hawking radiation}\lb{s:step2} 
In the step1, we have neglected the effect of the mass of the shell and 
considered only the Hawking radiation from the black hole inside the shell. 
In this step, we will take into account the effect of the mass and a pre-Hawking radiation. 

\begin{figure}[h]
\begin{center}
\includegraphics*[scale=0.22]{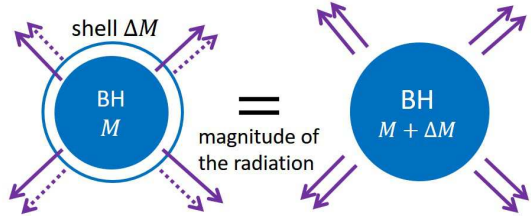}
\caption{Statement of step 2. 
The total magnitude of the pre-Hawking radiation (dashed arrows)
from the shell of $\D M$ and the usual Hawking radiation (solid arrows) from the black hole is equal to 
that of the usual Hawking radiation (solid arrows) from a larger black hole of mass $M+\D M$.} 
\label{f:shellBH}
\end{center}
\end{figure}

Suppose that we add a spherical thin null shell with a small but \textit{finite} 
energy $\Delta M$ to the black hole with mass $M$ evaporating as \Ref{da}
\footnote{Here, we consider the shell infinitely thin. }. 
From now, we will show that a pre-Hawking radiation is induced by this shell 
so that the total magnitude of the pre-Hawking radiation and the Hawking radiation from the black hole is equal to 
that of the Hawking radiation from a larger black hole of mass $M+\D M$ (up to $\MO\l( \f{\D M}{M^{3}} \r)$ corrections). 
See Fig.\ref{f:shellBH}. 
Therefore, the total system of the black hole and the shell behaves like a larger black hole. 
(This provides a more precise description of the right of the lower in Fig.\ref{f:collapse}.)

\subsubsection{Setup}
Although we will start a 4D complete analysis from section 3, 
here for simplicity we consider only s-waves of $N$ massless scalar fields. 
For example, by using the conservation law with 2D Weyl anomaly \ci{DFU,Fulling,BD,BH_model}, 
we can evaluate the outgoing flux from the center black hole as  
\begin{equation}\lb{J_2D}
J\equiv 4\pi r^2 \bra T_{uu}\ket= \f{N\hbar}{192\pi a(u)^2}.
\end{equation}
Comparing this to \Ref{da}, we see the intensity of s-wave Hawking radiation: 
\begin{equation}\lb{sigma_s}
\s_s = \f{N l_p^2}{96\pi}.
\end{equation}

\begin{figure}[h]
\begin{center}
\includegraphics*[scale=0.21]{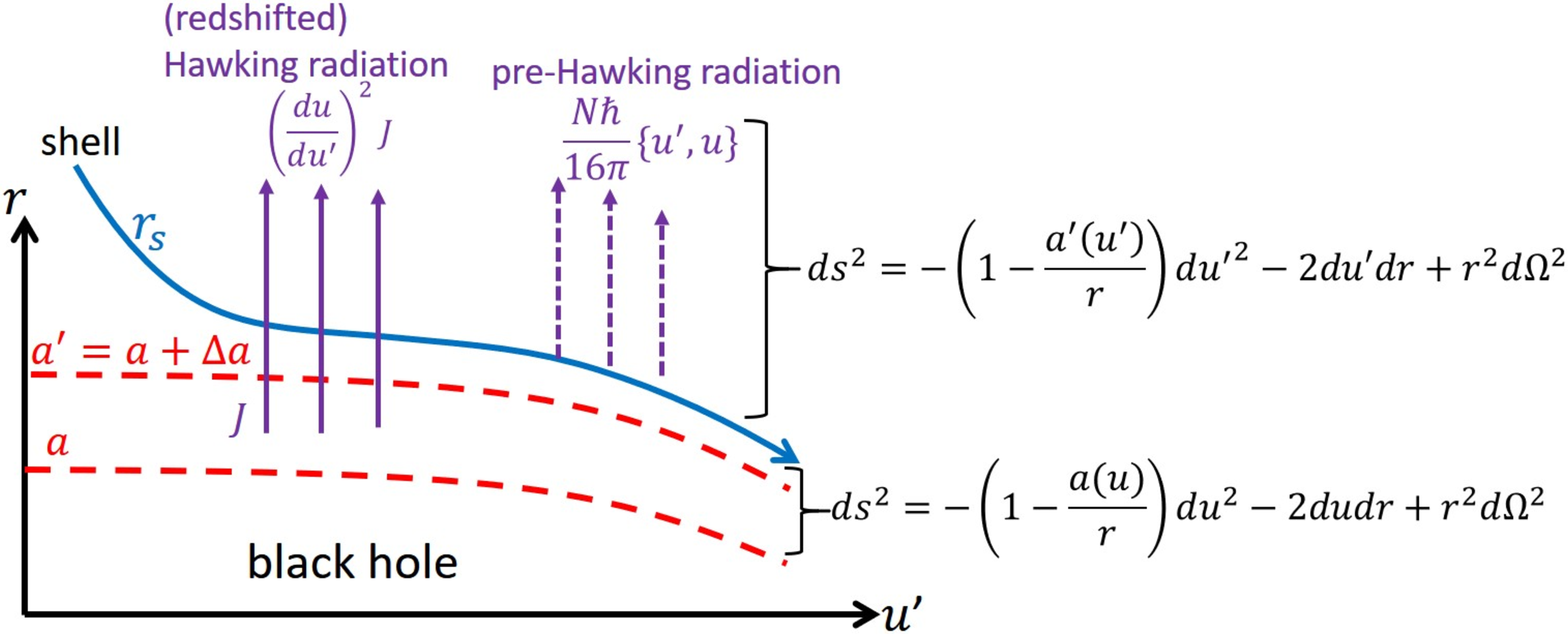}
\caption{The system of the shell and the evaporating black hole. 
The pre-Hawking radiation (dashed arrows) is induced by the shell and added to 
the Hawking radiation (solid arrows) from the black hole.} 
\label{f:coreshell}
\end{center}
\end{figure}

We are now interested in the situation 
where the shell comes to the black hole as in Fig.\ref{f:coreshell}. 
Because of the spherical symmetry, 
the region between the black hole and the shell is still described by \Ref{SchV} with \Ref{da} of \Ref{sigma_s}
\footnote{Here, we neglect the effect of scattering between the shell and Hawking radiation. 
This effect will be considered in subsection \ref{s:f}}, 
while the region above the shell is expressed by another Vaidya metric
\begin{equation}\lb{Vaidya_out}
ds'^2=-\l(1-\f{a'(u')}{r}\r)du'^2 -2 du' dr + r^2 d\Om^2. 
\end{equation}
Here, $a'\equiv 2G (M+\D M)$ is the Schwarzschild radius of the total mass, 
and the time $u'$ is different from $u$ due to the mass of the shell $\D M$. 

The evolution equation of $a'(u')$ is given by the energy conservation, 
$\f{da'}{du'}=-2G J'$, where $J'=4\pi r^2 \bra T_{u'u'}\ket|_{r\gg a}$ is the total flux coming out of the total system.  
In general, particles are created in a time-dependent spacetime of a collapsing matter. 
It is well-known that we can formulate the s-wave approximated energy flux of 
the particles as \cite{Fulling,Stro,Iso} 
\begin{equation}\lb{JJ'}
J=\f{N\hbar}{16\pi}\{u,U\},~~J'=\f{N\hbar}{16\pi}\{u',U\}.
\end{equation}
Here, $U$ is the outgoing null time of the flat spacetime before the collapse, 
and $\{x,y\}\equiv \f{\ddot y^2}{\dot y^2}-\f{2}{3}\f{\dddot y}{\dot y}$ 
is the Schwarzian derivative for $y=y(x)$. 
(See above \Ref{J_Sch} for a more explanation.)
Using a formula \ci{CFTbook}
\begin{equation}\lb{Sch_formula}
\{u',U\}=\l(\f{du}{du'}\r)^2\{u,U\}+\{u',u\},
\end{equation}
and applying \Ref{J_2D} and \Ref{sigma_s}, we have 
\begin{equation}\lb{J'}
J'= \l(\f{du}{du'}\r)^2 \f{\s_s}{2G a^2} + \f{\hbar N}{16\pi}\{u',u\}.
\end{equation}
This has a physical interpretation. See Fig.\ref{f:coreshell}. 
The first term represents the energy flux from the black hole, \Ref{J_2D}, 
that is redshifted due to the mass of the shell,
and the second one corresponds to the radiation induced by the shell. 
We call this term pre-Hawking radiation because any horizon structure is not relevant to $\{u',u\}$. 
Thus, the total of the redshifted Hawking radiation and the pre-Hawking radiation 
determines time evolution of the energy $\f{a'}{2G}$ of the total system: 
\begin{equation}\lb{a'eq_ano}
\f{da'}{d u'} = - \l(\f{du}{du'}\r)^2 \f{\s_s}{a^2}-12 \s_s\{u',u\},
\end{equation}
which is nothing but the $u'u'$-component of the semi-classical Einstein equation. 

To relate $u$ and $u'$ each other, 
we use a fact that the trajectory of the shell, $r=r_s(u)$, is null from both sides of the shell 
to get the connection conditions:\footnote{Israel's junction condition will be discussed in subsection \ref{s:step3}.}
\begin{equation}\lb{con_uu}
\f{r_s-a}{2r_s}du =-dr_s =\f{r_s-a'}{2r_s}du'.
\end{equation}
This is equivalent to \Ref{r_t} and 
\begin{equation}\lb{con_uu2}
\f{d u'}{d u}=\f{r_s-a}{r_s-a'}.
\end{equation}
Now, suppose that the shell is close to $a$ so that \Ref{r_t3} holds. 
Then, we can evaluate \Ref{con_uu2} as 
\begin{equation}\lb{con_uu3}
\f{d u'}{d u}=\l(1-\f{\D a}{r_s-a}\r)^{-1}\approx 1+\f{a\D a}{2\s_s},
\end{equation}
where we have used \Ref{r_t3} with $\s$ replaced by $\s_s$ and assumed that 
$\D a \equiv a'-a \ll \MO(\f{\s}{a})$. 
\subsubsection{Time evolution of the pre-Hawking radiation}
We examine how the pre-Hawking radiation occurs 
by solving directly the coupled time-evolution equations \Ref{a'eq_ano} and \Ref{con_uu3}. 
Here we consider linearized equations for $\D a$. 

In the linear order of $\D a$, from \Ref{con_uu3} we have 
\begin{equation}\lb{Sch_Da}
\{u',u\}=\f{a}{3\s_s} \D \ddot a\l(1+\MO\l(\f{\s_s}{a^2}\r)\r), 
\end{equation}
where the dot stands for the $u$ derivative (e.g. $\D \ddot a=\f{d^2 \D a}{du^2}$). 
Note that once the quantity is linear in $\D a$, 
we no longer need to distinguish $u$ derivative from $u'$ derivative because of \Ref{con_uu3}. 
In order to get \Ref{Sch_Da}, we have used an identity for the Schwarzian derivative 
\begin{equation}\lb{u'_eqs}
\{u',u\}
=-\f{1}{3}\l(\f{du}{du'}\r)^2( \dot q^2 -2 \ddot q),~~q(u)\equiv \log \f{du'}{du}
\end{equation}
and the fact that $\dot a \sim a\times \MO(\f{\s}{a^3})$ 
and $\D \dot a \sim \D a \times \MO(\f{1}{a})$, 
which will be checked in a self-consist manner below. 

Then, from \Ref{da} (with $\s$ replaced by $\s_s$), \Ref{a'eq_ano}, 
\Ref{con_uu3} and \Ref{Sch_Da}, we obtain 
\begin{equation}
\f{d\D a}{du'}=\f{da'}{du'}-\f{da}{du'}
=\f{du}{du'}\l(\f{du}{du'}-1\r)\f{da}{du}-12\s_s\{u',u\}
=\f{\D a}{2a}-4a \D \ddot a+\MO\l(\f{(\D a)^2}{\s_s},\f{\s_s \D a}{a^3}\r).
\end{equation}
As mentioned above, 
the $u'$ derivative in the l.h.s. can be replaced to $u$ derivative, and we reach 
\begin{equation}\lb{Da_eq}
\D \ddot a  +\f{1}{4a}\D \dot a -\f{1}{8a^2}\D a=0.
\end{equation}
This determines the time evolution of $\D a(u)$ and how the pre-Hawking radiation is emitted.  

Let us solve \Ref{Da_eq} in the time scale $\D u \sim a$, 
where $a(u)$ hardly changes because of \Ref{da}. 
Putting the ansatz $\D a(u)=C e^{\f{\g}{a} u}$ into \Ref{Da_eq}, we have
\begin{equation}
\g^2+\f{1}{4}\g - \f{1}{8}=0,
\end{equation} 
that is, $\g=-\f{1}{2},\f{1}{4}$. 
$\g=\f{1}{4}$ means that the energy of the shell increases exponentially in time, 
which is not accepted physically. 
Here, one might wonder why such an unphysical solution appears. 
The reason is that \Ref{a'eq_ano} is a higher derivative equation describing the back reaction of 
the radiation to be created in the time evolution. 
A similar problem occurs in ``Lorentz friction" (a recoil force on an accelerating charged particle 
caused by electromagnetic radiation emitted by the particle), 
where one must choose a physical solution by hand \ci{Landau_C}. 
In the present case, therefore, we select as the physical solution 
\begin{equation}\lb{Da_sol}
\D a(u)= \D a_0 e^{-\f{u}{2a}}.
\end{equation}
This indicates that, as the pre-Hawking radiation is emitted, 
the energy of the shell decreases exponentially in the time scale $\D u\sim 2a$
\footnote{This time scale is consistent with the life time of the whole black hole, $\D u \sim \f{a^3}{l_p^2}$, 
which is predicted from \Ref{da}. 
Imagine that a black hole with $a$ is made of $\MO(\f{a^2}{l_p^2})$ shells with $\D a \sim \f{l_p^2}{a}$ 
as in Fig.\ref{f:idea}. 
Then, it takes $\D u_{total} \sim 2a \times \f{a^2}{l_p^2}\sim\f{a^3}{l_p^2}$ for all the shells to evaporate according to \Ref{Da_sol}. }. 
In particular, this solution satisfies 
\begin{equation}\lb{Da_sol2}
\f{d}{du}\D a=-\f{1}{2a} \D a. 
\end{equation}

Now, we check the time evolution of $a'(u')$. 
Using $a'=a+\D a$, \Ref{da} with $\s_s$, \Ref{con_uu3}, and \Ref{Da_sol2}, 
we can evaluate 
\begin{align}\lb{total_J}
\f{da'}{du'} &=\f{du}{du'}\l(\f{da}{du}+\f{d \D a}{du} \r) \nn \\
 &= \l(1-\f{a \D a}{2\s_s} \r) \l(-\f{\s_s}{a^2}-\f{\D a}{2a} \r)+ \MO\l( \f{\s_s\D a}{a^{3}} \r)  \nonumber \\
 &=-\f{\s_s}{a^2}+ \f{\D a}{2a}- \f{\D a}{2a} + \MO\l( \f{\s_s\D a}{a^{3}} \r)\nonumber \\
 &=-\f{\s_s}{a^2}+ \MO\l( \f{\s_s\D a}{a^{3}} \r). 
\end{align}
This agrees with $\f{da'}{du'}=-\f{\s_s}{a'^2}$ up to $\MO\l( \f{\s_s\D a}{a^{3}} \r)$, 
which shows that Fig.\ref{f:shellBH} holds as a result of the time evolution. 
(In the next step we will show $\f{da'}{du'}=-\f{\s_s}{a'^2}$ including the correction of $\MO\l( \f{\s_s\D a}{a^{3}} \r)$.)
It should be noted that 
at the third line of \Ref{total_J}, 
the amount of the Hawking radiation reduced by the redshift is compensated 
by the pre-Hawking radiation\footnote{Note that $\f{\D a}{2a}$ is much larger than $\f{2\s_s}{a^3}\D a$, 
which is the difference between $\f{\s_s}{a'^2}$ and $\f{\s_s}{a^2}$.}. 
Therefore, the pre-Hawking radiation plays an essential role in \Ref{total_J}. 

\subsection{Step3: A multi-shell model}\lb{s:step3}
We have seen so far that the statement of Fig.\ref{f:shellBH} works 
up to $\MO(\f{\s}{a^2})$ corrections for \textit{any} part of the collapsing matter. 
Therefore, we can imagine that the collapsing matter consisting of many shells 
shrinks emitting the pre-Hawing radiation from each shell in the time evolution. 
In this last step, we will construct a dynamical model representing the situation and 
show under the s-wave approximation that the pre-Hawking radiation \textit{is} emitted with exactly as the same magnitude as the usual Hawking radiation. 
Then, we see that a surface pressure is induced on each shell by the pre-Hawking radiation 
and discuss its role in the time evolution. 
\subsubsection{Setup}
\begin{figure}[h]
\begin{center}
\includegraphics*[scale=0.18]{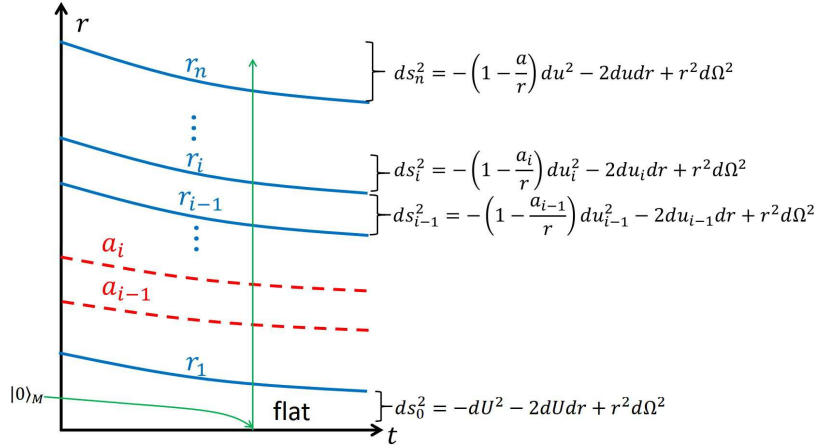}
\caption{A multi-shell model, 
which models the time evolution of the matter depicted by the left in Fig.\ref{f:idea}.}
\label{f:multi}
\end{center}
\end{figure}

\lb{s:multi}
We consider a spherical collapsing matter consisting of $n$ spherical thin null shells. 
See Fig.\ref{f:multi}, where $r_i$ represents the position of the $i$-th shell. 
Here, physically, some part of the radiation emitted from a shell can be scattered by 
the other shells or the gravitational potential, 
but we neglect the effect for simplicity. (We will introduce it in subsection \ref{s:f}.) 
Then, because of spherical symmetry, for $i=1\cdots n$, 
the region just outside the $i$-th shell can be described 
by the outgoing Vaidya metric: 
\begin{equation}\lb{Sch_i}
ds^2_i =-\l(1-\f{a_i(u_i)}{r}\r)du_i^2-2du_idr+r^2d\Omega^2.
\end{equation}
Here, $u_i$ is the local time, $a_i=2Gm_i\gg l_p$, and 
$m_i$ is the energy inside the $i$-th shell (including the contribution from the shell itself). 
For $i=n$, $u_n=u$ is the time coordinate at infinity, 
and $a_n=a=2GM$, where $M$ is the total mass. 
That is, the outside is given by \Ref{SchV}, but we do \textit{not} assume \Ref{da}. 
On the other hand, the center, which is below the $1$-st shell, is the flat spacetime,
\begin{equation}\lb{flatU}
ds^2=-dU^2-2dUdr+r^2d\Omega^2,
\end{equation}
because of the spherical symmetry. Therefore, we can regard that
\begin{equation}\lb{flat2}
a_0=0,~~~u_0=U.
\end{equation}

To set the time evolution equations of $a_i(u_i)$, 
we consider how particle creation occurs in this time-dependent spacetime \ci{Hawking}. 
Suppose that 
the quantum fields start in the Minkowski vacuum state $|0\ket_M$ from a distance. 
They come to and pass the center as the green arrow in Fig.\ref{f:multi}.  
Then, they will propagate through the matter and 
be excited by the curve metric to create particles, which corresponds to the pre-Hawking radiation. 
For example, 
by solving the field equation in the eikonal approximation and using the point-splitting regularization \ci{KMY,KY2}, 
we can evaluate the total outgoing energy flux observed just above the $i$-th shell as 
\begin{equation}\lb{J_Sch}
J_i\equiv 4\pi r^2 {}_M\bra 0|T_{u_iu_i}|0\ket_M = \f{N\hbar}{16\pi}\{u_i,U\}
\end{equation}
for $i=1,2,\cdots,n$. 
Therefore, we have for each $i$
\begin{equation}\lb{a_eq_i}
\f{da_i}{du_i}=-2G J_i=-12 \s_s \{u_i,U\}.
\end{equation}

To complete the setup, we need to connect time coordinates, $u_n=u, u_{n-1},\cdots, u_0=U$. 
Following the idea used to get \Ref{con_uu}, we have the connection conditions at $r=r_i$:
\begin{equation}\lb{connect_i}
\f{r_i-a_i}{r_i}du_i =-2dr_i =\f{r_i-a_{i-1}}{r_i}du_{i-1}~~~{\rm for}~i=1\cdots n.
\end{equation}
This is equivalent to 
\begin{equation}\lb{r_t_i}
\f{dr_i(u_i)}{du_i}=-\f{r_i(u_i)-a_i(u_i)}{2r_i(u_i)} 
\end{equation}
and 
\begin{equation}\lb{t_i}
\f{\rd u_i}{\rd u_{i-1}}=\f{r_i-a_{i-1}}{r_i-a_i}=1+\f{a_i-a_{i-1}}{r_i-a_i}.
\end{equation}
\Ref{r_t_i} determines $r_i(u_i)$ for a given $a_i(u_i)$, 
and then the solution $r_i(u_i)$ and \Ref{t_i} give the relation between $\{u_i\}$. 
Thus, the coordinates are connected smoothly. 
(Israel's junction condition will be checked later.)

\subsubsection{Check of the pre-Hawking radiation}\lb{s:multi}
In order to solve the coupled equations \Ref{a_eq_i} and \Ref{connect_i}, 
let us take the continuum limit by $\Delta M_i \equiv \f{a_i-a_{i-1}}{2G}\to0$. 
Especially, we focus on a configuration in which 
the following ansatz holds:\footnote{A more general case is discussed in \ci{Ho3}.} 
\begin{align}\lb{An_ai}
\f{da_i}{du_i}&=-\f{C}{a_i^2},\\
\lb{An_ri}
r_i &=a_i-2a_i \f{da_i}{du_i}=a_i+ \f{2C}{a_i}.
\end{align}
These can be justified by the result of step 2: 
From the outside of the $i$-th shell, 
the system consisting of the shell and its inside behaves 
like the ordinary evaporating black hole as in \Ref{da}, 
and the shell comes close to the asymptotic position as in \Ref{r_t3}. 

In the following, we show that the ansatz \Ref{An_ai} and \Ref{An_ri} 
gives a solution to the coupled equations to be solved, \Ref{a_eq_i}, \Ref{r_t_i} and \Ref{t_i}. 
First of all, \Ref{An_ri} is nothing but the asymptotic solution of \Ref{r_t_i} under the assumption of \Ref{An_ai}. 
As we will see below, there exists a $C$ that makes \Ref{a_eq_i} and \Ref{t_i} be satisfied. 

We first solve the equations \Ref{t_i}.
By introducing 
\begin{equation}\lb{def_eta}
\xi_i\equiv \log \f{\rd U}{\rd u_i},
\end{equation}
we have 
\begin{align}\lb{derive_xi}
\xi_i-\xi_{i-1} &=\log \f{ \f{\rd U}{\rd u_i} }{ \f{\rd U}{\rd u_{i-1}}} 
=- \log \f{\rd u_i}{\rd u_{i-1}}\nn\\
 &=-\log \l( 1+\f{a_i-a_{i-1}}{r_i-a_i} \r)  \nn\\
 &\approx -\f{a_i-a_{i-1}}{r_i-a_i} 
 =-\f{a_i-a_{i-1}}{\f{2C}{a_i}} \nn\\
 &\approx - \f{1}{4C} \l(a_i^2-a_{i-1}^2\r)~.
\end{align}
Here, at the second line, we have used \Ref{t_i}; 
at the third line, we have used \Ref{An_ri} and assumed $\f{a_i-a_{i-1}}{\f{2C}{a_i}}\ll 1$; 
and at the last line, we have approximated $2a_i\approx a_i+a_{i-1}$. 
These approximations become exact in the continuum limit. 
Then, with the condition \Ref{flat2}, we obtain
\begin{equation}\lb{xi_i}
\xi_i =-\f{1}{4C}a_i^2.
\end{equation}

Next, we consider \Ref{a_eq_i}. We first note that 
from the definitions of the Schwarzian derivative and $\xi_i$ \Ref{def_eta}, 
we have 
\begin{equation}\lb{Sch_xi_i}
\{u_i,U\}=\f{1}{3} \l(\f{d \xi_i}{d u_i}\r)^2 -\f{2}{3} \f{d^2 \xi_i}{d u^2_i}.
\end{equation}
On the other hand, from \Ref{xi_i} and \Ref{An_ai}, we have  
\begin{equation}\lb{dxi}
\f{d\xi_i}{du_i}=-\f{1}{2C}a_i \f{da_i}{du_i}=\f{1}{2a_i}.
\end{equation}
Therefore, we have 
\begin{equation}\lb{uiU}
\{u_i,U\}\approx \f{1}{12 a_i^2},
\end{equation}
where the higher terms in $\f{l_p}{a_i}$ have been neglected for $a_i\gg l_p$. 
Using this, \Ref{a_eq_i}, and \Ref{An_ai}, we find $C=\s_s$. 
That is, the solution tells that 
the system inside (including) each shell evaporates emitting the pre-Hawking radiation 
with the same magnitude as the usual Hawking radiation: 
\begin{equation}\lb{da_i}
\f{da_i}{du_i}=-\f{\s_s}{a_i^2}.
\end{equation}
Note again that this is not an assumption but the result of solving 
the semi-classical Einstein equation in the s-wave approximation \ci{KMY,KY2}. 
Thus, we have seen in the s-wave approximation that 
the collapsing matter becomes a dense object with the surface \Ref{R} and evaporates as in Fig.\ref{f:idea}. 
The pre-Hawking radiation is emitted from each shell, 
and the sum of them comes out of the surface of the object just like the usual Hawking radiation. 
In the following, we will use the term ``Hawking radiation" to mean the pre-Hawking radiation
\footnote{If we want to emphasize or distinguish between the two, we explicitly say ``pre-Hawking radiation".}. 
(In Appendix \ref{A:pre} we compare this pre-Hawking radiation to the usual Hawking radiation \ci{Hawking}.) 

Here, one might wonder how the Planck-like distribution with the Hawking temperature is obtained 
in this pre-Hawking radiation. 
Suppose that we want to evaluate the distribution of the particles created around a time $u=u_*$. 
Because $a(u)$ changes slowly as \Ref{da_i}, 
we can approximate it as $a(u)\approx a_* -\f{\s_s}{a^{2}_*}(u-u_*)$ (where $a_*=a(u_*)$), 
which leads to $a(u)^2\approx a^{2}_* -\f{2\s_s}{a_*}(u-u_*)$. 
Using this, \Ref{def_eta} and \Ref{xi_i} with $C=\s_s$, we have 
\begin{equation}\lb{affines}
\f{dU}{du}\approx e^{-\f{a^{2}_*}{4\s_s}+ \f{1}{2a_*}(u-u_*)}=D_* e^{\f{1}{2a_*}u}.
\end{equation}
Note that $U$ is the affine parameter for outgoing null modes in the initial flat space, 
and $u$ is that in the asymptotically flat region of the spacetime after the formation of the black hole.  
In general, such an exponential relation between the two affine parameters
plays an essential role in obtaining the Planck-like distribution of the Hawking temperature \ci{Barcelo1,Barcelo2}. 
In fact, \Ref{affines} leads to the Planck-like distribution of the 
time-dependent Hawking temperature \ci{KMY,KY2}
\footnote{In the usual derivation of Hawking radiation \ci{Hawking,BD}, 
the exponential part contains a factor like $-\f{1}{2a}u$. 
The corresponding part of \Ref{affines} has opposite sign, but indeed it gives \Ref{T_H}. See \ci{KMY,KY2}.}
\begin{equation}\lb{T_H}
T_H(u)=\f{\hbar}{4\pi a(u)}.
\end{equation}
Therefore, it is possible to discuss black-hole thermodynamics by using the pre-Hawking radiation
(without any horizon structure) and consider the entropy (see section \ref{s:entropy}). 

\subsubsection{Surface pressure induced by the pre-Hawking radiation}
We check the junction condition along the trajectory of each shell 
and study why the matter does not collapse. 
When two different metrics are connected at a null hypersurface, 
a surface energy-momentum tensor exists on it generically. 
By using the Barrabes-Israel formalism \ci{Israel_null,Poisson}, we can calculate 
the surface energy density $\e_{2d}^{(i)}$ and the surface pressure $p_{2d}^{(i)}$ 
on the $i$-th shell as (see Appendix F of \ci{KY2} for the derivation) 
\begin{equation}\lb{EMT_s_i}
 \epsilon_{2d}^{(i)} = \frac{a_i-a_{i-1}}{8\pi G r_i^2},~~~
 p_{2d}^{(i)} = -\frac{r_i}{4\pi G(r_i-a_i)^2}\l(\f{da_i}{du_i}-\l(\f{r_i-a_i}{r_i-a_{i-1}}\r)^2\f{da_{i-1}}{du_{i-1}}\r).
\end{equation}
Naturally, $\e_{2d}^{(i)}$ expresses the energy density of the shell with energy $\D M_i=\f{a_i-a_{i-1}}{2G}$ and area $4\pi r_i^2$. 
Using \Ref{a_eq_i} and \Ref{t_i} and then applying \Ref{Sch_formula} to $(U,u,u')\to(U,u_{i-1},u_i)$, 
we can express $p_{2d}^{(i)}$ as 
\begin{equation}\lb{p2d0}
 p_{2d}^{(i)} = \frac{3\s_sr_s}{\pi G(r_s-a')^2}\{u_i,u_{i-1}\},
\end{equation}
which shows through \Ref{J'} that the pressure is induced by the pre-Hawking radiation $\f{\hbar N}{16\pi}\{u_i,u_{i-1}\}$. 

Let us evaluate $p_{2d}^{(i)}$. 
We start with the expression of \Ref{EMT_s_i}. 
Using \Ref{An_ri} with $C=\s_s$ and \Ref{da_i} and performing a similar calculation to \Ref{total_J}, 
we obtain 
\begin{equation}\lb{p2d}
 p_{2d}^{(i)} \approx \f{a_i^2\D a_i}{16\pi G \s_s^2},
\end{equation}
for $\D a_i=a_i-a_{i-1}$. 
Therefore, the pressure is positive and strong (even for $\D a_i \sim \f{l_p^2}{a}$). 
This means that, as the shell contracts, 
the pressure works to resist the gravitational force while the pre-Hawking radiation is dissipated. 

We can understand the origin of this pressure from a 4D field-theoretic point of view. 
Indeed, the pressure appears naturally from conservation law and 4D Weyl anomaly \ci{KY3}.
In the following sections, 
we will calculate directly the expectation value of the energy-momentum tensor 
and show that the vacuum fluctuation of the bound modes with $l\gg1$ creates the pressure. 

\section{Construction of the candidate metric}\lb{s:metric}
From now, we start a full 4D self-consistent discussion,  
and show that the basic idea works as a solution of the Einstein equation \Ref{Einstein}. 
In the present and following sections (except for section \ref{s:energy}), 
we do \textit{not} employ the s-wave approximation used in the previous section, 
but we consider the full 4D dynamics of quantum fields. 

In this section, we use and generalize the multi-shell model in the previous section 
and construct a candidate metric for the interior \ci{KMY,KY1,KY2,KY3}.
Note that at this stage we do \textit{not} mind 
if the metric is the solution of \Ref{Einstein} or not. 
In section \ref{s:sol} we will show that indeed it is. 

\subsection{The interior metric}
We use the multi-shell model of Fig.\ref{f:multi} and construct a candidate metric for the interior of the object. 
To do that, we suppose again that in the continuum limit the ansatz 
\begin{equation}\lb{R_i}
\f{da_i}{du_i}=-\f{\s}{a_i^2},~~r_i =a_i+ \f{2\s}{a_i}
\end{equation}
hold for an intensity $\s$. 
Then, we can use these and \Ref{t_i} to obtain (like \Ref{derive_xi})
\begin{equation}\lb{xi_ii}
\xi_i=-\f{1}{4\s}a_i^2.
\end{equation}
Note that in the previous section we have used the s-wave approximated Einstein equation \Ref{a_eq_i} to obtain $C=\s_s$ in \Ref{da_i}, 
but we are now trying to use the full 4D Einstein equation 
to identify the self-consistent value of $\s$ (see section \ref{s:sol}). 

Now, the metric at a spacetime point $(U,r)$ inside the object 
is obtained by considering the shell that passes the point  
and evaluating the metric \Ref{Sch_i}. 
We have at $r=r_i$ 
\begin{align}\lb{evalu1}
\f{r-a_i}{r} &= \f{r_i-a_i}{r_i}=\f{\f{2\s}{a_i}}{r_i}\approx \f{2\s}{r^2} \\
\lb{evalu2}
\f{du_i}{dU} &=e^{-\xi_i}=e^{\f{a_i^2}{4\s}}\approx e^{\f{r^2}{4\s}}, 
\end{align}
where \Ref{R_i} and \Ref{xi_ii} have been used. 
From these, we can obtain the metric 
\begin{align}\lb{Ur}
ds^2 &=-\f{r-a_i}{r}du_i^2-2du_idr+r^2d\Omega^2  \nn\\
 &= -\f{r_i-a_i}{r_i}\l(\f{du_i}{dU} \r)^2dU^2-2\f{du_i}{dU}dUdr+r^2_id\Omega^2\nonumber \\
 &\approx - \f{2\s}{r^2}e^{\f{r^2}{2\s}} dU^2 -2e^{\f{r^2}{4\s}}dUdr+r^2 d\Omega^2. 
\end{align}
Note that this is static although 
each shell is emitting the pre-Hawking radiation and shrinking physically. 

\begin{figure}[h]
\begin{center}
\includegraphics*[scale=0.23]{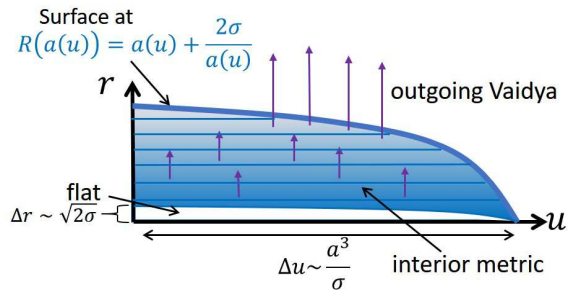}
\caption{Time evolution of the evaporating black hole.}
\label{f:evo}
\end{center}
\end{figure}

Thus, our candidate metric for the evaporating black hole is given by \ci{KMY} 
\begin{equation}\lb{metric1V}
d s^2=\begin{cases}
 - \f{r-a(u)}{r}d u^2-2 dudr + r^2 d \Omega^2,~~{\rm for}~~R(a(u))\leq r~,\\
-\f{2\s}{r^2} e^{- \f{R(a(u))^2-r^2}{2\s}} d u^2 -2 e^{- \f{R(a(u))^2-r^2}{4\s}} du dr + r^2 d \Omega^2,~~{\rm for}~~\sqrt{2\s}\lesssim r\leq R(a(u))~,
\end{cases}
\end{equation}
under the assumption that the scattering effects are neglected. See Fig.\ref{f:evo}. 
This metric is continuous at the null surface located 
at $r=R(a(u))=a(u) + \f{2\s}{a(u)}$, 
where the total Schwarzschild radius $a(u)$ decreases as \Ref{da}. 
Here, we have converted $U$ in \Ref{Ur} to $u$ in \Ref{SchV}
by the relation $dU=e^{-\f{R(a(u))^2}{4\s}}du$, 
which can be obtained by \Ref{def_eta} and \Ref{xi_ii}. 

In the construction, we have started from the vacuum spacetime and piled up many shells. 
The flat region around $r=0$ with width $\D r\sim \sqrt{\s}$ 
is extremely delayed due to the large redshift, and is still flat because of the spherical symmetry. 
Around $r=\sqrt{2\s}$, 
the interior metric \Ref{Tr} takes the same form as the flat metric \Ref{flat}. 
Then, for convenience, we assume that the region $0\leq r\lesssim \sqrt{2\s}$ is flat. 
Note that the choice of $r=\sqrt{2\s}$ is not essential to the following discussion 
because we can redefine the time coordinate $T$ (or $U$) 
and select another point, say, $r=\sqrt{4\s}$. 

As shown in Fig.\ref{f:evo}, 
this object evaporates emitting the Hawking radiation through the null surface 
although the interior is static. 
Thus, the whole system is time-dependent. 

This object has the surface at $r=R(a(u))$, instead of a horizon. 
We can also check that there is no trapped region inside. 
However, the redshift is exponentially large inside, and 
time is almost frozen in the region deeper than the surface by $\Delta r \gg \f{\s}{a}$.
Therefore, this object looks like the conventional black hole from the outside. 
Note also that 
because of this large redshift, 
only the Hawking radiation from the outermost region comes out 
although the radiation is emitted from each region inside \ci{KMY}. 

Next, let's consider a stationary black hole. 
Suppose that we put the evaporating object in the vacuum into the heat bath with temperature $T_H=\f{\hbar}{4\pi a}$. 
Then, the ingoing and outgoing radiations balance each other, and the system becomes stationary. 
The object has the surface at $r=R(a)$ for $a=const.$, 
which corresponds to a stationary black hole in the heat bath
\footnote{We will discuss the stationary black hole more in section \ref{s:entropy}.}. 
Then, the outside spacetime is described approximately by the Schwarzschild metric 
(instead of the Vaidya metric):
\begin{equation}\lb{Sch}
ds^2=-\f{r-a}{r}dt^2+\f{r}{r-a}dr^2+r^2d\Omega^2.
\end{equation}
By defining the time coordinate $T$ around the origin as 
\begin{equation}\lb{TU}
dT=dU+\f{r^2}{2\s}e^{-\f{r^2}{4\s}}dr,
\end{equation}
we can reexpress the interior metric \Ref{Ur} as 
\begin{equation}\lb{Tr}
ds^2=- \f{2\s}{r^2}e^{\f{r^2}{2\s}} dT^2 +\f{r^2}{2\s}dr^2+r^2 d\Omega^2. 
\end{equation}
Thus, by changing $T$ to $t$ through $dT=e^{-\f{R(a)^2}{4\s}}dt$, we have \ci{KMY}
\begin{equation}\lb{metric1}
d s^2=\begin{cases}
 - \f{r-a}{r}d t^2 + \f{r}{r-a}d r^2 + r^2 d \Omega^2,~~{\rm for}~~R(a)\leq r~,\\
-\f{2\s}{r^2} e^{- \f{R(a)^2-r^2}{2\s}} d t^2 + \f{r^2}{2\s} d r^2 + r^2 d \Omega^2,~~{\rm for}~~\sqrt{2\s}\lesssim r\leq R(a)~,
\end{cases}
\end{equation}
which is the candidate metric for the stationary black hole. 
This metric is continuous at $r=R(a)$ and connected to the flat region around $r=0$
\begin{equation}\lb{flat}
ds^2=-dT^2+dr^2+r^2d\Omega^2
\end{equation}
at $r\approx \sqrt{2\s}$ by the relation $dT=e^{-\f{R(a)^2}{4\s}}dt$. 

Note that the both interior metrics of \Ref{metric1V} and \Ref{metric1} are static and the same. 
In the following sections, 
we will be interested basically in the interior region. 
Therefore, we will often consider the case of $a=const.$ and use \Ref{metric1}. 

\subsection{Generalization of the metric}
The interior metric of \Ref{metric1} doesn't contain the effect of scattering which has been mentioned above \Ref{Sch_i}.
Actually, the metric has $-G^t{}_t=G^r{}_r=\f{1}{r^2}$ for $r\gg l_p$, 
which is, through \Ref{Einstein}, equivalent to $-\bra T^t{}_t\ket=\bra T^r{}_r\ket$. 
Here, $-\bra T^t{}_t\ket$ and $\bra T^r{}_r\ket$ are the energy density and radial pressure, respectively. 
This indicates that from a microscopic point of view, 
the collapsing matter and radiation move radially in a lightlike way without scattering. 
In this subsection, we introduce a phenomenological function 
representing the effect of the scattering 
and generalize the candidate metric \Ref{metric1} \ci{KY1}. 

\subsubsection{Another phenomenological function $\eta$}\lb{s:f}
We first examine the energy-momentum flow in the general static metric \Ref{AB}. 
The self-consistent energy-momentum flow must be time-reversal, which can be characterized by 
\begin{equation}\lb{T1}
-\bra T^{\mu \nu} \ket k_\nu=\kappa (l^\mu+(\eta-1) k^\mu),~~~-\bra T^{\mu \nu}\ket l_\nu=\kappa (k^\mu+(\eta-1) l^\mu)~.
\end{equation}
Here, $\kappa$ is a function. 
As we have seen that the interior of the object is very dense, 
we assume that $\eta=\eta(r)$ is a function of $\MO(1)$ which varies slowly 
\footnote{$\l|\f{d\eta}{dr}\r|l_p\ll1$. This point can be examined more by a phenomenological discussion \ci{KY1}.}. 
$\bl$ and $\bk$ are the radial outgoing and ingoing null vectors, respectively:
\begin{equation}\lb{l-k}
\bl = e^{-\f{A}{2}}\p_t+\f{1}{B}\p_r ,~~~\bk = e^{-\f{A}{2}}\p_t-\f{1}{B}\p_r.
\end{equation}
These transform under time reversal as $(\bl,\bk)\rightarrow (-\bk,-\bl)$. 
\Ref{T1} can be rewritten as   
\begin{equation}\lb{T2}
\bra T^{\bk \bk}\ket:\bra T^{\bl \bk}\ket=1:\eta-1,~~~\bra T^{\bk \bk}\ket=\bra T^{\bl \bl}\ket~,
\end{equation}
where $T^{\bk \bk}$ stands for $T^{\mu \nu}k_\mu k_\nu$, and so on. 
Furthermore, this can also be expressed in terms of  
the ratio between $-\bra T^t{}_t \ket$ and $\bra T^r{}_r \ket$: 
\begin{equation}\lb{T3}
\f{\bra T^r{}_r \ket}{- \bra T^t{}_t \ket}=\f{2-\eta}{\eta}.
\end{equation}
Therefore, $\eta$ must satisfy 
\begin{equation}\lb{eta}
1\leq \eta<2.
\end{equation}
Here, the first inequality is required by the fact that 
in \Ref{T2}, $\eta-1$ plays a role of the ratio between two energy flows, which must be positive. 
The second one is needed in order for the pressure to be positive under $- \bra T^t{}_t \ket>0$.

\begin{figure}[h]
 \begin{center}
 \includegraphics*[scale=0.27]{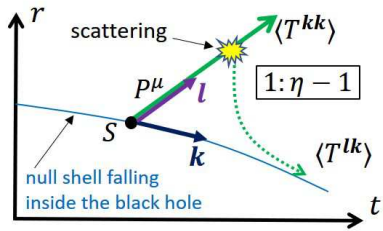}
 \caption{The meaning of a phenomenological function $\eta$.}
 \label{fmean}
 \end{center}
 \end{figure}

Now, we discuss the physical meaning of $\eta$ from a microscopic point of view. See Fig.\ref{fmean}.
We focus on one of the null shells that make up the black hole 
and consider the moment when the size is $r$, which is represented by $S$ in Fig.\ref{fmean}. 
The vector $P^\mu\equiv\bra T^{\mu \bk}\ket$ expresses the energy-momentum flow through the shell, 
which moves lightlike inward along $\bk$.  
If the radiated particle is massless and propagates outward along the radial direction without scattering, 
$P^\mu$ should be parallel to $l^\mu$, which means $\eta=1$. 
Therefore, $\eta-1$ represents the deviation from this ideal situation. 
$\eta-1$ can become non-zero if the massless particle is scattered in the ingoing direction 
by the gravitational potential or interaction with other matters\footnote{
If the radiated particle is massive, $P^\mu$ is timelike, and we have $\eta>1$. 
In this paper, however, we consider massless particles basically.}. 
This is because such a scattered particle comes back to the surface in the time scale of $O(a)$ 
according to \Ref{r_t}, and produces an energy-momentum flow along $\bk$ direction. 
In this way, $\eta$ is a phenomenological function that depends on the detail of microscopic dynamics.

\subsubsection{The candidate metric}\lb{s:can}
We determine $A(r)$ and $B(r)$ by considering \Ref{T3} for a given $\eta$. 
Here, we assume for simplicity that $\eta$ is a constant satisfying \Ref{eta} 
in $\sqrt{2\s}\lesssim r \leq R(a)$. 
We can expect that the introduction of $\eta (\neq1)$ should not change 
the functional forms of $A(r)$ and $B(r)$ in a drastically different way from those of \Ref{Tr}
\footnote{We can justify this expectation by a thermodynamical discussion \ci{KY1}.}.
Therefore, we can put 
\begin{equation}\lb{ABf0}
A(r)=C_1\f{r^2}{2\s},~~B(r)=C_2\f{r^2}{2\s},
\end{equation}
where $C_1$ and $C_2$ are some coefficients such that $C_1,C_2\to1$ in $\eta\to1$
\footnote{In $A$ of \Ref{ABf0}, we have dropped a term proportional to $R(a)^2$, 
which corresponds to considering $A$ in $ds^2=-\f{e^A}{B}dT^2+\cdots$. 
This does not affect the calculation \Ref{dA} because $G^t{}_t=G^T{}_T$. }. 

Then, using \Ref{T3} and the Einstein equation \Ref{Einstein}, we have for $r\gg l_p$
\begin{equation}\lb{dA}
\f{2}{\eta} =\f{G^r{}_r}{-G^t{}_t}+1=\f{r\p_r A}{B-1+r\p_r \log B}\approx \f{r\p_r A}{B}=\f{2C_1}{C_2}, 
\end{equation}
that is, $C_1=\f{1}{\eta}C_2$. 
Therefore, \Ref{ABf0} becomes 
\begin{equation}
A(r)=C_2\f{r^2}{2\s \eta},~~B(r)=C_2\f{r^2}{2\s}.
\end{equation}
At this stage, the intensity $\s$ is arbitrary, 
and we can redefine it and remove $C_2$ without losing generality to obtain 
\begin{equation}\lb{ABf}
A(r)=\f{r^2}{2\s\eta},~~B(r)=\f{r^2}{2\s}.
\end{equation}

Thus, introducing the scattering effect by $\eta$, \Ref{metric1} is generalized to \ci{KY1}
\begin{equation}\lb{metricf}
d s^2=\begin{cases}
 - \f{r-a}{r}d t^2 + \f{r}{r-a}d r^2 + r^2 d \Omega^2,~~{\rm for}~~R(a)\leq r,\\
-\f{2\s}{r^2} e^{- \f{R(a)^2-r^2}{2\s \eta}} d t^2 + \f{r^2}{2\s} d r^2 + r^2 d \Omega^2,~~{\rm for}~~\sqrt{2\s}\lesssim r\leq R(a).
\end{cases}
\end{equation}
This metric is again continuous at the surface at $r=R(a)\equiv a+\f{2\s}{a}$. 
The center is assumed again to be flat, which requires that $\eta=1$ there. 
The flat metric around $r=0$ \Ref{flat} is expressed in terms of $t$ approximately as 
\begin{equation}\lb{flat_t}
ds^2 = - e^{-\f{R(a)^2}{2\s}+1}dt^2 + dr^2+r^2d\Omega^2,~~{\rm for}~~0\leq r \lesssim \sqrt{2\s}.
\end{equation}
\Ref{metricf} and \Ref{flat_t} are our candidate metric\footnote{\lb{foot:inter}We consider this metric as a first approximation metric 
in that we have neglected the effect of dilute radiation outside the surface (see footnote \ref{foot:outside})
and connected the interior and exterior metrics directly at $r=R(a)$. 
For a more proper description, we would need to consider such a small effect 
and construct an ``interpolation" metric connecting the two metrics in a smooth manner. 
However, the most dominant effect of the back reaction of evaporation 
is incorporated into the interior of \Ref{metricf}.}. 
It depends on two parameters $(\s,\eta)$, which will be determined self-consistently later. 

The interior metric of \Ref{metricf} can also be applied to 
the evaporating black hole  because the interior is static. 
Therefore, \Ref{metric1V} is generalized to 
\begin{equation}\lb{metricfV}
d s^2=\begin{cases}
 - \f{r-a(u)}{r}d u^2-2 dudr + r^2 d \Omega^2,~~{\rm for}~~R(a(u))\leq r~,\\
-\f{2\s}{r^2} e^{- \f{R(a(u))^2-r^2}{2\s \eta}} d u^2 -2 e^{- \f{R(a(u))^2-r^2}{4\s \eta}} du dr + r^2 d \Omega^2,~~{\rm for}~~\sqrt{2\s}\lesssim r\leq R(a(u))~,
\end{cases}
\end{equation}

We check the form of the Einstein tensor. The interior metric of \Ref{metricf} has 
\begin{equation}\lb{Gtensor}
G^t{}_t=-\f{1}{r^2},~G^r{}_r=\f{2-\eta}{\eta}\f{1}{r^2},~G^\th{}_\th=G^\phi{}_\phi=\f{1}{2\s\eta^2}-\f{1}{\eta r^2},
\end{equation}
to $\MO(r^{-2})$ for $r\gg l_p$. 
This means through \Ref{Einstein} that 
the energy density and pressure are positive, 
but the angular pressure is so large (almost Planckian) that 
the dominant energy condition violates, as mentioned in section \ref{s:intro}. 

Next, we calculate the curvatures in $\sqrt{2\s}\lesssim r \leq R(a)$: 
\begin{align}\lb{curve1}
R&=-\f{1}{\eta^2\s}+\f{2}{r^2} \\
\lb{curve2}
R_{\mu\nu}R^{\mu\nu}&=\f{1}{2\eta^4\s^2}-\f{2}{\eta^3\s r^2}+\MO(r^{-4}),\\
\lb{curve3}
R_{\mu\nu\a\b}R^{\mu\nu\a\b}&=\f{1}{\eta^4\s^2}-\f{8}{\eta^3\s r^2}+\MO(r^{-4}).
\end{align}
This means that, 
if the metric is the solution of \Ref{Einstein} and \Ref{largeN} and \Ref{eta} are satisfied, 
the geometry has no singularity. 
Then, the Penrose diagram of the evaporating black hole is given by Fig.\ref{f:Penrose}. 

We here discuss Hawking radiation in this picture. 
We consider the total energy flux through an ingoing null shell along $\bk$ in Fig.\ref{fmean}: 
\begin{align}\lb{Jdef}
J &\equiv 4\pi r^2 \langle T^{\bm u \bk}\rangle  \nn\\
 &=4\pi r^2 \l(-B^{-1} \bra T^t{}_t \ket -e^{-\f{A}{2}}\bra T^r{}_t \ket \r), 
\end{align}
where \Ref{l-k} has been used, 
and $\bm u\equiv \f{1}{2} e^{-\f{A}{2}} \p_t$ is the vector of the local time 
(like $u_i$ in Fig.\ref{f:multi}). 
This is a generalized definition of the total flux. 
Applying \Ref{Jdef} to the interior metric of \Ref{metricf} (or \Ref{metricfV}) and using \Ref{Einstein} and \Ref{Gtensor}, 
we have $J=4\pi r^2 \l(-B^{-1} \f{1}{16\pi G} G^t{}_t \r)= \f{\s}{2Gr^2}$. 
For the exterior metric of \Ref{metricfV}, on the other hand, 
we obtain $G_{\bm u \bm k}\approx G_{uu}=-\f{\dot a}{r^2}$. 
Then, we have $J\approx \f{\s}{2G a^2}$ by using \Ref{Einstein} and \Ref{da}. 
Thus, these agree each other at $r=R(a)\approx a$, 
which means that the radiation emitted from the inside goes to the outside.

Finally, we argue that the interior metric of \Ref{metricf} is $AdS_2\times S^2$ locally. 
First, we can check that in general, a spherically symmetric metric 
$ds^2=-f(r)dt^2 + h(r)dr^2 + r^2 d\Omega^2$ is $AdS_2\times S^2$ spacetime with 
$ds^2 =\f{l^2}{z^2}(-dt^2+dz^2)+r(z)^2d\Omega^2$ if the condition 
\begin{equation}\lb{con_AdS}
\sqrt{h(r)}=\pm \f{l}{2}\p_r \log f(r)
\end{equation}
is satisfied
\footnote{Comparing $g_{tt}$ in the both metrics, 
we put $f(r)=\f{l^2}{z^2}$, from which we have $\p_r fdr=-2\f{l^2}{z^3}dz$. 
Then, we can obtain $dr=-\f{2}{z\p_r \log f} dz$ 
and see that $h(r) dr^2=\f{l^2}{z^2}dz^2 $ holds if \Ref{con_AdS} is satisfied.}. 
For the interior metric,  we have $h(r)=B(r)=\f{r^2}{2\s}$ and 
$\p_r \log f(r)\approx \p_r A(r)=\f{r}{\s \eta}$, 
where we have neglected the contribution from $B(r)$ of $g_{tt}=-\f{e^{A(r)}}{B(r)}$ for $r\gg l_p$. 
Then, \Ref{con_AdS} is satisfied for $l=\sqrt{2\s\eta^2}$ around the point $r$ we are focusing. 
Thus, the interior metric can be approximated locally as $AdS_2\times S^2$ geometry: 
\begin{equation}\lb{AdS}
ds^2 \approx (2\s \eta^2)\l[\f{1}{z^2}(-dt^2+dz^2)+\bar r^2 d\Omega^2\r],
\end{equation}
where $\bar r \equiv \f{r(z)}{\sqrt{2\s\eta^2}}$\footnote{
In fact, the Ricci scalar \Ref{curve1} is approximately $-\f{1}{\eta^2 \s}$, which is negative and constant.}. 
This means that the local symmetry for a fluid before the matter collapses becomes that for $AdS_2$ after the black hole is formed. 
It would be interesting to study this more in future.

\section{Field configurations}\lb{s:field}
We consider $N$ massless free scalar fields 
\begin{equation}\lb{S_phi}
S_{M}[\phi;g_{\mu\nu}]=-\f{1}{2}\sum_{a=1}^N \int dx^4\sqrt{-g} g^{\mu\nu}\p_\mu \phi_a\p_\nu\phi_a
\end{equation}
in the candidate metric \Ref{metricf} and \Ref{flat_t} and study their configurations to find a candidate state. 
This action leads to the equation of motion\footnote{For simplicity, we write $\phi_a$ as $\phi$ in the following.}
\begin{equation}\lb{p_eq}
0=\Box \phi(x)=\f{1}{\sqrt{-g}}\p_\mu(\sqrt{-g}g^{\mu\nu}\p_\nu\phi). 
\end{equation} 
\subsection{Classical effective potential and bound modes}\lb{s:classical}
Before going to quantum fields, we study the classical behaviour of matter 
to see intuitively what happens in the spacetime \Ref{metricf}. 
To do that, we analyze \Ref{p_eq} in the classical approximation to 
draw the effective potential for the partial waves of the fields. 

We first consider the general static metric \Ref{AB} and set
\begin{equation}\lb{pan}
\phi(x)={\mathcal N}_i \f{e^{-i\o t}}{\sqrt{C(r)}}\vp_i(r)Y_{lm}(\th,\phi),~~~C(r)\equiv r^2B(r)^{-1}e^{\f{A(r)}{2}},
\end{equation}
where $i=(\o,l)$ and the normalization factor ${\mathcal N}_i$ will be fixed in subsection \ref{s:WKB}. 
Then, the field equation \Ref{p_eq} becomes
\begin{align}\lb{p_eq2}
0 &=\p_r^2 \vp_i(r)+p_i^2(r)\vp_i(r), \\
\lb{p}
p_i^2(r) &\equiv B(r)\l(B(r)e^{-A(r)}\o^2 -\f{l(l+1)}{r^2} -{\mathcal M}(r) \r),~~~{\mathcal M}(r)\equiv\f{\p_r^2 \sqrt{C(r)}}{B(r)\sqrt{C(r)}}.  
\end{align}
This equation \Ref{p_eq2} takes the same form as the Schr\"{o}dinger equation with energy $E=0$ and the potential $V(r)=-p^2_i(r)$. 
Therefore, the classically allowed region is determined by the condition $p^2_i(r)\geq0$. 

In general, when one studies a field equation in a curved spacetime, 
he needs to take the curvature effect into account, 
which is expressed here as the ``mass term" ${\mathcal M}$, the two derivative term of the metric\footnote{
For example, in studying cosmological particle creation, 
an equation of the same form as \Ref{p_eq2} is used to analyze an adiabatic calculation \ci{BD,Parker}.}. 
For \Ref{metricf} and \Ref{flat_t}, $\MM$ takes\footnote{In fact, $\MM$ is the same order as $R$ inside the black hole. See \Ref{curve1}.}
\begin{equation}\lb{MM}
\MM=\begin{cases}
-\f{a^2}{4r^3(r-a)},~~{\rm for}~~R(a)\leq r,\\
\f{1}{8\s \eta^2}+\f{1}{2\eta r^2},~~{\rm for}~~\sqrt{2\s}\lesssim r\leq R(a),\\
0,~~{\rm for}~~0\leq r \lesssim \sqrt{2\s}.
\end{cases}
\end{equation}
However, ${\mathcal M}$ should not significantly affect the purely classical motion of matter 
because a classical particle equation (Hamilton-Jacobi equation) does not include derivative terms of the metric. 
Therefore, in this subsection, we ignore ${\mathcal M}$ for a while.  

Let us draw the classical effective potential $p_i^{(cl)}(r)^2$ for the whole spacetime 
of the static black hole with $a=$const. for simplicity. 
(This corresponds to the stationary black hole in the heat bath.) 
We apply \Ref{metricf} and \Ref{flat_t} to \Ref{p} (without ${\mathcal M}$) and obtain 
\begin{equation}\lb{pcl}
p_i^{(cl)}(r)^2=\begin{cases}
 \l(1-\f{a}{r}\r)^{-2} \o^2 - \l(1-\f{a}{r}\r)^{-1}\f{l(l+1)}{r^2},~~{\rm for}~~R(a)\leq r,\\
 \f{r^4}{4\s^2} e^{- \f{r^2-R(a)^2}{2\s \eta}} \o^2- \f{l(l+1)}{2\s},~~{\rm for}~~\sqrt{2\s}\lesssim r\leq R(a),\\
 e^{\f{R(a)^2}{2\s}-1}\o^2- \f{l(l+1)}{r^2},~~{\rm for}~~0\leq r \lesssim \sqrt{2\s}.
\end{cases}
\end{equation}
This is continuous\footnote{
In particular, it is continuous at $r=\sqrt{2\s}$ 
because the center region is flat, which means that $\eta=1$ for $r\lesssim \sqrt{2\s}$, 
as we have mentioned below \Ref{metricf}. }. 
Note that the frequency $\o$ is measured at $r\gg a$. 

\begin{figure}[h]
 \begin{center}
 \includegraphics*[scale=0.25]{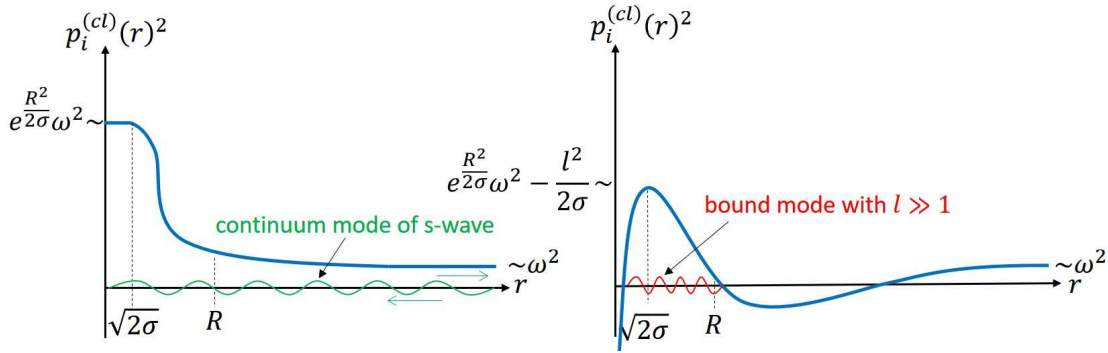}
 \caption{The classical effective potential $p_i^{(cl)}(r)^2$ for a given $a$.
 Left: $l=0$, and Right: $l\gg1$. 
 The region of $p_i^{(cl)}(r)^2>0$ is allowed classically. }
 \label{f:poten}
 \end{center}
 \end{figure}

For $l=0$, we have the left of Fig.\ref{f:poten}, 
which shows that the whole region is allowed classically;  
s-waves can enter the inside from the outside, pass the center, and come back to the outside, 
which takes an exponentially long time because of the large redshift. 
We refer to such modes as continuum modes 
in the following. 
For $l\gg1$, on the other hand, the classically allowed region consists of two disconnected domains as in the right of Fig.\ref{f:poten}. 
The outer one indicates that such modes coming from the outside are reflected by the barrier $\f{l(l+1)}{r^2}$ 
while the inner one shows that they are trapped inside, which we call bound modes. 
 
Now, we discuss the behavior of each mode in the formation process of the black hole. 
We first study the condition for a mode with $(\o,l)$ to enter the black hole with $a$ from the outside.
From \Ref{pcl}, for $r\gg a$ we have $p_i^{(cl)}(r)^2\approx \o^2-\f{l^2}{r^2}$, which becomes zero at $r=\f{l}{\o}$. 
This means that the mode is reflected at the point and returns to the outside. 
Therefore, the condition is $r=\f{l}{\o}<a$, that is, $l<\o a$. 
From this, we can also see that a mode with $(\o,l)$ composing $i$-th shell of the multi-shell model in Fig.\ref{f:multi} 
can enter inside if $l< \o a_i$. 
As we will see in section \ref{s:excite}, most of such modes have the energy $\hbar \o \sim \f{\hbar}{a_i}$. 
Then, the condition becomes $l< \o a_i\sim1$. 
Thus, we conclude that only the continuum modes of s-wave can enter the black hole from the outside. 

\begin{figure}[h]
 \begin{center}
 \includegraphics*[scale=0.25]{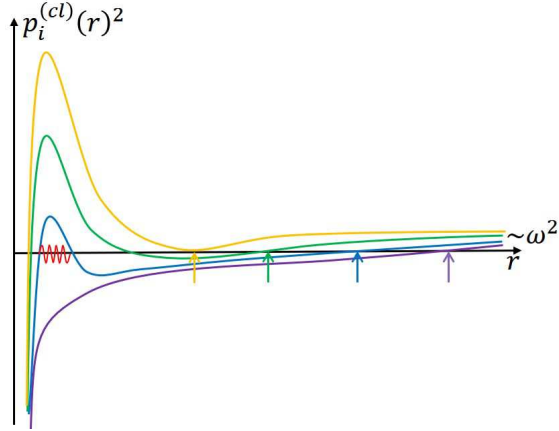}
 \caption{Change in the potential $p_i^{(cl)}(r)^2$ for a given $(\o,l)$ as $a$ increases.
 Each arrow indicates the outer zero point in each potential. }
 \label{f:bound}
 \end{center}
 \end{figure}

Although modes with $l\geq1$ cannot enter from the outside, they emerge inside the black hole in the formation process. 
To see it, we study how the potential $p_i^{(cl)}(r)^2$ changes for a given $(\o,l)$ as $a$ increases from zero. 
See Fig.\ref{f:bound}.  
Initially, there is no mass, $a=0$, so the spacetime is flat and the potential is given by the purple line.
When the mass becomes larger than $m_p$, the allowed region appears inside, which is described by the blue line. 
Then, the bound modes emerge there. 
As the mass increases, the potential grows up and the allowed region is broaden (see the green and yellow lines). 
Then, the outer zero point (which is depicted by the arrows) moves inward, 
which means that as the mass increases, the gravitational attraction increases, 
allowing the mode to overcome the centrifugal repulsion and penetrate more inside. 
In this way, many bound modes emerge inside the black hole independently of the initial state. 

\begin{figure}[h]
\begin{center}
\includegraphics*[scale=0.25]{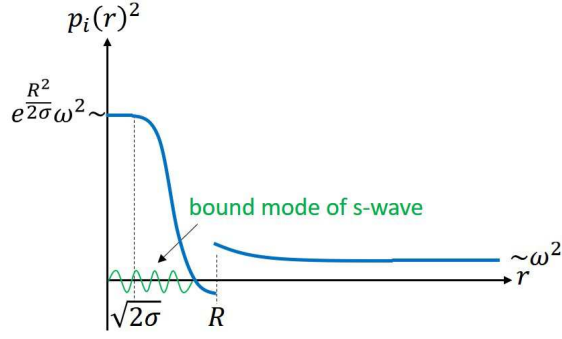}
\caption{The effective potential $p_{\o,l=0}(r)^2$ for $\o <\MO(\f{1}{a})$. 
Bound modes of s-wave can emerge due to the curvature $\MM$. 
$p_{\o,l=0}(r)^2$ has a gap at $r=R(a)$ because $\MM$ \Ref{MM} is not continuous there. 
See footnote \ref{foot:inter}.}
\label{f:s_bound}
\end{center}
\end{figure}

Finally, we note that if the curvature term $\MM$ is considered in \Ref{p}, 
the bound mode of s-wave can also emerge inside the black hole. 
In fact, for $\sqrt{2\s}\lesssim r \leq R(a)$, 
we have $p_{\o,l=0}^2|_{r=R(a)}=\f{R(a)^2}{2\s}\l(\f{R(a)^2}{2\s} \o^2 -\f{1}{8\s \eta^2} \r)$. 
This means that the s-waves with $\o< \MO(\f{1}{R(a)})$ are trapped inside $r=R(a)$. 
See Fig.\ref{f:s_bound}. 
Thus, the s-wave in the interior metric can be in a continuum mode as in the left of Fig.\ref{f:poten} 
or in a bound mode as in Fig.\ref{f:s_bound}. 

\subsection{WKB approximation}\lb{s:WKB}
We now consider quantum fields and try to solve \Ref{p_eq} 
by WKB method\footnote{
Note again that the field equation \Ref{p_eq2} takes the same form as 
the Schr\"{o}dinger equation with energy $E=0$ and the potential $V(r)=-p^2_i(r)$. 
Therefore, in order to solve \Ref{p_eq2}, 
we can use the same technique as the WKB approximation in quantum mechanics. 
We just mean it by the term ``WKB method (or approximation)".}. 
In the interior metric of \Ref{metricf}, the leading WKB solution 
is given, through \Ref{pan}, \Ref{p_eq2} and \Ref{p}, by \ci{Landau_QM} \footnote{
Note again that we are writing $\phi_a$ just by $\phi$, 
and $u_i(x)$ depends on the kind of fields.} 
\begin{align}\lb{WKB1}
\phi(x) &=\sum_i (a_i u_i(x)+ a_i^\dagger u_i^{\ast}(x)), \\
\lb{WKB2}
u_i(x) &=\MN_i\f{e^{-i\o_{i} t}}{\sqrt{C(r)}}\f{1}{\sqrt{p_{i}(r)}}\cos[\int^r dr' p_{i}(r')+\th_{i}] Y_{lm}(\th,\phi).
\end{align}
Here, $A(r)=\f{r^2-R(a)^2}{2\s\eta}$, $B(r)=\f{r^2}{2\s}$, $i=(\a,l,m)$, 
and $\th_{i}$ is a phase factor, where $\a$ labels the frequencies for each $l$ and $m$. 
For the bound modes, 
the frequencies are quantized, $\o_i=\o_{nl}$ ($n\in \mathbb{Z}$), and are approximately given by
\footnote{$\hbar$ does not appear in the l.h.s. of \Ref{BS_con} because $p_i$ is not momentum but wave number.}
\begin{equation}\lb{BS_con}
2\pi n=\oint dr p_{n l}(r)=2\int^{r_{n l}^+}_{r_{n l}^-}dr \sqrt{B\l(Be^{-A}\o_{nl}^2-\f{l(l+1)}{r^2}-{\mathcal M}\r)},
\end{equation}
where $p_{n l}(r)$ vanishes at $r= r_{n l}^+, r_{n l}^-$. 
The normalization is fixed as (see Appendix \ref{A:N})
\begin{equation}\lb{norm}
{\mathcal N}_i=\sqrt{\f{\hbar}{\pi}}\sqrt{\f{\p \o_{nl}}{\p n}}.
\end{equation}
On the other hand, the s-waves have continuum modes (as in the left of Fig.\ref{f:poten}), which we will discuss more in section \ref{s:excite}. 

We also have the commutation relation 
\begin{equation}\lb{pp}
[\phi(t,\bm x),\pi(t,\bm y)]=i \hbar \delta^3(\bm x-\bm y)
\end{equation}
and 
\begin{equation}\lb{aa}
[a_i,a_j^\dagger]=\d_{ij}, 
\end{equation}
where $\pi(t,\bm x)$ is the momentum conjugate to $\phi(t,\bm x)$ $(S_M=\int dt L_M)$:
\begin{equation}\lb{pi}
\pi(t,\bm x) \equiv \f{\p L_M}{\p (\p_t\phi(t,\bm x))}=-\sqrt{-g}g^{tt}\dot \phi.
\end{equation}
The ground state $|0\ket$ for all modes $\{u_i\}$ in the interior metric is characterized by 
\begin{equation}\lb{vac}
a_i |0\ket=0. 
\end{equation}

\subsection{A candidate state}\lb{s:ground}
We show that the bound modes are in the ground state 
while the continuum modes of s-wave are in an excited state. 
If $|\psi\ket $ is the ground state for the bound modes $\{u_i \}_{i\in B}$, it satisfies
\begin{equation}\lb{vacB}
a_i|\psi\ket =0~~{\rm for}~i\in B,
\end{equation}
where $B$ stands for the set of the bound modes. 
From Fig.\ref{f:bound}, \Ref{vacB} is independent of the initial state of the system 
because the bound modes emerge in the region disconnected from the outside.

In general, the ADM energy $M$ inside radius $r$ in a spherically-symmetric spacetime is given by \ci{Landau_C} 
\begin{equation}\lb{ADM}
M=4\pi \int^r_0 dr' r'^2 \bra-T^t{}_t\ket,
\end{equation}
where 
\begin{equation}\lb{EMT}
T_{\mu\nu}(x)\equiv\f{-2}{\sqrt{-g}}\f{\d S_{M}}{\d g^{\mu\nu}(x)}
=\sum_{a=1}^N \l(\p_\mu\phi_a(x)\p_\nu\phi_a(x)-\f{1}{2}g_{\mu\nu}(x)g^{\a\b}(x)\p_\a\phi_a(x)\p_\b\phi_a(x)\r).
\end{equation}
Then, we can define the ADM-energy increase $(\D M)_i$ 
of the first-excited state of the $i$-th bound mode (of a component of fields $\{\phi_a\}_{a=1}^N$),
 $|i; \psi\ket\equiv a_i^\dagger |\psi\ket$, by
\begin{equation}\lb{DM_def}
(\D M)_i\equiv 4\pi \int^{r_i^+}_{r_i^-} dr r^2 (\bra i;\psi|-T^t{}_t|i;\psi\ket-\bra \psi|-T^t{}_t|\psi\ket),
\end{equation}
where $|\psi \ket$ satisfies \Ref{vacB} and the integration interval is the allowed region of the bound mode $u_i$. 
This is finite because the UV divergence is subtracted by the second term. 
We estimate the order of this by using the WKB approximation.

We first check the condition for the bound mode to exist, 
which is determined by the quantization condition \Ref{BS_con}. 
That is, the quantum number of the $i$-th bound mode must be at least greater than 1: 
\begin{equation}
1\lesssim \pi n =\int^{r_i^+}_{r_i^-}drp_i\approx \int^{r_i^+}_{r_i^-}dr B e^{-\f{A}{2}}\o_i 
=e^{\f{R(a)^2}{4\s\eta}}\o_i \int^{r_i^+}_{r_i^-}dr \f{r^2}{2\s}e^{-\f{r^2}{4\eta\s}}=e^{\f{R(a)^2}{4\s\eta}}\o_i K\sqrt{\s},
\end{equation}
where we have considered only the most dominant term of \Ref{p} in the WKB approximation 
and evaluated the integration as $\int^{r_i^+}_{r_i^-}dr \f{r^2}{2\s}e^{-\f{r^2}{4\eta\s}}=K\sqrt{\s}$ 
with a constant $K=\MO(1)$\footnote{
The function $\f{r^2}{2\s}e^{-\f{r^2}{4\eta\s}}$ is like a Gaussian with width $\sim \sqrt{\s}$, 
and the integration has the dimension of the length. 
Therefore, we can have $\int^{r_i^+}_{r_i^-}dr \f{r^2}{2\s}e^{-\f{r^2}{4\eta\s}}=\MO(\sqrt{\s})$}.
Therefore, the frequency must satisfy 
\begin{equation}\lb{omega_con1}
\o_i \gtrsim \f{1}{K\sqrt{\s}} e^{-\f{R(a)^2}{4\s \eta}}.
\end{equation}
Similarly, we can calculate from \Ref{BS_con2} 
\begin{equation}\lb{omega_con2}
\f{\p \o_i^2}{\p n}=2\pi \l( \int^{r_i^+}_{r_i^-}dr \f{B^2 e^{-A}}{p_i} \r)^{-1} 
\approx 2\pi \o_i \l( \int^{r_i^+}_{r_i^-}dr B  e^{-\f{A}{2}} \r)^{-1}
=\f{2\pi \o_i}{K\sqrt{\s}}e^{-\f{R(a)^2}{4\s \eta}}.
\end{equation}

Next, we see $\dot \phi=\sum_j(-i\o_j)(a_ju_j-a_j^\dagger u_j^*)$ from \Ref{WKB1} and \Ref{WKB2}. 
Using \Ref{aa} and \Ref{vacB}, we can see easily 
$\bra i;\psi|\dot \phi^2 |i;\psi\ket=2\o_i^2 |u_i|^2 +\bra \psi|\dot \phi^2 |\psi\ket$. 
We check the form 
\begin{equation}\lb{Ttt}
T^t{}_t =-\f{1}{2}\sum_{a=1}^N\l(- g^{tt}\dot \phi^2_a+g^{rr}(\p_r \phi_a)^2 +g^{\th\th}(\p_\th\phi_a)^2+g^{\phi\phi}(\p_\phi \phi_a)^2\r).
\end{equation}
Then, we evaluate
\begin{align}\lb{DTtt}
 \bra i;\psi|-T^t{}_t|i;\psi\ket-\bra \psi|-T^t{}_t|\psi\ket&
 \approx \l(- g^{tt} \o_i^2 +g^{rr}p_i^2 + \f{l(l+1)}{r^2}\r)|u_i|^2 \nn\\
 &\approx\f{\hbar }{8\pi^2}\f{\p \o_i}{\p n}\f{2Be^{-A}\o_i^2-\f{1}{8\eta^2\s}}{r^2 B^{-1}e^{\f{A}{2}} \sqrt{ B\l(Be^{-A}\o^2_i -\f{l(l+1)}{r^2} -\f{1}{8\eta^2\s} \r) }} \nonumber \\
 &\approx \f{\hbar  }{8\pi^2 r^2}\f{\p \o_i^2}{\p n} Be^{-A},
\end{align}
Here, at the first line, we have applied $\p_r$ only to $\cos\int dr p_i$ in $u_i(x)$; 
at the second line, we have used \Ref{WKB2}, $\cos^2 \int dr p_i \approx \f{1}{2}$ and $|Y_{lm}|^2\approx \f{1}{4\pi}$; 
and at the last line, we have picked up only the most dominant terms because we are interested in the order estimation. 

Thus, we estimate \Ref{DM_def} as 
\begin{align}
(\D M)_i &\approx \f{\hbar  }{2\pi}\f{\p \o_i^2}{\p n} \int^{r_i^+}_{r_i^-}dr \f{r^2}{2\s}e^{\f{R(a)^2-r^2}{2\s \eta}}\nn \\
 &= \f{\hbar }{2\pi}\f{\p \o_i^2}{\p n} K' \sqrt{\s} e^{\f{R(a)^2}{2\s \eta}} \nonumber \\
 &\approx \hbar  \f{K'}{K} e^{\f{R(a)^2}{4\s \eta}} \o_i \nn\\
 &\gtrsim \hbar  \f{K'}{K} \f{1}{K\sqrt{\s}}.
\end{align}
Here, at the first line, we have used \Ref{DTtt}; 
at the second line, we have expressed $\int^{r_i^+}_{r_i^-}dr \f{r^2}{2\s}e^{-\f{r^2}{2\s \eta}}=K'\sqrt{\s}$ with $K'=\MO(1)$; 
at the third line, we have employed \Ref{omega_con2}; 
and at the last line, we have applied the condition \Ref{omega_con1}. 
That is, we have\footnote{
In this sense, the spectrum of the black hole is quantized and gapped. }
\begin{equation}\lb{M_excite}
(\D M)_i \gtrsim \MO \l(\f{m_p}{\sqrt{N}}\r).
\end{equation}

Here, as we will see in section \ref{s:excite}, 
the statistical fluctuation of the mass of the black hole is $\MO \l(m_p\r)$. 
Thus, \Ref{M_excite} means that 
when the number of excited bound modes exceeds $\MO(\sqrt{N})$, 
the excitation energy becomes larger than $\MO \l(m_p\r)$, 
and the object becomes different from the black hole. 
Therefore, we can regard that the bound modes are in the ground state 
because $\MO(\sqrt{N})$ is negligible compared to the number of total modes, 
which is order of $\MO(\f{a^2}{l_p^2})$ (see section \ref{s:excite}). 

On the other hand, the continuum modes of s-wave are not restricted by the condition \Ref{omega_con1} 
because they are not trapped inside. 
Therefore, those modes can enter and exit the black hole as an excitation 
that represents the collapsing matter and Hawking radiation.
Thus, the candidate state $|\psi\ket$ is a state in which 
the bound modes are in the ground state and the s-waves are excited, 
leading to \Ref{TTT}. 
(In section \ref{s:excite}, we will characterize $|\psi\ket$ more specifically.)

\section{Energy-momentum tensor in the ground state $|0\ket$}\lb{s:EMT}
In this section we evaluate $\bra 0|T_{\mu\nu}|0\ket$ of \Ref{TTT}, 
where $|0\ket$ is the ground state \Ref{vac}. 
The plan is as follows. 
In subsection \ref{s:bound}, we first study how the WKB approximation breaks down 
and solve the field equation in a different perturbation technique to obtain the leading solution of the bound modes. 
In subsection \ref{s:general}, we check the general procedure of the renormalization for 
the dimensional regularization. 
In subsection \ref{s:T0}, we evaluate the leading value of the renormalized energy-momentum tensor $\bra 0|T_{\mu\nu}|0\ket^{\prime (0)}_{ren}$. 
In subsection \ref{s:T1}, we study the subleading value $\bra 0|T_{\mu\nu}|0\ket^{\prime (1)}_{ren}$. 

\subsection{Solutions of the bound modes}\lb{s:bound}
We  construct the solutions of the bound modes in the interior, which will be used in subsection \ref{s:T0}. 
\subsubsection{Breakdown of the WKB approximation}\lb{s:break}
We first examine the validity of the WKB approximation in subsection \ref{s:WKB}. 
If we want to evaluate the energy-momentum tensor at a point $r=r_0$ inside the object, 
we need $u_i(x)$ at $r=r_0$ with various values of $i=(\o,l,m)$. 
Then, the proper frequency $\widetilde \o$ and the proper angular momentum $\wt L$ at $r=r_0$ are physically important:\footnote{
Here, we have dropped $R(a)$ in $A(r)$, which will not affect the followings.} 
\begin{equation}\lb{tilde_o}
\wt \o^2 \equiv \l(\f{\o}{\sqrt{-g_{tt}(r_0)}}\r)^2=\f{r_0^2}{2\s}e^{-\f{r_0^2}{2\s\eta}}\o^2
,~~~\wt L\equiv \f{l(l+1)}{r_0^2}.
\end{equation}
In terms of these, \Ref{p} is expressed as 
\begin{equation}\lb{p0}
p_i(r)^2 = \f{r_0^2}{2\s}\l[\l(\f{r}{r_0}\r)^4e^{-\f{r^2-r_0^2}{2\s\eta}}\wt \o^2 -\l(\wt L+\f{1}{2\eta r_0^2}\r) 
-\f{1}{8\s\eta^2}\l(\f{r}{r_0}\r)^2 \r].
\end{equation}
From this, we have 
$p_i(r_0)^2 = \f{r_0^2}{2\s}\l[\wt \o^2 -\l(\wt L+\f{1}{2\eta r_0^2}\r) -\f{1}{8\s\eta^2} \r]$.
This means that the WKB approximation is good at $r=r_0$ for $\wt \o^2 \gg \wt L$ 
because the semi-classical treatment is valid for a large wave number: $p_i(r_0)\gg1$ \ci{Landau_QM}. 
On the other hand, the approximation is bad for $\wt \o^2 \sim \wt L$; 
$r=r_0$ becomes the turning point. 
Although the WKB approach has a potential to reproduce the UV-divergent structure of the energy-momentum tensor properly, 
it cannot determine the finite values without ${\mathcal O}(1)$ errors\footnote{
Note that in subsection \ref{s:ground} 
we have used the WKB approximation only to evaluate the order of $(\D M)_i$ 
in the range where the approximation is good.}. 

For later analysis, we investigate more precisely how the approximation breaks down. 
In oder for the WKB analysis to be valid, the wave length $\lam_i(r)\equiv 1/p_i(r)$ must change slowly \ci{Landau_QM}. 
We pick up only the first term of \Ref{p0} because the exponential factor changes most drastically as a function of $r$. 
We then evaluate at $r=r_0-\D r$
\begin{equation}\lb{WKBv1}
\l.\f{d \lam_i}{d r}\r|_{r=r_0-\Delta r} \approx \l.\f{r}{2\s\eta}\f{\sqrt{2\s}}{r_0}\f{r_0^2}{r^2}e^{\f{r^2-r_0^2}{4\s \eta}}\f{1}{\wt \o} \r|_{r=r_0-\D r}
\approx \f{1}{\sqrt{2\s\eta^2}}e^{-\f{r_0}{2\s\eta}\Delta r}\f{1}{\wt \o},
\end{equation}
where the derivative has applied only to the exponential. 
If $\D r=0$, the approximation at $r=r_0$ is good for $\wt \o\gg1/\sqrt{\s\eta^2}$ 
and bad for $\wt \o \sim 1/\sqrt{\s\eta^2}$, 
which is consistent with the fact that the curvature radius is $\sim \sqrt{\s \eta^2}$ from \Ref{curve1}. 
The evaluation \Ref{WKBv1} tells more: 
even when $\wt \o\sim 1/\sqrt{\s\eta^2}$, the approximation is good at $r=r_0-\D r$ 
if $r=r_0-\D r$ is inside the turning point $r=r_0$ so that $\l.\f{d \lam_i}{d r}\r|_{r=r_0-\Delta r}\ll1$ is satisfied, that is,\footnote{
Note that $\sqrt{\s\eta^2}\sim \sqrt{N}l_p$, which is near the Planck scale, and the proper frequency should be smaller than it: 
$\wt \o \lesssim 1/\sqrt{\s\eta^2}$. Therefore, $\log (1/\wt \o\sqrt{\s\eta^2})>0$. }
\begin{equation}\lb{WKBv2}
\D r \gg \f{2\s \eta}{r_0}\log \f{1}{\wt \o\sqrt{2\s\eta^2}} \sim \f{\s}{r_0}.
\end{equation}

\subsubsection{A perturbative method and the leading exact solution}\lb{s:pert}
The finite values of the energy-momentum tensor are important for our self-consistent discussion. 
We here develop a new perturbation method to solve exactly 
the field equation \Ref{p_eq2} in the interior metric of \Ref{metricf}. 
Suppose that we want to determine $\vp_i(r)$ at a point $r=r_0$.
Motivated by \Ref{WKBv2}, we set 
\begin{equation}\lb{x}
r=r_0-x,~~~x=\MO\l(\f{\s}{r_0}\r)
\end{equation}
and expand \Ref{p0} as 
\begin{align}\lb{pexp}
p_i(r)^2 &= \f{r_0^2}{2\s}\l[\l(1-\f{x}{r_0}\r)^4e^{\f{r_0}{\s\eta}x}e^{-\f{x^2}{2\s\eta}}\wt \o^2 -\l(\wt L+\f{1}{2\eta r_0^2}\r) 
-\f{1}{8\s\eta^2}\l(1-\f{x}{r_0}\r)^2 \r] \nn\\
 &=\f{r_0^2}{2\s}\l[\l(e^{\f{r_0}{\s\eta}x}\wt \o^2- \wt L - \f{1}{8\s\eta^2}\r)
 +\l(-\l(\f{4x}{r_0}+\f{x^2}{2\s\eta}\r)e^{\f{r_0}{\s\eta}x}\wt \o^2 - \f{1}{2\eta r_0^2}+\f{1}{4\s\eta^2} \f{x}{r_0}  \r) 
 + \MO(r_0^{-4})\r]\nn\\
 &\equiv P_i^{(0)}(x)+s P_i^{(1)}(x)+\cdots, 
\end{align}
where $\wt \o$ and $\wt L$ are considered as $\MO(1)$, and $s$ is an expansion parameter. 
$P_i^{(0)}(x)$ is the leading potential of $\MO(r_0^2)$, and $P_i^{(1)}(x)$ is the subleading one of $\MO(1)$. 
Note here that $e^{\f{r_0}{\s\eta}x}$ cannot be expanded because its exponent is $\MO(1)$. 

Now, we can determine $\vp_i(r)$ around $r=r_0$ by the perturbative expansion with respect to $1/r_0^2$.  
We put it as 
\begin{equation}\lb{vpexp}
\vp_i(r)=\vp_i^{(0)}(x) + s \vp_i^{(1)}(x) + \cdots. 
\end{equation}
Combing this and \Ref{pexp}, then \Ref{p_eq2} becomes 
\begin{equation}
(\p_x^2 + P_i^{(0)}(x)+s P_i^{(1)}(x)+\cdots )(\vp_i^{(0)}(x) + s \vp_i^{(1)}(x) + \cdots)=0.
\end{equation}
This gives an iterative equation: 
\begin{align}\lb{eq0}
(\p_x^2 + P_i^{(0)}(x))\vp_i^{(0)}(x) &=0,\\
\lb{eq1}
(\p_x^2 + P_i^{(0)}(x))\vp_i^{(1)}(x) &=- P_i^{(1)}(x)\vp_i^{(0)}(x),\\
&\cdots.\nn
\end{align}
By construction, $\vp_i^{(0)}(x)$ is the leading exact solution in this perturbative expansion. 

Let us solve \Ref{eq0}. We can convert from $x$ to  
\begin{equation}\lb{xi}
\xi = \sqrt{2\s\eta^2}\wt \o e^{\f{r_0}{2\s\eta}x}
\end{equation}
and rewrite \Ref{eq0} as 
\begin{equation}\lb{eqB}
\l(\xi^2 \f{d^2}{d\xi^2}+\xi \f{d}{d\xi}+\xi^2-A^2 \r)\vp_i^{(0)}(\xi)=0,
\end{equation}
where 
\begin{equation}\lb{A}
A\equiv \sqrt{2\s\eta^2 \wt L +\f{1}{4}}.
\end{equation}
This is the Bessel equation, whose solutions are Bessel functions $J_A(\xi)$ and $J_{-A}(\xi)$. 
Using the boundary condition that the mode is bounded, 
we can choose $J_A(\xi)$ properly. 
Considering the evaluation \Ref{WKBv2} and the region where the WKB solution \Ref{WKB2} is valid, 
we can fix the normalization. See Appendix \ref{A:B}.
Thus, we obtain the leading solution
\begin{equation}\lb{vp0}
\vp_i^{(0)}(\xi)=\sqrt{\f{\pi\s\eta}{r_0}}J_A(\xi). 
\end{equation}
Substituting this and \Ref{norm} for \Ref{pan}, 
the leading bound-mode functions in the interior metric of \Ref{metricf} without $R(a)$ (which will not affect the following analysis) are given by 
\begin{equation}\lb{u0}
u_i^{(0)}(t,r,\th,\phi)= \sqrt{\f{\hbar\eta}{2r_0}}\sqrt{\f{\p \o_{nl}}{\p n}} e^{-i\o_{nl} t}e^{-\f{r^2}{8\s\eta}}J_A(\xi) Y_{lm}(\th,\phi).
\end{equation}
Note again that $\xi$ depends on $\o_{nl}$ and $A$ on $l$.

The subleading solution $\vp_i^{(1)}(x)$ can be determined from \Ref{eq1} and \Ref{vp0}, 
for example, by Green-function method. 
We leave it as a future task. 

Finally, we make a comment on the origin of the leading potential $P_i^{(0)}(x)$. 
We can also use the $AdS_2\times S^2$ metric \Ref{AdS} and focus a region around $r=r_0$ 
to show that \Ref{p_eq} becomes the Bessel equation. 
In this sense, the local $AdS_2\times S^2$ structure makes the field equation \Ref{p_eq} solvable. 
More generally, the condition \Ref{con_AdS} is the origin.
\subsection{Dimensional regularization and renormalization}\lb{s:general}
In general, when one considers composite operators such as $T_{\mu\nu}(x)$, he needs to regularize them. 
We use the dimensional regularization technique, 
which has an advantage that it is covariant, so is the UV-divergent part. 
Before going to the explicit evaluation of $\bra 0 |T_{\mu\nu}(x)|0 \ket_{ren}$, 
we here give the general discussion about the dimensional regularization and renormalization of the energy-momentum tensor. 

The bare action of our theory on a $d$-dimensional spacetime is 
\begin{align}\lb{S_B}
S_B[\phi_{aB},g_{B\mu\nu}]&
 =\int d^dx\sqrt{-g_B}\l( -\f{1}{2}\sum_{a=1}^N g_B^{\mu\nu}\p_\mu \phi_{Ba}\p_\nu\phi_{Ba}+ \f{1}{16\pi G_B}R_B\r.\nn\\
 &~~~~~~~~~~~~~~~~~~~~~\l.+ \a_B R_B^2 + \b_B R_{B\mu \nu }R^{\mu\nu}_B+ \g_B R_{B\mu \nu \a \b }R^{\mu\nu\a\b}_B\r),
\end{align}
where the index $B$ expresses the bare quantities. 

First, we don't consider quantization of gravity, and the quantum scalar fields are free. 
Therefore, the bare fields are the same as the renormalized ones: 
\begin{equation}\lb{f_ren}
\phi_{Ba}=\phi_a,~~~g_{B\mu\nu}=g_{\mu\nu}.
\end{equation}
Next, we introduce a renormalization point $\hbar \mu$. 
Because there is no mass in the theory, 
the renormalization of the Newton constant does not appear. 
Therefore, by dimensional analysis, we can put 
\begin{equation}\lb{mu1}
\f{1}{G_B}=\f{\mu^\e}{G},
\end{equation}
where $d=4+\e$, and $G$ is the 4D physical Newton constant, which is independent of $\mu$.  
$G_B$ agrees with $G$ in $\e\to0$. 
On the other hand, $\a_B$, $\b_B$ and $\g_B$ need to be renormalized. 
In the limit $\e\to0$, they take the form 
\begin{equation}\lb{mu2}
\a_B=\mu^\e\l(\f{\hbar N}{1152\pi^2\e}+\a(\mu)\r),~\b_B=\mu^\e\l(-\f{\hbar N}{2880\pi^2\e}+\b(\mu)\r),~
\g_B=\mu^\e\l(\f{\hbar N}{2880\pi^2\e}+\g(\mu)\r).
\end{equation}
Here, we choose the counter terms with $\f{1}{\e}$ as those which are required to 
renormalize the UV-divergent part of the effective action of $N$ massless free scalar fields 
by the minimal subtraction scheme \ci{BD,Parker,TV}. 
$\a(\mu)$, $\b(\mu)$ and $\g(\mu)$ are the renormalized coupling constants at a renormalization point $\hbar \mu$. 

Then, the equation of motion for $g_{\mu\nu}$, $\f{\d S_B}{\d g^{\mu\nu}}$=0, is given by
\begin{align}\lb{geq1}
G_{\mu\nu}&=8\pi G \l[\mu^{-\e}T_{\mu\nu}-2\l(\f{\hbar N}{1152\pi^2\e}+\a(\mu)\r)H_{\mu\nu}\r.\nn\\
  &\l.-2 \l(-\f{\hbar N}{2880\pi^2\e}+\b(\mu)\r)K_{\mu\nu}
  -2 \l(\f{\hbar N}{2880\pi^2\e}+\g(\mu)\r)J_{\mu\nu}
  \r].
\end{align}
Here, $T_{\mu\nu}$ is the regularized energy-momentum tensor operator, which is formally given 
by \Ref{EMT} with $S_M$ replaced by the $d$-dimensional matter action $S_B^{matter}$ of \Ref{S_B}. 
The other tensors are proportional to the identity operator, which are given by  
\begin{align}
\lb{Htensor}
H_{\mu\nu} &\equiv \f{1}{\sqrt{-g}}\f{\d}{\d g^{\mu\nu}}\int d^dx\sqrt{-g}R^2=-\f{1}{2}g_{\mu\nu}R^2 +2RR_{\mu\nu}-2\N_\mu\N_\nu R+2g_{\mu\nu}\Box R,\\
\lb{Ktensor}
K_{\mu\nu} &\equiv \f{1}{\sqrt{-g}}\f{\d}{\d g^{\mu\nu}}\int d^dx\sqrt{-g}R_{\a\b}R^{\a\b} \nn\\
 &=-\f{1}{2}g_{\mu\nu}R_{\a\b}R^{\a\b} +2R_{\mu\a\nu\b}R^{\a\b}+\Box R_{\mu\nu}+\f{1}{2}g_{\mu\nu}\Box R-\N_\mu\N_\nu R,\\
\lb{Jtensor}
J_{\mu\nu} &\equiv \f{1}{\sqrt{-g}}\f{\d}{\d g^{\mu\nu}}\int d^nx\sqrt{-g}R_{\a\b\g\d}R^{\a\b\g\d} \nn\\
 &=-\f{1}{2}g_{\mu\nu}R_{\a\b\g\d}R^{\a\b\g\d} +2R_{\mu\a\b\g}R_\nu{}^{\a\b\g}+4R_{\mu\a\nu\b}R^{\a\b}-4R_{\mu\a}R_\nu{}^\a
 +4\Box R_{\mu\nu}-2\N_\mu\N_\nu R.
\end{align}

From this, we can obtain the precise expression of the Einstein equation \Ref{Einstein}:  
\begin{equation}\lb{Einstein2}
G_{\mu\nu}= 8\pi G \bra \psi|T_{\mu\nu}|\psi \ket_{ren}^\prime.
\end{equation}
Here, we have defined 
\begin{equation}\lb{geq2}
\bra \psi|T_{\mu\nu}|\psi \ket_{ren}^\prime
\equiv \bra \psi|T_{\mu\nu}|\psi \ket_{ren}(\mu)-2 (\a(\mu)H_{\mu\nu}+\b(\mu)K_{\mu\nu}+\g(\mu)J_{\mu\nu}),
\end{equation}
where the 4D renormalized energy-momentum tensor at energy scale $\hbar \mu$ is 
\begin{equation}\lb{Tren}
\bra \psi|T_{\mu\nu}|\psi \ket_{ren}(\mu)\equiv \mu^{-\e}\bra \psi|T_{\mu\nu}|\psi \ket_{reg}
-\f{\hbar N}{1440\pi^2\e}\l(\f{5}{2}H_{\mu\nu}-K_{\mu\nu}+J_{\mu\nu}\r).
\end{equation}
This is finite in the limit $\e\to0$, as we will see explicitly below.

The second term in the r.h.s. of \Ref{geq2} is a finite renormalization which plays a role in choosing the theory. 
The point is that 
the $\mu$-dependence of the renormalized coupling constants $\a(\mu),\b(\mu),\g(\mu)$ must be chosen so that 
the bare coupling constants $\a_B,\b_B,\g_B$ are independent of $\mu$. 
From \Ref{mu2}, the condition $\f{d\log\a_B}{d\log\mu}=0$ 
provides $\f{d\log\a(\mu)}{d\log\mu}=-\f{\hbar N}{1152\pi^2}$ in $\e\to 0$. 
We can do the same procedure for the other couplings.  
From these, we obtain 
\begin{equation}\lb{ab_flow}
\a(\mu^2)=\a_0-\f{\hbar N}{2304\pi^2}\log\l(\f{\mu^2}{\mu_0^2}\r),~
\b(\mu^2)=\b_0+\f{\hbar N}{5760\pi^2}\log\l(\f{\mu^2}{\mu_0^2}\r),~
\g(\mu^2)=-\f{\hbar N}{5760\pi^2}\log\l(\f{\mu^2}{\mu_0^2}\r).
\end{equation}
Here, $\a_0$ and $\b_0$ fix a 4D theory at energy scale $\hbar\mu_0$ 
while we have chosen $\g_0=0$ because of the 4D Gauss-Bonnet theorem. 
We will see later that 
\Ref{geq2} with \Ref{ab_flow} is independent of $\mu$, 
which is consistent with the l.h.s. of \Ref{Einstein2}. 

\subsection{Leading terms of the energy-momentum tensor}\lb{s:T0} 
Now, we evaluate the leading value of the renormalized energy-momentum tensor \Ref{geq2} for $|0\ket$, 
$\bra 0|T_{\mu\nu}|0\ket_{ren}^{\prime(0)}$. 
Generically, the continuum modes of s-wave also appear 
in the mode expansion of a field $\phi(x)$. 
However, as we will see in subsection \ref{s:Tren}, 
the leading value $\bra 0|T_{\mu\nu}|0\ket_{reg}^{(0)}$ becomes $\MO(1)$ by integrating both $\o$ and $l$, 
which means that the sum of such s-wave modes (integration only over $\o$) can contribute 
to at most $\MO(r^{-2})$. 
Therefore, in this subsection we keep only the bound modes \Ref{u0} 
in the expansion of the leading solution of $\phi(x)$.

\subsubsection{Fields in the dimensional regularization}\lb{s:phi_dim}
To use dimensional regularization, 
we consider the $(4+\e)$-dimensional spacetime manifold $M\times \mathbb{R}^\e$, 
where $M$ is our 4D physical spacetime and $\mathbb{R}^\e$ is $\e$-dimensional flat spacetime \ci{KN}.
That is, we take
\begin{equation}\lb{met_reg}
ds^2=-\f{2\s}{r^2} e^{\f{r^2}{2\s \eta}} d t^2 + \f{r^2}{2\s} d r^2 + r^2 d \Omega^2+\sum_{a=1}^\e (dy^a)^2.
\end{equation}
In this metric, the leading bound mode function \Ref{u0} becomes 
\begin{equation}\lb{u0reg}
u_i^{(0)}(t,r,\th,\phi,y^a)= 
\sqrt{\f{\hbar\eta}{2r_0}}\sqrt{\f{\p \o_{nl}}{\p n}} e^{-i\o_{nl} t}e^{-\f{r^2}{8\s\eta}}J_A(\xi) Y_{lm}(\th,\phi) \f{e^{ik\cdot y}}{(2\pi)^{\e/2}}.
\end{equation}
The plane wave part $e^{ik\cdot y}$ makes a shift $\MM \to \MM +k^2$ in \Ref{p}, 
and the definition of the label $A$ changes from \Ref{A} to  
\begin{equation}\lb{Areg}
A\equiv \sqrt{2\s\eta^2 (\tilde L +k^2) +\f{1}{4}}.
\end{equation}
In terms of \Ref{u0reg}, we express the leading solution of $\phi(x)$ as  
\begin{equation}\lb{phireg}
\phi(x)=\sum_i (a_iu_i^{(0)}(x)+a_i^\dagger u_i^{(0)\ast}(x)),
\end{equation}
where $\sum_i=\sum_{n,l,m}\int d^\e k$ and $a_i$ satisfies \Ref{vac}.

\subsubsection{Renormalized energy-momentum tensor}\lb{s:Tren} 
From \Ref{phireg}, we can obtain the leading values of the regularized energy-momentum 
tensor (see Appendix \ref{A:reg} for the detailed calculation)\footnote{Note that we don't impose the Einstein equation \Ref{Einstein2} here.}:
\begin{equation}\lb{T0reg}
\mu^{-\e}\bra0|T^\mu{}_\nu|0\ket_{reg}^{(0)}
=\left(
\begin{array}{cccc}
1 & & &  \\
 & 1& &  \\
 & & -1&  \\
 & & & -1
\end{array}
\right)\mu^{-\e}\bra0|T^t{}_t|0\ket_{reg}^{(0)}
+
\left(
\begin{array}{cccc}
0 & & &  \\
 &0 & &  \\
 & & 1&  \\
 & & &1 
\end{array}
\right)\f{\hbar N}{1920\pi^2\eta^4\s^2},
\end{equation}
where the components are in the order of $(t,r,\th,\phi)$ and 
\begin{equation}\lb{Tttreg}
\mu^{-\e}\bra0|T^t{}_t|0\ket_{reg}^{(0)}
=\f{\hbar N}{960\pi^2\eta^4\s^2}\l[\f{1}{\e}+ \f{1}{2}\l( \g+\log\f{1}{32\pi \eta^2\s \mu^2}\r) + c\r].
\end{equation}
Here, $\g$ is Euler's constant and $c$ is the non-trivial finite value for $|0\ket$:
\begin{equation}\lb{c}
c=0.055868.
\end{equation}
Note here that, 
as shown in Appendix \ref{A:reg}, 
this leading value of $\MO(1)$ is obtained as a result of the integration over $\o$ \textit{and} $l$, 
and the 4D dynamics is important. 

Next, we can check that the poles $\f{1}{\e}$ of \Ref{T0reg} are cancelled by the counter terms in \Ref{Tren}, 
and $\bra0|T^\mu{}_\nu|0\ket_{ren}^{(0)}(\mu)$ is indeed finite. 
Here, we have used the explicit form of $H_{\mu\nu}$, $K_{\mu\nu}$ and $J_{\mu\nu}$ 
for the interior metric of \Ref{metricf}: 
\begin{align}\lb{HL_form}
  H^\mu{}_\nu&
=\f{1}{2\eta^4\sigma^2}\left(
\begin{array}{ccccc}
1 & & & &  \\
 & 1& & &  \\
 & & -1& &  \\
 & & & -1&  
\end{array}
\right)
+\f{1}{\eta^3\s r^2}\left(
\begin{array}{ccccc}
0 & & & &  \\
 &-4 & & &  \\
 & &2 & &  \\
 & & &2 &  
\end{array}
\right)+\MO(r^{-4}),
\nn\\
K^\mu{}_\nu&
=\f{1}{4\eta^4\sigma^2}\left(
\begin{array}{ccccc}
1 & & & &  \\
 & 1& & &  \\
 & & -1& &  \\
 & & & -1&  
\end{array}
\right)
+\f{1}{\eta^3\s r^2}\left(
\begin{array}{ccccc}
-1 & & & &  \\
 &-3 & & &  \\
 & &2 & &  \\
 & & &2 &  
\end{array}
\right)+\MO(r^{-4}),\nn\\
J^\mu{}_\nu&
=\f{1}{2\eta^4\sigma^2}\left(
\begin{array}{ccccc}
1 & & & &  \\
 & 1& & &  \\
 & & -1& &  \\
 & & & -1&  
\end{array}
\right)
+\f{1}{\eta^3\s r^2}\left(
\begin{array}{ccccc}
-4 & & & &  \\
 &-8 & & &  \\
 & &6 & &  \\
 & & &6 &  
\end{array}
\right)+\MO(r^{-4}).
\end{align}

Finally, we use \Ref{geq2} with the running coupling constants \Ref{ab_flow} 
to obtain the renormalized energy-momentum tensor with a finite renormalization: 
\begin{equation}\lb{T0ren}
\bra0|T^\mu{}_\nu|0\ket_{ren}^{\prime(0)}
=\left(
\begin{array}{cccc}
1 & & &  \\
 & 1& &  \\
 & & -1&  \\
 & & & -1
\end{array}
\right)\bra0|T^t{}_t|0\ket_{ren}^{\prime(0)}
+
\left(
\begin{array}{cccc}
0 & & &  \\
 &0 & &  \\
 & & 1&  \\
 & & &1 
\end{array}
\right)\f{\hbar N}{1920\pi^2\eta^4\s^2},
\end{equation}
where 
\begin{equation}\lb{T0tren}
\bra0|T^t{}_t|0\ket_{ren}^{\prime(0)}
=\f{\hbar N}{1920\pi^2\eta^4\s^2}\l[2c+\g + \log\f{1}{32\pi \eta^2\s \mu^2_0}-\f{960\pi^2}{\hbar N }(2\a_0+\b_0) \r].
\end{equation}
Here, the $\mu$-dependence disappears.  
In section \ref{s:sol}, we will see that $\a_0$ and $\b_0$ should be tuned properly 
in order to have a self-consistent solution of the form of \Ref{metricf}.

The point is that the leading value of the trace is fixed independently of $(\a_0,\b_0)$: 
\begin{equation}\lb{trace0}
\bra0|T^\mu{}_\mu|0\ket_{ren}^{\prime(0)}=\f{\hbar N}{960\pi^2\eta^4\s^2}. 
\end{equation}
We can see that this comes from the UV-divergent structure 
and is essentially the 4D Weyl anomaly \ci{BD,Parker,Duff}
although the matter fields are \textit{not} conformal. 
\Ref{trace0} was obtained by first renormalizing the divergences and then taking the trace.
We can reverse the order to see clearly how the trace part appears.
If we first take the trace of $\bra0|T_{\mu\nu}|0\ket_{reg}^{(0)}$, we have, through \Ref{EMT}, \Ref{req3}, \Ref{theq3} and \Ref{yeq3},  
\begin{align}\lb{ano1}
\bra0|T^\mu{}_\mu|0\ket_{reg}^{(0)} 
 &=N\bra0|g^{\mu\nu}\p_\mu \phi \p_\nu \phi -\f{1}{2}(4+\e)g^{\mu\nu}\p_\mu \phi \p_\nu \phi|0\ket_{reg}^{(0)} \nn\\
 &=-N\l(1+\f{\e}{2}\r) \bra0|g^{tt}(\p_t \phi)^2+g^{rr}(\p_r \phi)^2+g^{\th\th}(\p_\th \phi)^2+g^{\phi\phi}(\p_\phi \phi)^2+\sum_{a=1}^\e (\p_{y^a} \phi)^2 |0\ket_{reg}^{(0)}  \nonumber \\
 &=-N\l(1+\f{\e}{2}\r) \bra0|g^{tt}(\p_t \phi)^2+g^{tt}(\p_t \phi)^2+(-2+\e)g^{tt}(\p_t \phi)^2-\e g^{tt}(\p_t \phi)^2 |0\ket_{reg}^{(0)}  \nonumber \\
 &= 0
\end{align}
before taking the limit $\e\to0$. 
However, the trace of the counter terms in \Ref{Tren} makes a non-trivial contribution:
\begin{align}\lb{ano2}
&-\f{\hbar N}{1440\pi^2\e}\l(\f{5}{2}H^\mu{}_\mu-K^\mu{}_\mu+J^\mu{}_\mu\r)\nn\\
 &=-\f{\hbar N}{1440\pi^2\e}\f{-\e}{2}\l(\f{5}{2}R^2-R_{\a\b}R^{\a\b}+R_{\a\b\mu\nu}R^{\a\b\mu\nu} \r)+\MO(r^{-4})  \nonumber \\
 &= \f{\hbar N}{960\pi^2}\l(\f{1}{\eta^4\s^2}-\f{2}{\eta^3\s r^2}\l(\f{5}{3}\eta +1\r) \r)+\MO(r^{-4}),
\end{align}
where we have used \Ref{curve1}, \Ref{curve2}, \Ref{curve3} and 
\begin{align}\lb{tr_1}
H^\mu{}_\mu &=-\f{\e}{2}R^2+(6+2\e)\Box R,~~K^\mu{}_\mu =-\f{\e}{2}R_{\a\b}R^{\a\b}+\l(2+\f{\e}{2}\r)\Box R,\\
\lb{tr_2}
J^\mu{}_\mu &=-\f{\e}{2}R_{\mu\nu\a\b}R^{\mu\nu\a\b}+2\Box R,\\
\lb{box_R}
\Box R&=\MO(r^{-4}).
\end{align}
Here, the last equation has been evaluated by using the interior metric of \Ref{metricf}. 
The first term of \Ref{ano2} agrees with \Ref{trace0}\footnote{
Because $\a(\mu),\b(\mu),\g(\mu)$ don't have the poles $\f{1}{\e}$, 
the finite renormalization terms in \Ref{geq2} make $\Box R=\MO(r^{-4})$ and don't contribute here.}. 

\subsection{Sub-leading terms of the energy-momentum tensor}\lb{s:T1} 
The sub-leading bound-mode solution $\vp_i^{(1)}(x)$ of \Ref{eq1}, which we have not found yet, determines 
the subleading value $\mu^{-\e}\bra 0|T_{\mu\nu}|0\ket_{reg}^{(1)}$ completely \footnote{
The continuum modes of s-wave may contribute to this. }. 
In this subsection, we instead use the conservation law to show that 
$\bra 0|T_{\mu\nu}|0\ket_{ren}^{\prime(1)}$ is expressed in terms of two parameters. 

In the interior metric of \Ref{metricf}, 
the conservation law $\N_\mu T^\mu{}_\nu=0$ is expressed as 
\begin{align}\lb{con_eq}
0&=\p_r T^r{}_r+\p_r \log\sqrt{-g_{tt}}(-T^t{}_t+T^r{}_r)+\f{2}{r}(T^r{}_r-T^\th{}_\th) \nn\\
&=\p_r T^r{}_r+\f{r}{2\eta\sigma}(-T^t{}_t+T^r{}_r)+\f{1}{r}(T^t{}_t+T^r{}_r-2T^\th{}_\th), 
\end{align}
where $T^\mu{}_\nu$ is assumed to be static and spherically symmetric: $T^\mu{}_\nu(r)$ and $T^\th{}_\th=T^\phi{}_\phi$\footnote{
\Ref{con_eq} corresponds to the Tolman-Oppenheimer-Volkoff equation without $T^r{}_r=T^\th{}_\th$.}. 

$H_{\mu\nu}$, $K_{\mu\nu}$ and $J_{\mu\nu}$ are conserved, and their explicit forms are given by \Ref{HL_form}, 
which are negative power polynomials in $r^2$ starting from a constant. 
Also, the leading value $\bra0|T^\mu{}_\nu|0\ket_{ren}^{'(0)}$ is constant. 
Motivated by these facts, we can set the ansatz as
\begin{equation}\lb{Eansatz}
\bra 0 |T^t{}_t|0 \ket_{ren}'=a_0+\f{a_1}{r^2}+\cdots,~~
\bra 0 |T^r{}_r|0 \ket_{ren}'  =b_0+\f{b_1}{r^2}+\cdots,~~
\bra 0 |T^\th{}_\th|0 \ket_{ren}'  =c_0+\f{c_1}{r^2}+\cdots. 
\end{equation}
Then, we substitute this into \Ref{con_eq} and solve it for each order of $r$ to get
\begin{equation}\lb{con_sol}
b_0 =a_0,~~c_0=a_0+\f{-a_1+b_1}{4\eta \sigma},~~c_1 =\f{a_1-b_1}{2}+\f{-a_2+b_2}{4\eta\sigma},~~\cdots.
\end{equation}
In fact, \Ref{T0ren} satisfies the first equation of \Ref{con_sol}.

From \Ref{geq2}, we put  
\begin{align}\lb{Tt1ren}
\bra 0| T^t{}_t|0\ket_{ren}^{\prime (1) } 
 &= \l[\mu^{-\e}\bra 0| T^t{}_t|0\ket_{reg}^{(1)} 
+\f{\hbar N}{480\pi^2\eta^3\s r^2 \e}-\f{\hbar N}{960\pi^2 \eta^3\s r^2} \log\f{\mu^2}{\mu_0^2} \r]+\f{2\b_0}{\eta^3\s r^2} \nonumber \\
 &\equiv \f{\hbar N}{480\pi^2 \eta^3\s r^2}\wt a_1 +\f{2\b_0}{\eta^3\s r^2},
\end{align}
where we have used \Ref{ab_flow} and \Ref{HL_form}. This means
\begin{equation}\lb{a1}
a_1= \f{\hbar N}{480\pi^2 \eta^3\s }\wt a_1 +\f{2\b_0}{\eta^3\s }.
\end{equation}
By construction, $\wt a_1$ should be a $\mu$-independent finite number of $\MO(1)$, as we have seen in \Ref{T0tren}. 
We suppose that such a $\wt a_1$ is given. 
By using \Ref{T0ren}, \Ref{T0tren}, the second equation of \Ref{con_sol} and \Ref{a1}, we can obtain
\begin{align}\lb{Tr1ren}
\bra 0|T^r{}_r|0\ket^{\prime(1)}_{ren}&=\f{b_1}{r^2}=\f{1}{r^2}(a_1+4\eta\s(c_0-a_0))\nn\\
 &= \f{\hbar N}{480\pi^2 \eta^3 \s r^2} \l(\wt a_1 +\f{960\pi^2}{\hbar N}\b_0 +1\r.\nn\\ 
 &~~~~~~~~~~~~~~~~~~~~~~\l.-2\l[2c+\g +\log\f{1}{32\pi \eta^2 \s \mu_0^2}-\f{960\pi^2}{\hbar N}(2\a_0+\b_0) \r]  \r).
\end{align}

Finally, we write the subleading value of the trace part $\bra 0|T^\mu{}_\mu|0\ket^{\prime(1)}_{ren}$ as
\begin{equation}\lb{Ttrace1ren}
\bra 0|T^\mu{}_\mu|0\ket^{\prime(1)}_{ren}=
\f{\hbar N}{240\pi^2 \eta^3 \s r^2}\wt \tau_1, 
\end{equation}
where $\wt \tau_1$ is a finite value 
independent of the finite renormalization in \Ref{geq2} because of \Ref{HL_form}. 
We suppose that such a $\wt \tau_1$ is obtained. 
This together with \Ref{Tt1ren} and \Ref{Tr1ren} determines $\bra 0|T^\th{}_\th|0\ket^{\prime(1)}_{ren}$. 

Thus, we can express all the components of $\bra 0|T^\mu{}_\nu|0\ket^{\prime}_{ren}$ 
to $\MO(r^{-2})$ in terms of $(\wt a_1,\wt \tau_1)$.

\section{Contribution of s-wave excitations to the energy-momentum tensor}\lb{s:excite}
In this section we first consider the generic configuration 
of excitations of the continuum modes of s-wave 
from a statistical point of view and characterize the candidate state $|\psi \ket$ more. 
Then, we study the excitation energy in the interior of the black hole 
and find the energy density $-T^{(\psi)t}{}_t$. 
From this and the conservation law, we infer $T^{(\psi)}_{\mu\nu}$ of \Ref{TTT}.
\subsection{Fluctuation of mass of the black hole}\lb{s:formation}
We consider the black holes with mass $M=\f{a}{2G}$ that are most likely to be formed statistically. 
Such black holes are created in the most generic manner. 
When making a black hole in several operations, the more energy given per operation, 
the more specific and less generic the formed black hole is. 
Therefore, the most generic black hole should be created by using a collection of quanta with as small energy as possible. 
According to Bekenstein's idea \ci{Bekenstein}, 
in order for a wave to enter the black hole, 
the wavelength $\lambda$ must be smaller than $a$: $\lambda \lesssim a$. 
This means that the energy of the wave, $\e\sim \f{\hbar}{\lambda}$, must satisfy $\e\gtrsim\f{\hbar}{a}$. 
As discussed in subsection \ref{s:classical}, only continuum modes of s-wave can play a role in the black-hole formation. 
Therefore, a quantum required is such a s-wave with the minimum energy, 
\begin{equation}\lb{e_mini}
\e\sim \f{\hbar}{a}.
\end{equation}

Suppose that we form a black hole of size $a$ by injecting such quanta many times. 
Because the wavelength of each quantum is almost the same as the size $a'$ of the black hole at each stage ($\lambda\sim a'$), 
the wave enters the black hole with a probability of 1/2 and is bounced back with a probability of 1/2. 
Here, we model the formation as a stochastic process according to binomial distribution. 
Then, the average number of trials $\MN$ is given 
by dividing $M=\f{a}{2G}$ by the energy $\sim \f{\hbar}{a}$ that can enter at one time: 
\begin{equation}\lb{Ntrial}
\MN\sim \f{a}{2G}\times \f{1}{\hbar/a}\sim \f{a^2}{l_p^2}.
\end{equation}
Therefore, the statistical fluctuation of mass $M$ is evaluated as \ci{Landau_SM}
\begin{equation}\lb{M_fluc}
\D M \sim \f{M}{\sqrt{\MN}}\sim m_p.
\end{equation}
This means that 
all the black holes with mass $\in[M,M+m_p]$ are not statistically distinguished, 
and they are considered as the same macroscopically.

In this way, our candidate state $|\psi\ket$ is one of the states $\{|\psi\ket \}$ that 
represent microscopically different continuum s-wave configurations 
and have the energy-momentum tensor in the form of \Ref{TTT}. 
The number of those states reproduces the area law of the entropy, as shown in section \ref{s:entropy}. 

\subsection{S-wave excitation inside the black hole and $T^{(\psi)t}{}_t$}\lb{s:Tpsitt}
To determine the functional form of $T^{(\psi)t}{}_t(r)$, 
we consider how the s-wave excitations used to form the black hole 
are distributed in the interior metric of \Ref{metricf}. 
First, suppose that an s-wave having a proper wavelength $\lambda_{local}$ 
is excited at a certain point $r$ inside the black hole. 
It has the proper energy $\e_{local}\sim\f{\hbar}{\lambda_{local}}$. 
Here, the local Hamiltonian is given by $\D H_{local}=4\pi r^2 \D l~T^{(\psi)\tau \tau}$, 
where $\D l$ is the proper length of a width $\D r$ around $r$, 
$T^{(\psi)}_{\mu\nu}$ is the energy-momentum tensor from the contribution of the excitation, 
and $\tau$ is the proper time of the local inertial frame at $r$. 
Therefore, 
by considering the number of fields $N$, 
we can set $\D H_{local}=N \e_{local}$ and $\D l = \lambda_{local}$ and obtain 
\begin{equation}\lb{Ttautau}
T^{(\psi)\tau \tau}\sim \f{N\hbar}{4\pi r^2 \lambda_{local}^2}.
\end{equation}
Here, $T^{(\psi)\tau \tau}=-T^{(\psi)\tau}{}_\tau$ holds 
because we have $g^{\tau\tau}=-1$ in the local inertial frame. 
Furthermore, in the static metric \Ref{metricf}, we have $T^{(\psi)\tau}{}_\tau=T^{(\psi)t}{}_t$. 
Thus, we reach
\begin{equation}\lb{Ttt_formula}
-T^{(\psi)t}{}_t \sim \f{N\hbar}{4\pi r^2 \lambda_{local}^2}.
\end{equation}

Using this and \Ref{ADM}, the ADM-mass contribution from this excitation can be expressed as 
\begin{align}
(\D M)^{(\psi)} &=4\pi r^2 \D r (-T^{(\psi)t}{}_t)/N \nn\\
&\sim \f{\hbar}{\lambda_{local}^2}\D r\nn\\
&\sim \f{\hbar\sqrt{\s}}{r\lambda_{local}}.
\end{align}
Here, at the first line, we have divided it by $N$ 
because the s-wave is an excitation of one kind of field 
and $T^{(\psi)}_{\mu\nu}$ contains all the contribution from $N$ kinds of fields; 
and at the last line, we have used $\lambda_{local}=\D l =\sqrt{g_{rr}}\D r= \f{r}{\sqrt{2\s}}\D r $ 
for the interior metric of \Ref{metricf}. 
According to the formation process in the previous subsection, 
each s-wave excitation corresponds to the ADM energy of \Ref{e_mini}\footnote{
One might wonder that bound modes of s-waves with $\o\sim \MO(\f{1}{a})$ also 
contribute to this excitation as discussed in Fig.\ref{f:s_bound}. 
As shown in subsection \ref{s:ground}, however, such modes cannot be excited inside the black hole, 
and only the continuum modes of s-wave are responsible for \Ref{Tpsi_tt}.}. 
Therefore, the condition $(\D M)^{(\psi)} \sim \f{\hbar}{a}$ indicates 
\begin{equation}\lb{lam_local}
\lambda_{local}\sim \sqrt{\s},
\end{equation}
and then \Ref{Ttt_formula} becomes 
\begin{equation}\lb{Tpsi_tt}
-T^{(\psi)t}{}_t \sim \f{1}{4\pi G r^2 },
\end{equation}
where we have used $\s\sim Nl_p^2$. 
This should be the self-consistent form. 
In section \ref{s:energy}, we will discuss how to obtain this functional form 
from the field equation. 

We here discuss the entropy from this point of view. 
We consider a unit with width $\D l\sim \lambda_{local}\sim \sqrt{\s}$ inside the black hole. 
It contains $N$ waves with $\e\sim \f{\hbar}{a}$ 
because \Ref{lam_local} corresponds to $\D r =\f{\D l}{\sqrt{g_{rr}}} \sim \f{\s}{r}$ 
and the ADM mass of the unit is evaluated from \Ref{Tpsi_tt} as 
$4\pi r^2 \D r (-T^{(\psi)t}{}_t) \sim \f{\s}{G r}\sim \f{N\hbar}{a}$. 
Therefore, the entropy per proper radial length, $s$, can be evaluated as 
\begin{equation}\lb{s_local}
s\sim \f{N}{\D l} \sim \f{\sqrt{N}}{l_p}, 
\end{equation}
which means that $\MO(\sqrt{N})$ bits of information are contained per proper length. 
On the other hand, the proper length of the interior of the black hole is evaluated from \Ref{metricf} as 
\begin{equation}
l_{BH}=\int^{R(a)}_0 dr \sqrt{g_{rr}(r)} \approx \f{a^2}{2\sqrt{2\s}},
\end{equation}
which is much longer than $a$. 
Thus, the entropy is evaluated as 
\begin{equation}
S_{BH}\sim s \times l_{BH}\sim \f{\sqrt{N}}{l_p}\f{a^2}{\sqrt{\s}}\sim \f{a^2}{l_p^2}. 
\end{equation}

\subsection{The form of $T^{(\psi)}_{\mu\nu}$}\lb{s:Tpsi}
Using the form of $T^{(\psi)t}{}_t$ \Ref{Tpsi_tt} and the conservation law inside the black hole \Ref{con_eq}, 
we express all the components of $T^{(\psi)}_{\mu\nu}$ in terms of a single parameter $\wt a_\psi$. 
First, by the same reason as $\bra 0| T^\mu{}_{\nu}|0\ket_{ren}^\prime$, we can set the ansatz for $T^{(\psi)\mu}{}_\nu$ as \Ref{Eansatz}:
\begin{equation}
T^{(\psi)t}{}_t=a^{(\psi)}_0+\f{a^{(\psi)}_1}{r^2}+\cdots,~
T^{(\psi)r}{}_r=b^{(\psi)}_0+\f{b^{(\psi)}_1}{r^2}+\cdots,~
T^{(\psi)\th}{}_\th=c^{(\psi)}_0+\f{c^{(\psi)}_1}{r^2}+\cdots.
\end{equation}
Putting this into the conservation law \Ref{con_eq}, we can obtain (like \Ref{con_sol})
\begin{equation}\lb{con_psi}
b^{(\psi)}_0=a^{(\psi)}_0,~c^{(\psi)}_0=a^{(\psi)}_0+\f{-a^{(\psi)}_1+b^{(\psi)}_1}{4\eta\s},~
c^{(\psi)}_1=\f{a^{(\psi)}_1-b^{(\psi)}_1}{2}+\f{-a^{(\psi)}_2+b^{(\psi)}_2}{4\eta \s},\cdots.
\end{equation}

Now, \Ref{Tpsi_tt} shows that 
$T^{(\psi)t}{}_t$ is at most order of $\MO(r^{-2})$, 
which means $a^{(\psi)}_0=0$.
The first equation of \Ref{con_psi} indicates $b^{(\psi)}_0=a^{(\psi)}_0=0$. 
Next, as seen in subsection \ref{s:Tren}, 
the leading order $\bra0|T^\mu{}_{\nu}|0\ket_{ren}^{(0)\prime}$ appears as a result of the integration over $(\o,l)$. 
Therefore, $T^{(\psi)\mu}{}_\nu$, which represents the contribution of the s-wave excitation, 
should be smaller than $\MO(1)$. 
This requires $c^{(\psi)}_0=0$, 
and then the second one of \Ref{con_psi} leads to $b^{(\psi)}_1=a^{(\psi)}_1$. 
Thus, all the components are at most order of $\MO(r^{-2})$, 
which can be expressed as 
\begin{align}\lb{T_psi}
T^{(\psi)t}{}_t=a^{(\psi)}_1\l(\f{1}{r^2}+\f{4\kappa_t \eta \s}{r^4}\r)+\MO(r^{-6}),&~
T^{(\psi)r}{}_r=a^{(\psi)}_1\l(\f{1}{r^2}+\f{4\kappa_r \eta \s}{r^4}\r)+\MO(r^{-6}),\nn\\
T^{(\psi)\th}{}_\th&=c^{(\psi)}_1\l(\f{1}{r^2}+\f{4\kappa_\th \eta \s}{r^4}\r)+\MO(r^{-6}).
\end{align}
Here, $\kappa_t,\kappa_r,\kappa_\th$ are 
the $\MO(1)$ ratios between terms of $r^{-2}$ and $r^{-4}$ 
which are determined by dynamics of the s-waves including the contribution from the negative energy flow (see section \ref{s:energy}). 

Then, the third one of \Ref{con_psi} determines $c^{(\psi)}_1$ as 
\begin{equation}\lb{c_psi}
c^{(\psi)}_1=a^{(\psi)}_1 (-\kappa_t+\kappa_r).
\end{equation}
Thus, all the components of $\MO(r^{-2}) $ are expressed in terms of the parameter $a^{(\psi)}_1$ 
under a given $(\kappa_t,\kappa_r)$. 
Note that $a^{(\psi)}_1$ corresponds to the square of the amplitude of the classical field that 
represents the collapsing matter and radiation. 
For a later convenience, we represent $a_1^{(\psi)}$ in terms of a $\MO(1)$ parameter $\wt a_\psi$ as 
\begin{equation}\lb{wt_a_psi}
a_1^{(\psi)}=\f{\hbar N}{480\pi^2\eta^3\s}\wt a_\psi
\end{equation}
In section \ref{s:sol}, we will determine $\wt a_\psi$ by the self-consistent equation \Ref{Einstein2}. 

\section{Self-consistent solution}\lb{s:sol}
We have obtained all the ingredients for the self-consistent analysis: 
the candidate metric \Ref{metricf}, which is written in terms of $(\s,\eta)$, 
the energy-momentum tensor for the ground state \Ref{T0ren}, \Ref{Tt1ren} and \Ref{Ttrace1ren}, 
which depend on $(\a_0,\b_0)$, 
and the energy-momentum tensor for the s-wave excitation contribution \Ref{T_psi}, 
which is controlled by $\wt a_\psi$.
In this section, we solve the Einstein equation \Ref{Einstein2} 
and determine the self-consistent values of $(\s,\eta,\wt a_\psi)$
for a certain class of $(\a_0,\b_0)$. 
We then examine the consistency. 

\subsection{Self-consistent values of $(\s,\eta,\wt a_\psi)$ for a class of $(\a_0,\b_0)$}
First, we examine the trace part of \Ref{Einstein2}, 
$G^\mu{}_\mu=8\pi G (\bra 0| T^\mu{}_\mu|0\ket_{ren}^\prime + T^{(\psi)\mu}{}_\mu )$, to $\MO(r^{-2})$. 
It is obtained from \Ref{curve1}, \Ref{trace0}, \Ref{Ttrace1ren}, and \Ref{T_psi} 
(with \Ref{c_psi} and \Ref{wt_a_psi}) as 
\begin{equation}\lb{trace_eq1}
\f{1}{\eta^2 \s}-\f{2}{r^2}
=8\pi G\l(\f{\hbar N}{960\pi^2} \l[\f{1}{\eta^4\s^2} +\f{4}{\eta^3\s r^2}\wt \tau_1 \r]  
+ \f{\hbar N }{240\pi^2 \eta^3\s r^2}\wt a_\psi (1-\kappa_t+\kappa_r) \r)
\end{equation}
Equating the terms of $\MO(1)$ on the both sides, we have 
$\f{1}{\eta^2 \s}=8\pi G\f{\hbar N}{960\pi^2\eta^4\s^2}$, that is, 
\begin{equation}\lb{sigma}
\s = \f{N l_p^2}{120\pi \eta^2}.
\end{equation}
This is indeed proportional to $N l_p^2$, 
which is consistent with the assumption we have put just below \Ref{da}. 
Also ,$\eta$ appears in the denominator, and $\s$ with $\eta >1$ is smaller than that with $\eta=1$, 
which is consistent with the meaning of $\eta$ in Fig.\ref{fmean}. 
Then, the terms of $\MO(r^{-2})$ in \Ref{trace_eq1} together with \Ref{sigma} bring 
$-\f{2}{r^2}= \f{4}{r^2\eta}\l[\wt \tau_1+  \wt a_\psi (1-\kappa_t + \kappa_r)\r] $, 
which means 
\begin{equation}\lb{eta1}
\eta = -2 \wt \tau_1 -2 \wt a_\psi (1-\kappa_t+\kappa_r).
\end{equation}

Next, we construct $G^t{}_t=8\pi G (\bra 0| T^t{}_t|0\ket_{ren}^\prime + T^{(\psi)t}{}_t)$ to $\MO(r^{-2})$. 
From \Ref{Gtensor}, \Ref{T0tren}, \Ref{Tt1ren}, and \Ref{T_psi} (with \Ref{wt_a_psi}), we have 
\begin{align}\lb{tt_eq}
-\f{1}{r^2} &= 8\pi G \l(\f{\hbar N}{1920\pi^2\eta^4\s^2}\l[2c+\g + \log\f{1}{32\pi \eta^2\s \mu^2_0}-\f{960\pi^2}{\hbar N }(2\a_0+\b_0) \r]\r.\nn\\
&\l.+\f{\hbar N }{480\pi^2\eta^3\s r^2}\l[\wt a_1 + \wt a_\psi +\f{960 \pi^2}{\hbar N}\b_0 \r]\r).
\end{align}
First, the terms of $\MO(1)$ are balanced on the both sides if $\a_0$ and $\b_0$ are tuned such that 
\begin{equation}\lb{ab_con}
2\a_0+\b_0=\f{\hbar N }{960\pi^2}\l(\g+2c + \log\f{1}{32\pi \eta^2\s \mu^2_0}\r).
\end{equation}
Then, the terms of $\MO(r^{-2})$ leads together with \Ref{sigma} to 
\begin{equation}\lb{eta2}
\eta = -2\l(\wt a_1 + \wt a_\psi +\f{960 \pi^2}{\hbar N}\b_0 \r).
\end{equation}
Note that the $rr$-component and $\th \th$-component of the Einstein equation hold automatically 
because of the conservation law and the trace equation. 

Now, using \Ref{eta1} and \Ref{eta2}, we can determine 
\begin{equation}\lb{eta3}
\eta = -2 \f{\wt \tau_1+(-1+\kappa_t-\kappa_r)\l(\wt a_1 +\f{960 \pi^2}{\hbar N}\b_0 \r)}{\kappa_t-\kappa_r},
\end{equation}
\begin{equation}\lb{a_psi}
\wt a_\psi = \f{\wt \tau_1-\l(\wt a_1 +\f{960 \pi^2}{\hbar N}\b_0 \r)}{\kappa_t-\kappa_r}.
\end{equation}
Here, $\b_0$ should be chosen so that \Ref{eta} is satisfied. 
Note that $\eta$ is actually constant, as we have assumed in subsection \ref{s:can}. 

Thus, we have determined the self-consistent values of $(\s,\eta,\wt a_\psi)$ 
in terms of $(\wt a_1,\wt \tau_1,\kappa_t,\kappa_r)$ that can be fixed in principle 
by dynamics (see \Ref{Tt1ren}, \Ref{Ttrace1ren} and \Ref{T_psi} for their definitions). 
Therefore, the interior metric of \Ref{metricf} and the state $|\psi\ket$ 
are the self-consistent solution of the Einstein equation \Ref{Einstein}. 
Note that this is a non-perturbative solution w.r.t. $\hbar$ because we cannot take $\hbar \to 0$ 
in the solution metric \Ref{metricf} with \Ref{sigma} and \Ref{eta3}. 

\subsection{Consistency check}\lb{s:check}
First, we check the curvature ${\mathcal R}$. 
From \Ref{curve1}, \Ref{curve2}, \Ref{curve3}, and \Ref{sigma}, we can evaluate 
\begin{equation}\lb{Rresult}
{\mathcal R}\sim \f{1}{Nl_p^2},
\end{equation}
which is smaller than the Planck scale if $N$ is large, \Ref{largeN}\footnote{
See e.g. \ci{Park1,Park2} for a Planckian-energy scale correction to the geometry.}. 
Thus, no singularity exists inside the black hole\footnote{See e.g. \ci{Bolokhov} for another semi-classical resolution of singularity.   } and the Penrose diagram is actually given by Fig.\ref{f:Penrose}. 
Note that this result is so robust because \Ref{sigma} is not affected by the values of $(\a_0,\b_0)$. 

Second, we study the energy density. 
From \Ref{T0tren}, \Ref{Tt1ren}, \Ref{T_psi} (with \Ref{wt_a_psi}), 
\Ref{eta3} and \Ref{a_psi}, we have 
\begin{equation}\lb{e_den}
-\bra \psi|T^t{}_t|\psi \ket_{ren}^\prime =\f{1}{8\pi G r^2},
\end{equation}
to which \textit{both} $\bra 0|T^t{}_t|0 \ket_{ren}^\prime$ and $T^{(\psi)t}{}_t$ contribute. 
Integrating this over the volume in \Ref{ADM}, we can reproduce the total energy $M$. 
We also see from \Ref{Jdef} that 
\Ref{e_den} also leads to the Hawking radiation consistently. 
Therefore, both the bound modes and s-waves play a role in the energy inside the black hole. 

Third, we discuss the tangential pressure. The self-consistent solution has 
\begin{equation}\lb{p_tan}
\bra \psi|T^\th{}_\th |\psi \ket_{ren}^\prime = \bra 0|T^\th{}_\th |0 \ket_{ren}^{\prime(0)}+\MO(r^{-2})
=\f{15}{2G Nl_p^2}+\MO(r^{-2}),
\end{equation}
which can be checked by \Ref{T0ren}, \Ref{T_psi}, \Ref{sigma} and \Ref{ab_con}. 
From this, we can see that this near-Planckian pressure comes 
from the vacuum fluctuation of the bound modes in the ground state. 
The origin is the same as the 4D Weyl anomaly 
because the value \Ref{p_tan} comes from the second term of \Ref{T0ren}, 
which appears from the pole $\f{1}{\e}$ as shown in \Ref{Ath}. 
Indeed, twice of \Ref{p_tan} is equal to the trace \Ref{trace0}. 
Therefore, the pressure is very robust in that it is independent of the state. 
Note also that this large pressure supports the object against the strong gravity, 
which can be seen by noting that in \Ref{con_eq} $T^\th{}_\th$ 
and $\p_r \log \sqrt{-g_{tt}}$ are balanced for the self-consistent solution. 

This large tangential pressure associated with the anomaly 
should be universal for any kind of matter fields. 
In the previous work \ci{KY3} only conformal matter fields were considered, and 
the 4D Weyl anomaly was used to obtain the self-consistent solution 
with the large tangential pressure.  
In the present work, however, 
we reach the same picture of the black hole without using the conformal property, 
which means that the assumption of conformal matter is not essential. 
It is related to the fact that any matter behaves ultra-relativistic near the black hole, 
as we have discussed around \Ref{r_t}.  
Therefore, our picture of the black hole should work for any kinds of matter fields. 
Even for a massive field it should if the mass is smaller than $\MO(m_p)$. 

The surface pressure $p_{2d}^{(i)}$ in \Ref{EMT_s_i} also comes from this tangential pressure. 
When a shell in Fig.\ref{f:multi} approaches so that \Ref{r_t3} is satisfied, 
the interior metric becomes \Ref{metricf} and 
the shell itself forms a new surface, 
whose position is represented by the point $R$ in Fig.\ref{f:poten}. 
Then, the vacuum fluctuation of the bound modes which reach the point $R$ 
produces the tangential pressure \Ref{p_tan}, 
which corresponds to the continuum version of $p_{2d}^{(i)}$. 
Note that 
the mode that makes the pressure is different from the mode that constitutes the shell  
while \Ref{EMT_s_i} has been obtained by a geometrical analysis. 
Therefore, the two modes are combined to form the consistent picture.

Finally, we discuss the finite renormalization. 
In this solution, there are no quantities with dimension other than $\s$. 
\Ref{sigma} implies that the renormalization point $\hbar \mu_0$ should be chosen near the Planck scale: 
\begin{equation}\lb{mu0}
\mu_0^2 \sim \f{1}{Nl_p^2}.
\end{equation}
Then, \Ref{ab_con} means that $2\a_0 + \b_0 \sim \MO(1) N\hbar$. 
On the other hand, \Ref{eta3} together with \Ref{eta} indicates that $\b_0\sim \MO(1)N\hbar$.  
Therefore, we have 
\begin{equation}
\a_0,~\b_0\sim \MO(1) N \hbar.
\end{equation}
From this and \Ref{Rresult}, we can see that 
all the four terms of gravity in \Ref{S_B} are $\sim \MO(\f{1}{G Nl_p^2})$. 
Therefore, the $R^2,R_{\a\b}R^{\a\b},R_{\a\b\mu\nu}R^{\a\b\mu\nu}$ terms cannot be much larger than the $R$ term, 
which means that the three terms would not change the picture of the black hole drastically 
even if we consider them from the beginning\footnote{Such terms may change each numerical coefficient up to $\MO(1)$, 
but they will not make a significant change to the basic picture.}.
Thus, our argument based on the Einstein-Hilbert term is self-consistent. 

As we have seen, the coupling constants $\a_0$ and $\b_0$ should be related by \Ref{ab_con}. 
However, there is a possible way to relax it. 
Instead of the interior metric in \Ref{metricf}, 
we consider $B(r)=C_1 r^{2+\zeta}$ and $A(r)=C_2 r^{2+\f{\zeta}{2}}$, where $\zeta$ is a small number. 
Because this ansatz satisfies \Ref{con_AdS}, 
we can solve the wave equation \Ref{p_eq} locally as discussed in the end of subsection \ref{s:pert}. 
Then, we can evaluate the renormalized energy-momentum tensor $\bra 0 |T^\mu{}_\nu|0\ket^\prime_{ren}$, 
which should contain $\zeta$-dependent terms. 
Therefore, we should be able to choose $\zeta$ so that the terms satisfy the Einstein equation \Ref{Einstein2} 
for a given value of $(\a_0,\b_0)$. 
It would be exciting to find another class of solutions in this direction. 

\section{Black hole entropy}\lb{s:entropy}
We count the number of states of the s-waves inside the black hole 
to evaluate the entropy more quantitatively than the discussion in subsection \ref{s:Tpsitt}. 
\subsection{Adiabatic formation in the heat bath}
Let us consider the adiabatic formation of the black hole in the heat bath \ci{KY1}. 
Specifically, suppose that we grow a small black hole to a large one very slowly 
by changing the temperature and size of the heat bath properly\footnote{
We can discuss the relaxation time, 
which is estimated as \Ref{t_scat}. See section 2-E of \ci{KY2}.}. 
Then, we obtain the stationary black hole, whose metric is given by \Ref{metricf} with $a=$const., 
and consider it as consisting of excitations of the s-wave continuum modes, as discussed in section \ref{s:excite}.

\begin{figure}[h]
\begin{center}
\includegraphics*[scale=0.28]{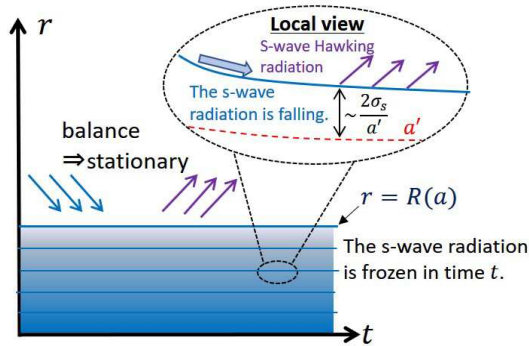}
\caption{Stationary black hole in the heat bath.}
\label{f:bath}
\end{center}
\end{figure}

This black hole is \textit{not} in equilibrium in an exact sense \ci{KY2}. 
Let us see the above formation process from a microscopic view. See Fig.\ref{f:bath}.
At a stage where the black hole has the radius $a'$, 
s-wave ingoing radiation from the bath comes to the black hole, 
balancing s-wave Hawking radiation of $T_H=\f{\hbar}{4\pi a'}$ with 
the s-wave intensity $\s_s$ \Ref{sigma_s}. 
The ingoing radiation approaches to 
\begin{equation}\lb{r_a'}
r=a'+\f{2\s_s}{a'}
\end{equation}
as we have shown \Ref{r_t3}. 
Then, the local temperature that the radiation feels there is 
\begin{equation}\lb{T_loc}
T_{local}=\l.\f{\f{\hbar}{4\pi a'}}{\sqrt{1-\f{a'}{r}}}\r|_{r=a'+\f{2\s_s}{a'}}\approx \f{\hbar}{4\pi\sqrt{2\s_s}},
\end{equation}
which is near the Planck temperature $\sim \f{\hbar}{\sqrt{N}l_p}$ because of \Ref{sigma_s}. 
After this, the subsequent radiations pile up and cover the radiation. 
Then, it is redshifted exponentially,  
almost frozen in time $t$, and suspends there. 
From the local point of view, however, the radiation keeps falling with the initial information. 
Therefore, the black hole is not in equilibrium in the usual sense. 
Rather, the ingoing radiation from the bath and the Hawking radiation are just balanced, 
and therefore the system is stationary. 

This property allows us to use \Ref{T_loc} as the local temperature at each radius $r$ inside the black hole. 
Because of the exponentially large redshift, 
the outgoing Hawking radiation there is frozen in time $t$.
Even if we wait for an exceptionally long time which is longer than the lifetime of the universe, 
the Hawking radiation of this temperature \Ref{T_loc} will not go outside 
to exchange energy with the radiation from the heat bath. 
This indicates that Tolman's law \ci{Landau_SM} doesn't hold inside the black hole\footnote{
If we apply Tolman's law to the interior metric of \Ref{metricf} naively, 
the local temperature would be $T_{loc}(r)\sim e^{\f{R^2-r^2}{4\s\eta}}\f{\hbar}{\sqrt{N}l_p}$, 
which is exponentially large (for $r \ll R- \f{2\s \eta}{R}$) compared to the Planck scale. 
This is not allowed physically.}. 
As we will see below, this is consistent with the area law of the entropy.

\subsection{Micro-counting of the s-wave states}
We count the number of possibilities of these s-wave configurations 
to find the entropy density $\hat s(r)$ per radius $r$ inside the black hole. 
First, we evaluate the number $\D n(r)$ 
of the s-waves with frequency $\leq \o$ in the width $\D r$ near the surface when the black hole of $a'$ is formed. 
In the semi-classical approximation \ci{Landau_QM}, 
using \Ref{BS_con} for $l=0$, $\MM=0$, $A=0$, and $B=\f{r}{r-a'}$, we can evaluate for $N$ kinds of fields 
\footnote{The s-waves are trapped in the heat bath with a finite size, 
and we can use approximately the condition \Ref{BS_con} to count the quantum number $n$ \ci{brick}.}
\begin{equation}
\D n (r)= \f{N}{\pi} \D r \f{1}{\sqrt{1-\f{a'}{r}}}\f{\o}{\sqrt{1-\f{a'}{r}}}
\approx \f{N}{\pi} \f{r}{\sqrt{2\s_s}} \wt \o \D r.
\end{equation}
Here, $\wt \o =\f{\o}{\sqrt{-g_{tt}}}=\f{\o}{\sqrt{1-\f{a'}{r}}}$ is 
the blueshifted frequency when approaching the surface, 
and we have 
approximated 
$\f{1}{\sqrt{1-\f{a'}{r}}}\approx \f{a'}{\sqrt{2\s_s}}\approx \f{r}{\sqrt{2\s_s}}$ near \Ref{r_a'}. 
Thus, the number of modes for $[\wt \o, \wt \o+\D \wt \o ]$ is given by 
\begin{equation}\lb{d_mode}
\f{d \D n(r)}{d \wt \o} \D \wt \o = \f{N}{\pi} \f{r}{\sqrt{2\s_s}} \D \wt \o\D r 
\equiv {\mathcal D}(\wt \o,r) \D \wt \o\D r. 
\end{equation}

Next, we review the entropy of a harmonic oscillator with frequency $\o$ 
in the heat bath of temperature $\b^{-1}$ \ci{Landau_SM}. 
The canonical distribution is given by $e^{-\b\l(a^\dagger a + \f{1}{2}\r)\hbar \o}Z(\b)^{-1}$, 
where 
\begin{equation}
Z(\b)=\tr e^{-\b\l(a^\dagger a + \f{1}{2}\r)\hbar \o} = 2\sinh \l(\f{1}{2}\hbar \b \o\r).
\end{equation}
Therefore, the entropy is evaluated as 
\begin{equation}\lb{S_osc}
S=\b^2 \f{\p}{\p \b}\l(-\b^{-1} \log Z(\b) \r)=g_s(\hbar \b \o),
\end{equation}
where we have defined a function
\begin{equation}\lb{gs}
g_s(x)\equiv \f{x}{e^x-1}-\log (1-e^{-x}).
\end{equation}

From \Ref{d_mode} and \Ref{S_osc}, we obtain the entropy density:
\begin{equation}\lb{s_formula}
\hat s(r)=\f{1}{2}\int_0^\infty d \wt \o {\mathcal D}(\wt \o,r)  g_s(\hbar \b_{local} \wt \o), 
\end{equation}
with $1/\b_{local}=T_{local}$ \Ref{T_loc}. 
The reason of the factor $\f{1}{2}$ is that because of the time-freezing effect
only ingoing modes carry the information, and 
the integration only over ingoing direction momenta should be performed. 
This can be evaluated as 
\begin{align}\lb{s_formula2}
\hat s(r) &=\f{N}{2\pi} \f{r}{\sqrt{2\s_s}} \f{1}{\hbar \b_{local}} \int^\infty_0 dx g_s(x)  \nn\\
 &= \f{N}{2\pi} \f{r}{\sqrt{2\s_s}} \f{1}{4\pi \sqrt{2\s_s}} \f{\pi^2}{3}  \nonumber \\
 &= \f{Nr}{48 \s_s}\nonumber \\
 &= \f{2\pi}{l_p^2} r,
\end{align}
where at the third line we have used \Ref{sigma_s}\footnote{From this, we can evaluate the entropy per the proper length $\D l =\sqrt{g_{rr}} \D r =\f{r}{\sqrt{2\s}} \D r $ as 
$s(r)= \f{\hat s(r)}{\sqrt{g_{rr}}}=\f{2\pi \sqrt{2\s}}{l_p^2}\sim \f{\sqrt{N}}{l_p}$,
which is consistent with \Ref{s_local} and a thermodynamical discussion \ci{KY2}.
Therefore, it is consistent to assume that \Ref{T_loc} is kept at each $r$ inside the black hole. }. 
Integrating this over the interior reproduces the area law: 
\begin{equation}
S=\int^{R(a)}_0 dr \hat s(r)= \f{\pi R(a)^2}{l_p^2}=\f{\pi a^2}{l_p^2}+\MO(1)\approx \f{A}{4l_p^2},
\end{equation}
where $R(a)^2=\l(a+\f{2\s}{a}\r)^2=a^2+4\s + \f{4\s^2}{a^2}$ and $A\equiv4\pi a^2$ 
have been used\footnote{See \ci{Yagi} for another approach to the black-hole entropy from the interior region.}. 
This implies that 
the information is stored inside the black hole as excitations of the s-waves. 

\section{Mechanism of energy decrease of the collapsing matter}\lb{s:energy}
In this section, we examine how the energy of the collapsing matter decreases 
and discuss how $T^{(\psi)\mu}{}_\nu$, \Ref{T_psi}, is obtained by dynamics. 
The point is to consider the contribution 
from the vacuum fluctuation of s-waves. 
\subsection{Setup}
\begin{figure}[h]
\begin{center}
\includegraphics*[scale=0.27]{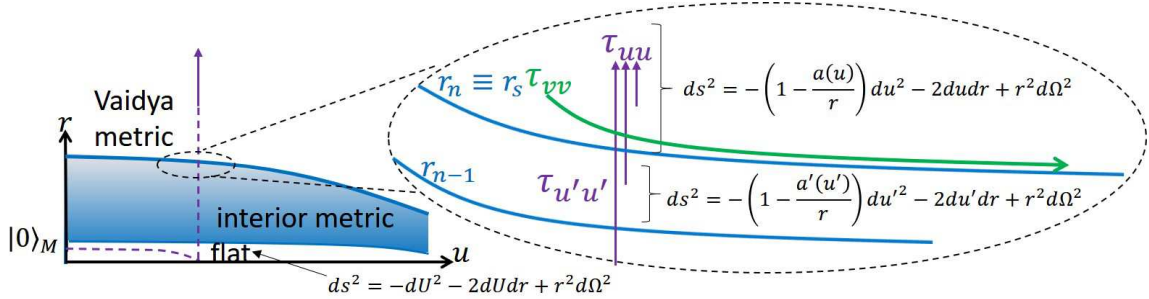}
\caption{Collapsing matter in the outermost region and energy flow $\tau_{uu},\tau_{vv}$.
The shell is described as the outermost shell of the multi-shell model of Fig.\ref{f:multi}.}
\label{f:2dmodel}
\end{center}
\end{figure}

Suppose that a classical matter collapses to the black hole which is described by the self-consistent metric. 
To analyze the time evolution of this matter, we can focus just on the part of the matter around the surface, 
since the deeper region is frozen in time due to the exponentially large redshift. 
We consider the matter as consisting of many shells like the multi-shell model of Fig.\ref{f:multi} and 
study the energy of the outermost shell along $r=r_n(u)\equiv r_s(u)$ as in Fig.\ref{f:2dmodel}. 

We use the Vaidya metric \Ref{SchV} in the $(u,v)$ coordinate to describe the outside region: 
\begin{align}\lb{Vaidya}
ds^2 &=-\l(1-\f{a(u)}{r}\r)du^2 -2dudr +r^2 d\Omega^2 \nn \\
 &=-e^{\vp(u,v)}dudv+r(u,v)^2 d\Omega^2,
\end{align}
where $a(u)$ follows \Ref{da} and each function is given by 
\begin{equation}\lb{vpr}
\vp(u,v)=-\int^u_{-\infty}d\tilde u\f{a(\tilde u)}{2r(\tilde u,v)^2},~~\p_ur(u,v)=-\f{r(u,v)-a(u)}{2r(u,v)},~~\p_vr(u,v)=\f{1}{2}e^{\vp(u,v)}.
\end{equation}
We label the outermost shell by $v_s$ so that its location is given by $r_s(u)=r(u,v_s)$. 

\subsection{S-wave model of quantum fields}
So far we have considered the collapsing matter as an excitation of s-waves, 
and we have also seen that only s-waves can enter and exit the black hole. 
Therefore, 
in order to study quantum fields propagating in this time-dependent background spacetime, 
we can use the s-wave approximation to analyze $N$ massless free scalar fields 
as we have done in section \ref{s:idea} \footnote{\lb{foot:swave}Even s-waves can have a tangential pressure, but 
the largest tangential pressure of $\MO(1)$ in our picture comes from the bound modes 
with $l\gg1$ as we have seen in section \ref{s:sol}. 
Therefore, we can expect that the tangential pressure from s-waves does not 
make a significantly important effect. }.  
In this subsection, we explain the general setup \ci{BD,CFTbook}. 

We consider s-waves of $N$ kinds of scalar fields, $\phi_a=\phi_a(x^i)$, 
in a spherically symmetric metric, 
where $x^i$ is the 2D part of the coordinate. 
The 4D action can be reduced to the 2D one:
\begin{align}\lb{2dS}
S &=-\f{1}{2}\sum_{a=1}^N\int d^4x \sqrt{-g}g^{\mu\nu}\p_\mu \phi_a \p_\nu \phi_a \nn   \\
 &=-\f{1}{2}\sum_{a=1}^N 4\pi \int d^2x_{2d} \sqrt{-g_{2d}}r^2 g^{ij} \p_i \phi_a \p_j \phi_a \nonumber \\
 &\approx -\f{1}{2}\sum_{a=1}^N\int d^2x_{2d} \sqrt{-g_{2d}}g^{ij} \p_i \Phi_a \p_j \Phi_a,
\end{align}
where $r$ is the areal radius, and the derivative $\p_i$ is applied only to $\phi(x^i)$ as the approximation.
Here, we have defined the 2D field as 
\begin{equation}\lb{2dphi}
\Phi_a\equiv \sqrt{4\pi}r\phi_a.
\end{equation}
The 2D energy momentum tensor is given by 
\begin{equation}\lb{tau}
\tau_{ij}\equiv \f{-2}{\sqrt{-g_{2d}}}\f{\d S}{\d g^{ij}}
=\sum_{a=1}^N\l(\p_i \Phi_a \p_j \Phi_a -\f{1}{2} g_{ij}g^{i'j'}\p_{i'}\Phi_a\p_{j'}\Phi_a\r),
\end{equation}
which is related to the 4D energy-momentum tensor $T_{ij}$ through \Ref{2dphi} as 
\begin{equation}\lb{tauT}
\tau_{ij}\approx 4\pi r^2 T_{ij}.
\end{equation}
Classically, this 2D part is traceless, $\tau^i{}_i=0$. 
However, the 2D Weyl anomaly contributes to the trace part \ci{BD,CFTbook}: 
\begin{equation}\lb{2d_ano}
\tau^i{}_i=\f{N\hbar}{24\pi}R_{2d},
\end{equation}
where $R_{2d}$ is the 2D Ricci scalar. 

We here consider a 2D metric of the form $ds_{2d}^2 =-e^{\vp(u,v)}dudv$. 
Then, the 2D action \Ref{2dS} becomes $S=\sum_{a=1}^N\int dudv\p_u \Phi_a \p_v \Phi_a$, and 
the equation of motion $\f{\d S}{\d \Phi_a}=0$ is 
\begin{equation}\lb{2deom}
\p_u \p_v \Phi_a=0.
\end{equation}
Each component of the energy-momentum tensor is given from \Ref{tau} and \Ref{2d_ano} by 
\begin{equation}\lb{tau_uv}
\tau_{uu}=\sum_{a=1}^N(\p_u\Phi_a)^2,~~\tau_{vv}=\sum_{a=1}^N(\p_v\Phi_a)^2,~~\tau_{uv}=-\f{N \hbar}{24\pi }\p_u\p_v \vp \equiv 2\g \p_u\p_v \vp,
\end{equation}
where we have used $R_{2d}=4e^{-\vp}\p_u\p_v\vp$. 
The conservation law $\N^i_{(2d)}\tau_{ij}=0$ is given by
\begin{align}\lb{con_u}
\p_v\tau_{uu}+e^\vp \p_u(e^{-\vp}\tau_{uv}) &=0, \\
\lb{con_v}
\p_u\tau_{vv}+e^\vp \p_v(e^{-\vp}\tau_{vu}) &=0.
\end{align}

In the following, we consider $\tau_{ij}$ as the expectation value 
with respect to a state $|\psi\ket$. 
The information of $|\psi\ket$ is introduced as a boundary condition of $\tau_{ij}$.

\subsection{Hawking radiation}\lb{s:Hawking}
Now, let us find $\tau_{uu}$ in our setup of Fig.\ref{f:2dmodel}. 
Initially $(u=-\infty)$, there is only the collapsing matter with $\tau_{vv}^{(0)}(v)$. 
Therefore, $\tau_{uu}$ appears as a result of dynamics of the quantum fields. 

First, we can use the third of \Ref{tau_uv} to express \Ref{con_u} as 
$\p_v \tau_{uu}+2\g\p_v\l(\p_u^2\vp-\f{1}{2}(\p_u\vp)^2 \r)=0.$
From this, we obtain 
\begin{equation}\lb{uu1}
\tau_{uu}=\g \l((\p_u \vp)^2-2\p_u^2 \vp \r)+\tau^{(0)}_{uu}(u).
\end{equation}

Next, we determine $\tau^{(0)}_{uu}(u)$ by the boundary condition given above \Ref{J_Sch}. 
That is, the quantum fields have started in the Minkowski vacuum state $|0\ket_M$ from a distance 
(see the left of Fig.\ref{f:2dmodel})\footnote{
Note that this vacuum is different from the ground state $|0\ket$ in the interior metric \Ref{vac}.}. 
This corresponds to the condition that $\tau^{(0)}_{UU}(U)=0$, 
where $U$ is the outgoing null time in the flat region around $r=0$. 
In general, in order for $\tau_{uu}$ to transform covariantly under $u\to U(u)$, 
this holomorphic term must change as \ci{CFTbook,Iso}
\begin{equation}\lb{Sch_trans}
\tau^{(0)}_{uu}(u)=\l(\f{dU}{du}\r)^2\tau^{(0)}_{UU}(U) + \f{N\hbar}{16\pi}\{u,U\}.
\end{equation}
Thus, \Ref{uu1} becomes 
\begin{equation}\lb{uu2}
\tau_{uu}=\g \l((\p_u \vp)^2-2\p_u^2 \vp \r)+\f{N\hbar}{16\pi}\{u,U\}.
\end{equation}

The shell we are now considering corresponds to the outermost one in the multi-shell model of section \ref{s:multi}. 
Therefore, we can use the result \Ref{xi_ii}, that is, 
\begin{equation}\lb{xi_uu}
\xi \equiv \log \f{dU}{du}=-\f{a(u)^2}{4\s}.
\end{equation} 
Calculating the first two terms of \Ref{uu2} from \Ref{vpr} 
and evaluating $\{u,U\}$ with \Ref{xi_uu} in the same way as in deriving \Ref{uiU} from \Ref{xi_i}, 
then \Ref{uu2} becomes 
\begin{equation}\lb{uu3}
\tau_{uu}=-\f{\hbar N}{192\pi}\l[\f{a^2}{r^4}+\f{4a}{r^3}\l(1-\f{a}{r}\r) +\f{4\dot a}{r^2}\r]+\f{\hbar N}{192\pi}\l(\f{1}{a^2}+\f{4\dot a}{a^2}\r),
\end{equation}
where the last two terms come from $\tau^{(0)}_{uu}(u)= \f{N\hbar}{16\pi}\{u,U\}$. 
This formula can be applied to any point outside the black hole. 

At $r\gg a$, this becomes 
\begin{equation}\lb{uu4}
\tau_{uu}\xrightarrow{r \gg a}\f{\hbar N}{192\pi a^2},
\end{equation}
which is consistent through \Ref{tauT} with \Ref{da_i}
\footnote{\lb{foot:2Dself}We can also show \Ref{J'} in this general framework. 
By replacing $u$ and $U$ with $u'$ and $u$ in \Ref{Sch_trans} respectively, 
we have 
$J'=\tau_{u'u'}|_{r\gg a}=\tau^{(0)}_{u'u'}(u')
=\l(\f{du}{du'}\r)^2\tau^{(0)}_{uu}(u) + \f{N\hbar}{16\pi}\{u',u\}$, 
where $\tau^{(0)}_{uu}(u) $ is given by \Ref{uu4}. 
This provides \Ref{J'}.}. 

On the other hand, $\tau_{uu}$ vanishes near the surface at $r=a+\f{2\s}{a}$: 
\begin{equation}\lb{uu5}
\tau_{uu}\xrightarrow{r \to a+\f{2\s}{a}} 0,
\end{equation}
which is consistent with the literature \ci{BD, DFU, BH_model, Fulling}. 

Thus, the Hawking radiation of the s-waves 
is created at $r\sim (1+k)a$ ($k=\MO(1)$) and emits the energy of the system. 
As we will see in the next subsection, 
a negative energy flow with the same magnitude as \Ref{uu4} 
is induced near $r=a+\f{2\s}{a}$, 
and the energy inside $r=a+\f{2\s}{a}$ decreases as $\f{da}{du}=-\f{\s_s}{a^2}$. 
This fact justifies the use of the Vaidya metric 
in studying the motion of a particle near $r=a+\f{2\s}{a}$. 

\subsection{Ingoing negative energy flow near the surface}
In turn, we determine the ingoing energy flow $\tau_{vv}$ by using \Ref{con_v}. 
From \Ref{vpr}, we have $\p_v \p_u \vp =\f{a}{r^3}\p_v r =\f{a}{2r^3}e^\vp$, 
which gives $\tau_{uv}=\g\f{a}{r^3}e^\vp$ through the third of \Ref{tau_uv}. 
From this and \Ref{vpr}, we can express \Ref{con_v} as 
$\p_u \tau_{vv}=-e^\vp \p_v\l(\g\f{a}{r^3} \r)=\f{3\g a}{2r^4}e^{2\vp}$.
Therefore, we obtain the general expression
\begin{equation}\lb{vv1}
\tau_{vv}=\f{3\g}{2}\int^u_{-\infty}d\tu \f{a(\tu)}{r(\tu,v)^4}e^{2\vp(\tu,v)}+\tau^{(0)}_{vv}(v). 
\end{equation}

Let us focus $\tau_{vv}$ at $v\sim v_s$. 
Then, the boundary term $\tau^{(0)}_{vv}(v)$ represents the energy of the outermost shell 
that comes from a distance at $u=-\infty$. 
The shell is classical, 
and the configuration $\Phi^{(cl)}_a(v)$ is an ingoing solution of \Ref{2deom} 
so that $\tau_{vv}^{(cl)}(v)=\sum_{a=1}^N(\p_v \Phi^{(cl)}_a(v))^2$ is non-zero only around $v\sim v_s$ and zero otherwise. 
We then put this classical ingoing energy flow as an idealization 
\begin{equation}\lb{v_s}
\tau_{vv}^{(0)}(v)=W\d(v-v_s),
\end{equation}
where $W$ is a positive constant. 
This indicates that the energy of the matter is kept along a line of $v=v_s$.  

Now, suppose that this shell 
comes close to $r=a+\f{2\s}{a}$ at a time $u=u_*$. 
Then, we can evaluate 
\begin{equation}\lb{vp_vs}
\vp(u,v_s) \approx \begin{cases}
 \log \f{r_s(u)-a_*}{r_s(u)},~~{\rm for}~~u\lesssim u_*,\\
 \log\f{2\s}{a_*^2}+\f{a(u)^2-a_*^2}{4\s},~~{\rm for}~~u_* \lesssim u,
\end{cases}
\end{equation}
and obtain for $u\gtrsim u_*$ 
\begin{equation}\lb{vv2}
\tau_{vv}(u,v_s)\approx -\f{N\hbar}{192\pi a(u)^2}+W\d(v-v_s)|_{v\to v_s},
\end{equation}
where $a_*=a(u_*)$ (see Appendix \ref{A:vv} for the derivation). 
The first negative term has the same absolute value as the outgoing energy flux at infinity \Ref{uu4} \ci{BD, DFU}. 
Also we can check that the 4D energy density of this negative part is order of $\MO(a^{-2})$ (see \Ref{den_vac_A}).

The mechanism for reducing the energy of the system is as follows (see Fig.\ref{f:2dmodel}). 
First, the collapsing matter comes with the positive ingoing energy $\tau_{vv}^{(0)}$. 
As it approaches to the black hole, 
the negative energy flow is created from the vacuum fluctuation of the s-waves 
and superposed on the matter. 
At the same time, the same amount of the positive energy flow is emitted to infinity. 
Thus, the negative energy cancels the positive energy of the matter locally, 
and the energy of the system decreases. 
We can also see how this occurs in the interior metric (see section 5 of \ci{KY1} and section 4-E of \ci{KY2})
\footnote{See e.g. \ci{Honega,Ho_vac1,Ho_vac2,Ho_vac3} for attempts to construct a metric with vacuum energy.}.

\subsection{Configuration of s-wave and $T^{(\psi)\mu}{}_\nu$}\lb{s:swave}
Here, by considering the above mechanics of the energy decrease, 
we discuss how to obtain $-T^{(\psi)t}{}_t\sim \f{1}{r^2}$ in \Ref{T_psi} 
directly from dynamics of fields. 

We first point out that 
the energy of the outermost shell in the multi-shell model 
decreases exponentially in the time $\D u \sim 2a$ \ci{KY2} (see the right of Fig.\ref{f:2dmodel}). 
This is the same statement as \Ref{Da_sol} (or \Ref{Da_sol2}) in Fig.\ref{f:coreshell}.   
Here, note that in Fig.\ref{f:coreshell} and Fig.\ref{f:2dmodel}, the roles of $(u,a)$ and $(u',a')$ are reversed!
Therefore, by almost the same calculation as \Ref{total_J}, we have in the case of Fig.\ref{f:2dmodel} 
\begin{align}\lb{evo_da}
\l(\f{d\D a}{du}\r)_{self-consistent} &=\f{da}{du}-\f{du'}{du}\f{da'}{du'} \nn \\
 &=-\f{\s_s}{a^2}+\f{r_s-a}{r_s-a'}\f{\s_s}{a'^2} \nonumber \\
 &\approx -\f{\s}{a^2}+\l(1-\f{a}{2\s_s}\D a\r)\f{\s}{a^2}\l(1+\f{2\D a}{a}\r) \nonumber \\
 &\approx -\f{1}{2a}\D a.
\end{align}
This is the self-consistent time evolution of the energy 
of the matter that has come from the outside and become a part of the evaporating black hole.

The behavior \Ref{evo_da} can be understood as a combination of 
the increase of the energy of the classical matter 
and the decrease of the energy by the negative energy flow: 
\begin{equation}\lb{evo_da_deco}
\l(\f{d\D a}{du}\r)_{self-consistent}=\l(\f{d\D a}{du}\r)_{classical}-2G \D J=\f{1}{2a}\D a - \f{1}{a}\D a =-\f{1}{2a}\D a, 
\end{equation}
where the first and second terms are the classical and quantum contributions, respectively. 
We explain these terms below. 

First, it is natural that, 
when we throw a particle to the evaporating black hole, 
the energy of the particle itself increases if we don't consider any quantum effect. 
This is because when viewed from a distant, 
the approaching particle is pulled 
by the gravitational force of the ``escaping" surface of the black hole 
and thus gains the extra kinetic energy just like gravity assist for spacecraft.
In fact, we can consider the contribution only from the ingoing energy $\tau_{vv}^{(0)}(v)$ of 
the classical shell \Ref{v_s} and evaluate the Bondi mass as (see Appendix \ref{A:Bondi})\footnote{This exponential increase of the classical energy also appears in the interior metric. 
The excitation part of the energy density \Ref{DTtt} 
takes the same form as the energy density of the outermost shell, \Ref{T_cl2}. 
This is natural because taking the continuum limit of the multi-shell model 
has led to the interior metric.}
\begin{equation}\lb{dBon}
(\D M)_{classical} \approx e^{-\f{a(u)^2-a_*^2}{4\s}}W, 
\end{equation}
from which we have the classical evolution of the mass
\begin{equation}\lb{evo_da_cl}
\l(\f{d\D a}{du}\r)_{classical}=-\f{a}{2\s}\f{da}{du}\D a =\f{1}{2a}\D a,
\end{equation}
where we have used \Ref{da}. This gives the first term in \Ref{evo_da_deco}.

Second, we evaluate the contribution of the negative energy flow that directly reduces the energy of the shell. 
As we have seen in the previous subsections, 
the amount of the negative energy flow near the black hole is the same as that of the positive energy flow going out. 
Thus, the contribution we should take is equal to the difference $\D J$ 
between the energy flows $\tau_{uu}$ and $\tau_{u'u'}$ exiting the systems with mass $\f{a}{2G}$ and $\f{a'}{2G}$, 
respectively (see the right of Fig.\ref{f:2dmodel}). 
More precisely, it is given by 
\begin{equation}
\D J \equiv \l.\l(\tau_{uu}-\l(\f{du'}{du}\r)^2 \tau_{u'u'} \r)\r|_{r=\infty},
\end{equation}
where we have considered the two redshift factors in order to translate $\tau_{u'u'}$ in terms of $u$. 
This can be evaluated as 
\begin{align}\lb{evo_da_qm}
\D J&=\f{N\hbar}{192\pi a^2}-\l(\f{r_s-a}{r_s-a'}\r)^2 \f{N\hbar}{192\pi a'^2}\nn\\
&\approx\f{N\hbar}{192\pi}\l(\f{1}{a^2} - \l(1-\f{a}{\s}\D a\r) \f{1}{a^2}\l(1+2\f{\D a}{a}\r)\r) \nonumber \\
&\approx \f{Nl_p^2}{2G\cdot 96\pi } \f{\D a}{\s a} \nonumber \\
&= \f{1}{2Ga}\D a.
\end{align}
Here, at the first line, \Ref{uu4} has been applied to both $\tau_{uu}$ and $\tau_{u'u'}$ because \Ref{xi_uu} holds for both $u$ and $u'$; 
at the second line, \Ref{r_t3} has been used; 
and at the last line, \Ref{sigma_s} has been considered. 
This provides the second term of \Ref{evo_da_deco}.

Now, we discuss the energy density. 
As shown in Appendix \ref{A:Bondi}, the energy density corresponding to \Ref{dBon} is given by 
\begin{equation}\lb{T_cl2}
-T^{(cl)u}{}_{u}\approx \f{W}{4\pi \s}e^{-\f{r^2-a_*^2}{2\s}} \d(v-v_s),
\end{equation}
while the energy density corresponding to the negative energy flow in \Ref{vv2} is evaluated as
\begin{equation}\lb{den_vac_A}
-T^{(vac)u}{}_{u}\approx -\f{1}{8\pi G r^2}  e^{-\f{r^2-a_*^2}{2\s}},
\end{equation}
where we express these in the $(u,r)$ coordinate of \Ref{Vaidya}\footnote{
\Ref{den_vac_A} shows that the order of the negative energy density from the vacuum 
is $\MO(a^{-2})$ in the outermost region, which is consistent with a general result \ci{Honega}.}. 
These are not consistent with $-T^{(\psi)t}{}_t\sim \f{1}{r^2}$ in \Ref{T_psi}. 

However, the above observation \Ref{evo_da_deco} implies that 
there should be a way to include the effect of the negative energy flow in solving the field equation of the s-wave. 
Such an analysis could be similar to ``Lorentz friction".
While the radiation is emitted \textit{directly} from the particle in the case of the Lorentz friction, 
the negative energy flow in our case is induced 
\textit{through} the change of the metric by the motion of the shell. 
In this sense, these might be qualitatively different. 
It should be useful to consider from a field-theoretic point of view 
the discussion of subsection \ref{s:step2}, 
which determines the time-evolution of both the spacetime and the energy and motion of the shell. 
Then, we should be able to obtain the proper functional form $T^{(\psi)t}{}_t\sim \f{1}{r^2}$, 
and the form of $T^{(\psi)\mu}{}_\nu$ \Ref{T_psi} will be determined 
together with $\kappa_t,\kappa_r,\kappa_\th$. 
We will consider this problem in future (see also the end of section \ref{s:concl}).

\section{Conformal matter fields}\lb{s:conformal}
As a special case, we consider conformal fields as the matter fields in \Ref{Einstein}\footnote{
There are various studies that examine 4D black holes using conformal fields. 
See e.g. \ci{Fulling, HKMY}.}. 
In the previous work \ci{KY3}, we have investigated this case 
and constructed the self-consistent metric.  
Here, we perform further analysis to completely determine $(\s,\eta)$. 
\subsection{Self-consistent values of $(\s,\eta)$ for conformal matter fields} 
The key equation is the trace part of the Einstein equation \Ref{Einstein}. 
The trace of the energy-momentum tensor 
is given by the 4D Weyl anomaly, independently of the state $|\psi\ket$ \ci{BD,Parker,Duff}: 
\begin{equation}\lb{conf_eq}
G^\mu{}_\mu=8\pi G \hbar \l(c_W {\mathcal F}-a_W {\mathcal G} +b_W \Box R \r),
\end{equation}
with 
\begin{equation}\lb{FG}
{\mathcal F}\equiv C_{\a\b\g\d}C^{\a\b\g\d}=R_{\a\b\g\d}R^{\a\b\g\d}-2R_{\a\b}R^{\a\b}+\f{1}{3}R^2,~~
{\mathcal G}\equiv R_{\a\b\g\d}R^{\a\b\g\d}-4R_{\a\b}R^{\a\b}+R^2,
\end{equation}
where $C_{\a\b\g\d}$ is the Weyl curvature and ${\mathcal G}$ is the Gauss-Bonnet density. 
The coefficients $c_W$ and $a_W$ are fixed only by the matter content of the action, 
as we will give explicitly below, while $b_W$ also depends on the finite renormalization. 

We can use the interior metric of \Ref{metricf} 
because the multi-shell model is independent of the matter content. 
Using \Ref{curve1}, \Ref{curve2}, \Ref{curve3} and \Ref{box_R}, 
the self-consistent equation \Ref{conf_eq} becomes 
\begin{equation}\lb{conf_eq1}
\f{1}{\eta^2\s}-\f{2}{r^2}=8\pi l_p^2\l[c_W \f{1}{3\eta^4\s^2}-\f{4}{\eta^2\s r^2}\l(c_W\l(\f{1}{\eta}+\f{1}{3}\r)-a_W \r)\r] +\MO(r^{-4}) 
\end{equation}
to $\MO(r^{-2})$. Note that $b_W$ does not contribute here because of $\Box R=\MO(r^{-4})$ from \Ref{box_R}. 
Equating the terms of $\MO(1)$ on the both sides, we reach \ci{KY1,KY3}
\begin{equation}\lb{sigma_conf}
\s = \f{8\pi l_p^2 c_W}{3\eta^2},
\end{equation}
which is similar to \Ref{sigma}. 
Next, comparing the terms of $\MO(r^{-2})$ and using \Ref{sigma_conf}, we can obtain easily 
\begin{equation}\lb{eta_conf}
\eta=\l(\f{a_W}{c_W}-\f{1}{6}\r)^{-1},
\end{equation}
which is a new result
\footnote{If a perturbation changes $B(r)$ as $B(r)\to \f{r^2}{2\s}+b$, where $b$ is a constant of $\MO(1)$, 
$b$ would appear as terms of $\MO(r^{-2})$ in \Ref{conf_eq1}, and $\eta$ would contain $b$. 
This change corresponds to a shift $r\to r+k\f{\s}{r},~(k=\MO(1))$, 
and it should be the leading perturbation for an expansion w.r.t $r\gg l_p$. 
(Note that such a constant shift of $A(r)$ could be removed by redefining a time coordinate.)
However, we can use a thermodynamical discussion to identify $B(r)=\f{r^2}{2\s}$ 
with the accuracy of $\D r =\MO\l(\f{l_p^2}{a}\r)$ \ci{KY1}. 
Therefore, we can compare the both sides of \Ref{conf_eq1} properly to $\MO(r^{-2})$.}. 
Note here that 
this result is not influenced by a finite renormalization 
because $H^\mu{}_\mu,K^\mu{}_\mu=\MO(r^{-4})$ from \Ref{tr_1} and \Ref{box_R}.
Thus, the interior metric has been determined completely only by the matter content, 
independently of the state and the theory of gravity. 

\subsection{A new constraint to the matter content}
We discuss the meaning of \Ref{eta_conf}. 
First, the explicit values of $c_W$ and $a_W$ are given by 
\begin{equation}\lb{ca_value}
c_W=\f{1}{1920\pi^2}(N_S+6N_F+12N_V),~~
a_W=\f{1}{5760\pi^2}(N_S+11N_F+62N_V),
\end{equation}
where $N_S$, $N_F$ and $N_V$ are respectively the numbers of scalars, 
spin-$\f{1}{2}$ Dirac fermions and vectors in the theory \ci{Duff}\footnote{
The graviton effect should be included here, 
but we ignore it because there seems no available result 
that satisfies the consistency condition of the trace anomaly \ci{BD,G_anomaly1,G_anomaly2,G_anomaly3}.}. 
Using this, \Ref{eta_conf} can be expressed as 
\begin{equation}\lb{eta_conf2}
\eta=6\f{N_S+6N_F+12N_V}{N_S+16N_F+112N_V}.
\end{equation}
For example, we can see 
\begin{equation}\lb{eta_ex}
\eta=\begin{cases}
 6~~~~{\rm scalar},\\
\f{9}{4}~~~~{\rm Dirac~ferminon},\\
\f{9}{14}~~~~{\rm vector}.
\end{cases}
\end{equation}

The consistency condition of $\eta$ is \Ref{eta}: $1\leq \eta <2$. 
The lower limit has come from the positive definiteness of the ratio of the scattered part of the radiation as in Fig.\ref{fmean},
while the upper limit has been derived by the positive definiteness of the radial pressure in \Ref{T3}
\footnote{If we relax the condition of $\eta$ more, at least $\eta$ must be non-negative: $\eta>0$.
If $\eta\leq0$, the redshift $g_{tt}=-\f{2\s}{r^2}e^{-\f{R(a)^2-r^2}{2\s\eta}}$ 
would make no sense.}. None of the above examples meet the condition. 

Let's examine whether various theories satisfy the condition \Ref{eta}. 
The first one is the Standard Model with right-handed neutrino\footnote{
Here, we assume that the Standard Model is a kind of conformal field theory and study what happens to \Ref{eta_conf2}.}. 
It has $N_S=4$, $N_F=24$, and $N_V=12$, and then \Ref{eta_conf2} becomes 
\begin{equation}
\eta=\f{438}{433}\approx1.012,
\end{equation}
which satisfies \Ref{eta}. 
On the other hand, if we remove right-handed neutrino, 
$N_F=24$ changes to $N_F=24-\f{1}{2}\times 3$, and then \Ref{eta_conf2} gives 
\begin{equation}
\eta=\f{849}{854}\approx0.994,
\end{equation}
which is outside \Ref{eta}. 
This result suggests that right-handed neutrino should exist.

Next, we consider $\MN=4$ supersymmetric theory. 
In this case, we have $N_S=6$, $N_F=2$ and $N_V=1$, 
and \Ref{eta_conf2} lead to 
\begin{equation}
\eta=\f{6}{5}=1.2,
\end{equation}
which satisfies \Ref{eta}. 
Instead, if we apply \Ref{eta_conf2} to $\MN=1$ supersymmetric theory with vector-multiplet, 
which has $N_S=0, N_F=\f{1}{2},N_V=1$, we get 
\begin{equation}
\eta=\f{3}{4}=0.75,
\end{equation}
which does not meet \Ref{eta}. 

In this way, the condition \Ref{eta} with the formula \Ref{eta_conf2}, 
which is a result of the self-consistent solution of the evaporating black hole,  
can provide a constraint to the matter content and classify effective field theories. 
The weak-gravity conjecture tells that 
black holes should evaporate without global symmetry charge \ci{Swamp1,Swamp2}.
Therefore, \Ref{eta} and \Ref{eta_conf2} may play a similar role
\footnote{See section 3-D of \ci{KY2} for a discussion on the non-conservation of baryon number in our picture.}. 
It would be exciting to study phenomenology using them. 

\section{Conclusion and discussions}\lb{s:concl}
In this paper, we have solved time evolution of the collapsing matter 
taking the full dynamics of the 4D quantum matter fields into account 
in the semi-classical Einstein equation. 
Then, we found the self-consistent metric $g_{\mu\nu}$  and state $|\psi\ket$. 
This solution  tells that 
the black hole is a compact dense object which has the surface (instead of horizon) and 
evaporates without a singularity\footnote{
Even if we consider the evaporating black hole as a closed trapped region, 
which is the conventional model e.g.\ci{Hayward}, 
the collapsing matter is \textit{below} the timelike trapping horizon 
just by a proper Planckian distance \ci{HMY1,HMY2}. 
This is similar to our picture in that the matter is \textit{above} 
the would-be horizon by the Planckian distance \Ref{dl}. 
These imply that such a conventional model is eventually very close to ours. }. 
It consists of three parts.
One is the vacuum fluctuation of the bound modes in the ground state, 
which produces the large tangential pressure supporting the object.  
Another one is the excitation of the s-waves composing the collapsing matter, 
which carries the information responsible for the area law of the entropy. 
The other one is the excitation of the s-waves producing the pair of 
the outgoing Hawking radiation and the ingoing negative energy flow, 
which decreases the energy of the matter. 
These three energy-momentum flows are conserved independently if there is no interaction among them. 

\begin{figure}[h]
\begin{center}
\includegraphics*[scale=0.22]{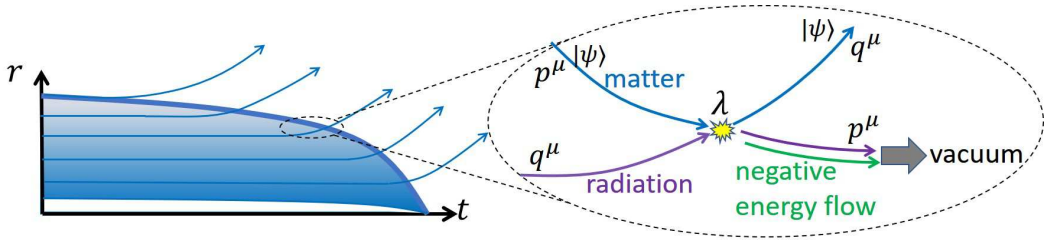}
\caption{Time evolution of the evaporating black hole with interaction.
The matter with the information $|\psi\ket$ and energy-momentum $p^\mu$ comes 
and is scattered with a certain probability by the radiation with energy-momentum $q^\mu$. 
Then, the ingoing radiation should approach to a vacuum state through interaction with 
the negative energy flow. }
\label{f:interact}
\end{center}
\end{figure}

Now, we would like to ask: How is the information $|\psi \ket$ of the collapsing matter 
reflected in the Hawking radiation and recovered in this picture? 
There are two remarkable points to answer this. 
One point is interaction between the outgoing Hawking radiation and the ingoing collapsing matter. 
As described in the basic idea of section \ref{s:idea} and below \Ref{e_den}, 
the Hawking radiation is emitted outwards from each region inside the black hole. 
Then, the collapsing matter may collide with the outgoing radiation coming from below, go outwards, and exit from the black hole 
together with the information. See Fig.\ref{f:interact}. 
Because the energy scale there is near the Planck scale (from \Ref{Rresult}, \Ref{mu0} and \Ref{T_loc}), 
some string-theory effect may be relevant and any kind of matter may interact with the radiation universally
\footnote{The introduction of interaction is natural from a thermodynamic point of view: 
black holes are thermodynamical objects \ci{GH}, and 
thermal equilibrium is reached by interaction \ci{Landau_SM}.
Also, higher-dimensional matter terms in effective action may 
induce scattering that occurs with a high probability inside the black hole \ci{firePM}.}.
Actually, we can describe this idea by a classical $\phi^4$-model and evaluate the scattering time scale as 
\begin{equation}\lb{t_scat}
\D t \sim a \log{\f{a}{\lambda N l_p}},
\end{equation}
where $\lambda$ is the dimensionless coupling constant (see section 3 of \ci{KY2}). 
Then, the scattering is an elastic collision in the radial direction, 
and the two particles swap their momenta (the right of Fig.\ref{f:interact}).
Therefore, we expect that 
the energy distribution of the coming-back matter follows the Planck distribution of $T_H=\f{\hbar}{4\pi a(t)}$ 
with a small correction that depends on the initial state.

This scattering process is stochastic. 
If an outgoing matter in the deep region goes out while the other remains inside, 
the matter would collide many times before going out, so the probability is very small. 
Therefore, 
the information returns in order from the surface region during the evaporation 
as if one peels off an onion\footnote{If there is no interaction, 
the earlier the matter parts start to go inward, the sooner they return. 
This is the reverse order of the case of Fig.\ref{f:interact}.}
(left of Fig.\ref{f:interact}).

The other point is the role of the negative energy induced by the vacuum. 
If the black hole evaporates and the information comes out by the above mechanism, 
the state along an ingoing line should approach to a vacuum state (see the right of Fig.\ref{f:interact}). 
When the negative energy is superposed on the matter and decreases the energy, 
interaction between them should play a role in this evolution of the state. 
This point may be related to the problem of finding the correct configuration of 
the s-waves discussed in subsection \ref{s:swave}. 
We will study this scenario in future. 
\section*{Acknowledgement}
We thank Pei-Ming Ho for valuable discussions.
Y.Y. is partially supported by Japan Society of Promotion of Science (JSPS), 
Grants-in-Aid for Scientific Research (KAKENHI) Grants No.\ 18K13550 and 17H01148. 
Y.Y. is also partially supported by RIKEN iTHEMS Program.

\appendix
\section{Derivation of \Ref{r_t3}}\lb{A:particle}
For a self-contained discussion, 
we review the derivation of \Ref{r_t3} \ci{KMY,KY1,KY2,KY3}. 
\Ref{r_t2} can be solved exactly by coefficient change method.
The general solution is given by 
\begin{equation}
\Delta r(u)=C_0 e^{-\int^u_{u_0}d u' \f{1}{2a(u')}} + \int^u_{u_0}d u' \l(-\f{d a}{d u}(u') \r)e^{-\int^{u}_{u'}d u''\f{1}{2a(u'')}}, \nn
\end{equation}
where $C_0$ is an integration constant. 
The typical time scale of this equation is $\MO(a)$ from the exponent, 
while that of $a(u)$ is $\MO(a^3/\s)$ from \Ref{da}. 
Therefore, $a(u)$ and $\f{d a(u)}{d u}$ can be considered as constants in the time scale of $\MO(a)$,
and the second term can be evaluated as 
\begin{align*}
 &\int^u_{u_0}d u' \l(-\f{d a}{d u}(u') \r)e^{-\int^{u}_{u'}d u''\f{1}{2a(u'')}}\\
 &\approx -\f{d a}{d u}(u) \int^u_{u_0}d u' e^{-\f{u-u'}{2a(u)}}=-2\f{d a}{d u}(u) a(u)(1-e^{-\f{u-u_0}{2a(u)}}). 
\end{align*}
Therefore, we have 
\begin{equation}
\Delta r(u)\approx C_0 e^{-\f{u-u_0}{2a(u)}} -2\f{d u}{d u}(u)a(u) (1-e^{-\f{u-u_0}{2a(u)}}), \nn 
\end{equation}
and obtain through \Ref{da}
\begin{align}\lb{r_fullsol}
r_s(u)&\approx a(u)-2a(u)\f{d a}{d u}(u)+Ce^{-\f{u}{2a(u)}} \nn \\
 &= a(u)+\f{2\s}{a(u)}+Ce^{-\f{u}{2a(u)}}\longrightarrow  a(u)+\f{2\s}{a(u)},
\end{align}
which is \Ref{r_t3}\footnote{By a numerical method, 
we can also solve \Ref{r_t} exactly and show \Ref{r_t3}. See Appendix B in \ci{KY2}.}
\footnote{The above analysis is based on the classical motion of particles. 
We can also show that this result is valid even if we treat them quantum mechanically. 
See section 2-B and appendix A in \ci{KY2}.}.
\section{Behavior of $r_s$ in the limit $a\to 0$}
\begin{figure}[h]
\begin{center}
\includegraphics*[scale=0.22]{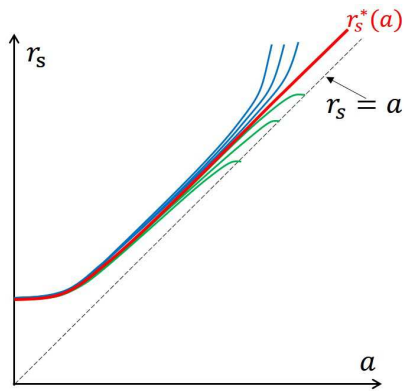}
\caption{Integral curves of \Ref{ra_eq} that start from points satisfying $r_s\sim a\gg l_p$. 
As \Ref{ra_eq} shows, their slopes are positive for $r_s>a$ and become zero at $a=0$. 
They are attracted by an asymptotic curve $r_s^\ast(a)$. 
We can check this easily by a numerical analysis. }
\label{f:ra_sol}
\end{center}
\end{figure}

\lb{A:r_a}
In this Appendix, we \textit{assume} the Vaidya metric \Ref{SchV} with \Ref{da} 
to study the behavior of $r_s(u)$ at the moment when $a(u)$ vanishes\footnote{
Note again that such a final stage of the evaporation is beyond the semi-classical approximation 
and should be investigated more properly, say, by string theory.}. 
Here, $r_s(u)$ follows \Ref{r_t}. 
Combining \Ref{da} and \Ref{r_t}, we obtain 
\begin{equation}\lb{ra_eq}
\f{dr_s}{da}=\f{a^2(r_s-a)}{2\s r_s}.
\end{equation}
This provides the relation of $r_s$ and $a$.
Note that time evolution is in the direction in which $a$ decreases. 
Here, we focus on the integral curves that start from points satisfying $r_s\sim a\gg l_p$. 
See Fig.\ref{f:ra_sol}. 
The curves are attracted by an asymptotic curve $r_s^\ast(a)$, 
which is the curve starting from $r_s\sim a\to \infty$ (red curve in Fig.\ref{f:ra_sol}). 
By numerical calculation we can check that 
\begin{equation}\lb{ra0}
r_s(a=0)\approx 2.33\sqrt{\s}
\end{equation}
for various initial values. 
This means that when $a$ becomes zero, $r_s$ is not zero. 

In Fig.\ref{f:ra_sol}, the blue curves are of interest to us. 
We are now considering a particle of those which compose
the outermost part (shell) of the collapsing matter. 
Therefore, the particle comes from the ouside of $r=a(u)$, 
which the blue curves describe. 
On the other hand, the green curves are not physical. 
They start from points on $r_s=a$ and would correspond to particles coming 
from the inside of $r=a(u)$ 
although the metric \Ref{SchV} can be applied only to the region of $r>a(u)$. 

Let us construct approximately the asymptotic curve $r_s^*(a)$. 
In the stage where $a\gtrsim \sqrt{\s}$, 
\Ref{r_fullsol} means that particles near $a(u)$ approach to, in the time scale $\D u\sim 2a$ 
\begin{equation}\lb{r*1}
r_s^*(a)=a+\f{2\s}{a}~~{\rm for}~~a\geq \sqrt{2\s}.
\end{equation}
Here, for simplicity, the lower limit value $\sqrt{2\s}$ 
has been taken as one giving the minimum value of $a+\f{2\s}{a}$: 
$r_s^*(\sqrt{2\s})=2\sqrt{2\s}$. 
On the other hand, when $a\to0$, we can approximate \Ref{ra_eq} as 
\begin{equation}\lb{ra_eq1}
\f{dr_s}{da}\approx\f{a^2}{2\s},
\end{equation}
where we have taken $r_s-a\approx r_s$ in the numerator of \Ref{ra_eq} 
for $r_s^*(a=0)>0$. Then, we obtain 
\begin{equation}\lb{r*2}
r_s^*(a)=\f{a^3}{6\s}+\f{5}{3}\sqrt{2\s}~~{\rm for}~~0\leq a\leq \sqrt{2\s},
\end{equation}
where the integration constant has been chosen s.t. 
this is connected to \Ref{r*1} at $r=\sqrt{2\s}$. 
Thus, \Ref{r*1} and \Ref{r*2} give approximately the asymptotic curve $r_s^*(a)$. 
Actually, from \Ref{r*2}, we have 
\begin{equation}\lb{r*0}
r_s^*(a=0)=\f{5}{3}\sqrt{2\s}\approx 2.36\sqrt{\s},
\end{equation}
which is close to \Ref{ra0}. 

\begin{figure}[h]
\begin{center}
\includegraphics*[scale=0.22]{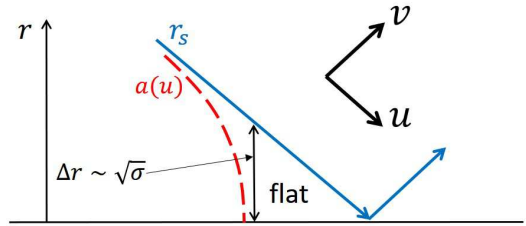}
\caption{The final stage of the evaporation in a lightlike coordinate $(u,v)$. } 
\label{f:ra_final}
\end{center}
\end{figure}

Based on this result, we can describe 
the picture around the final stage of the evaporation 
in a lightlike coordinate $(u,v)$. 
See Fig.\ref{f:ra_final}. 
The trajectory $r_s(u)$ is drawn with a constant $v$ line. 
The Schwarzschild radius $a(u)$ moves spacelike because from \Ref{SchV} and \Ref{da}, 
$ds^2|_{r=a(u)}=-2\dot a du^2 =\f{2\s}{a^2}du^2>0$.  
When $a(u)$ becomes zero, $r_s(u)$ is still positive by \Ref{ra0}. 
After that, $r_s(u)$ propagates in the flat space and reaches $r=0$.
Thus, we conclude that in \Ref{SchV} 
$r_s(u)$ and $a(u)$ do not become zero at the same time. 

\section{The radiation in our model and the usual Hawking radiation}\lb{A:pre}
We reconsider how to obtain the pre-Hawking radiation in the s-wave approximation model, 
which has been given by subsection \ref{s:step3}, 
and then discuss the difference from the usual Hawking radiation \ci{Hawking}. 

Let's recall that the formula of the energy flux for particle creation is given by
\begin{equation}\lb{J_again}
J=\f{\hbar N}{48\pi}\l(\l(\f{d \xi}{du}\r)^2-2 \f{d^2\xi}{du^2}\r),~~\xi\equiv \log \f{dU}{du}, 
\end{equation}
which comes from \Ref{J_Sch} and \Ref{Sch_xi_i}.  

In our model, we have from \Ref{xi_i} and $C=\s_s$
\begin{equation}\lb{xi_pre}
\xi(u) =-\f{a(u)^2}{4\s_s},
\end{equation}
where $a(u)$ satisfies \Ref{da} with $\s$ replaced by $\s_s$. 
This makes \Ref{J_again} into 
\begin{equation}\lb{J_again2}
J=\f{\hbar N}{192\pi a^2}
\end{equation}
at the leading order in $\f{l_p}{a}\ll1$. 

On the other hand, in the case of the usual Hawking radiation \ci{Hawking},  
the relation, $U\propto e^{-\f{u}{2a}}$, 
between the Kruskal coordinate $U$ and the Eddington-Finkelstein coordinate $u$ 
is used to derive the radiation for $a=$const.\footnote{Note that this $U$ is the same as that of the flat space before the collapse, 
and we can identify \Ref{xi_Haw}.}. 
Then, we have 
\begin{equation}\lb{xi_Haw}
\xi(u) =-\f{u }{2a},
\end{equation}
which also leads to \Ref{J_again2}
\footnote{Note that other approaches to a pre-Hawking radiation (e.g. \ci{Barcelo1,Barcelo2}) 
use almost the same derivation as this.}. 

Although the both cases give the same value of the energy flux $J$, 
the property of $\xi$ is quite different. 
\Ref{xi_pre} is controlled only by $a$ and does not depend explicitly on $u$. 
Therefore, we can add a shell with $\D a$ to the black hole 
and derive the pre-Hawking radiation \Ref{J_again2} with $a$ replaced by $a'=a+\D a$ 
in the recursive manner of subsection \ref{s:step3}. 
On the other hand, \Ref{xi_Haw} depends explicitly on $u$, 
and such a recursive argument does not work. 

The physical picture of particle creation is also different. 
First, we consider the radiation in our model. 
Imagine that an outgoing null geodesic starts from 
the flat region around $r=0$, passes through the dense object, and comes out of the surface, 
as the green arrow in Fig.\ref{f:multi}. 
It takes an exponentially long time due to the large redshift of the interior metric of \Ref{metricf}. 
At each stage where the line passes each shell, 
the pre-Hawking radiation is induced and added to compensate the reduction by the redshift 
(recall the discussion below \Ref{total_J}). 
The sum of the radiations is emitted from the object and the system evaporates. 
The point is that there is no trans-Planckian physics anywhere if $N$ is large, \Ref{largeN}. 

Next, we recall how the usual Hawking radiation occurs \ci{Hawking}. 
Quantum fields start in $|0\ket_M$ and pass the center and the collapsing matter, which is not so dense. 
Then, because of \Ref{xi_Haw}, 
the fields that propagates exponentially close to the horizon are excited to create 
particles as the radiation. 
In this process, modes with exponentially short wavelength are involved, 
which is a trans-Planckian problem \ci{trans1,trans2,trans3}.

\section{Normalization factor \Ref{norm}}\lb{A:N}
We derive the normalization for the bound modes, \Ref{norm}. 
This can be obtained from the relation \Ref{pp}:
\begin{equation}\lb{pp1}
[\phi(t,r,\th,\phi), \pi(t,r',\th',\phi')]=i\hbar\d(r-r')\d(\th-\th')\d(\phi-\phi').
\end{equation}

First, \Ref{WKB1} and \Ref{pi} give 
\begin{equation}\lb{pi2}
\pi(x)=i\sqrt{-g}g^{tt}\sum_i \o_i(a_i u_i(x)-a^{\dagger}_iu_i^*(x)), 
\end{equation}
and then \Ref{pp1} is equivalent through \Ref{aa} to 
\begin{equation}\lb{pp2}
-\sqrt{-g}g^{tt}\sum_i \o_i(u_i(t,r,\th,\phi)u_i^*(t,r',\th',\phi')+u_i^*(t,r,\th,\phi)u_i(t,r',\th',\phi'))
=\hbar\d(r-r')\d(\th-\th')\d(\phi-\phi').
\end{equation}
This is satisfied if we find $\{ \bar u_i(x)\}$ s.t. 
\begin{align}\lb{baru1}
u_i &=\sqrt{\f{\hbar}{2\o_i\sqrt{-g}(-g^{tt})}}\bar u_i \\
\lb{baru2}
\sum_i \bar u_i(t,r,\th,\phi) \bar u_i^*(t,r',\th',\phi') &= \d(r-r')\d(\th-\th')\d(\phi-\phi').
\end{align}

We here note that \Ref{baru2} is equivalent to 
\begin{equation}
\d_{ij}=(\bar u_i, \bar u_j)\equiv \int dr d\th d\phi \bar u_i \bar u_j^*.
\end{equation}
We use \Ref{baru1} and the leading WKB solution \Ref{pan} with $\vp_i(r)=\f{1}{\sqrt{p_i}}\cos \int^rdr'(p_i+\th_i)$ to have
\begin{equation}
\bar u_i={\mathcal N}_i\sqrt{\f{2\o_i}{\hbar}} B e^{-\f{A}{2}} 
\f{1}{\sqrt{p_{i}(r)}}\cos[\int^r dr' p_{i}(r')+\th_{i}] e^{-i\o_it} \sqrt{\sin \th}Y_{lm}.
\end{equation}
Therefore, the normalization condition is given by 
\begin{align}\lb{baru3}
1 &=(\bar u_i,\bar u_i)\nn \\
 &= \f{2\o_i}{\hbar} {\mathcal N}_i^2 \int^{r^+_i}_{r^-_i} dr B^2 e^{-A} \f{1}{p_i} 
 \cos^2 \l( \int^r dr' (p_i +\th_i)\r)\int d\th d\phi \sin \th |Y_{lm}|^2 \nonumber \\
 &= \f{2\o_i}{\hbar} {\mathcal N}_i^2 \int^{r^+_i}_{r^-_i} dr B^2 e^{-A} \f{1}{p_i} \times \f{1}{2},
\end{align}
where we have used $\int d\th d\phi \sin \th Y_{lm} Y_{l'm'}^*=\d_{ll'}\d_{mm'}$ 
and replaced $\cos^2\int dr p_i$ to the average $\f{1}{2}$ because the WKB approximation is good for a high frequency. 
Now, we take derivative of \Ref{BS_con} with respect to $n$, use the property $p_i(r_i^\pm)=0$ and obtain 
\begin{equation}\lb{BS_con2}
\f{\p \o^2_{nl}}{\p n}=2\pi  \l[\int^{r^+_{nl}}_{r^-_{nl}} dr \f{B^2 e^{-A}}{p_{nl}}\r]^{-1}.
\end{equation}
Thus, \Ref{baru3} and \Ref{BS_con2} provide \Ref{norm}. 

\section{Derivation of the leading solution \Ref{vp0}}\lb{A:B}
First, two Bessel functions $J_A(\xi), J_{-A}(\xi)$ satisfy \Ref{eqB}. 
Because we are now interested in the bound modes, the solution should vanish as $r-r_0=-x\to \infty$. 
From \Ref{xi}, this corresponds to $\xi \to 0$, in which $J_{\pm A}(\xi)$ behave as
\begin{equation}
J_{\pm A}(\xi)\xrightarrow{\xi \to 0} \xi^{\pm A}.
\end{equation}
Because $A$ is positive, we can find 
\begin{equation}\lb{k}
\vp_i^{(0)}(\xi)=k J_A(\xi).
\end{equation}

Next, we fix the normalization $k$. 
\Ref{WKBv2} means that the WKB approximation becomes better as $r-r_0=-x\to -\infty$, 
which corresponds through \Ref{xi} to $\xi \to \infty$. 
Then, the leading behavior of $J_A(\xi)$ is 
\begin{equation}\lb{Jlim}
J_A(\xi) \xrightarrow{\xi \to \infty} \sqrt{\f{2}{\pi\xi}}\cos\l(\xi-\f{\pi}{4}(1+2A)\r)=
\sqrt{\f{2}{\pi\sqrt{2\s\eta^2}\tilde \o  }}e^{-\f{r_0}{4\s\eta}x}\cos\l(\sqrt{2\s\eta^2}\tilde\o e^{\f{r_0}{2\s\eta}x}-\f{\pi}{4}(1+2A)\r),
\end{equation}
where \Ref{xi} has been used. 
In 
\Ref{WKB2}, on the other hand, we have for $x\to \infty$ 
\begin{equation}
p_i(r)^2 \approx P_i^{(0)}(x)\approx \f{r_0^2}{2\s}e^{\f{r_0}{\s\eta}x}\tilde \o^2,~~~
\int^rdr'p_i(r')\approx -\int^xdx'\f{r_0}{\sqrt{2\s}}e^{\f{r_0}{2\s\eta}x'}\tilde \o 
=-\sqrt{2\s\eta^2}\tilde\o e^{\f{r_0}{2\s\eta}x},
\end{equation}
where we have considered a region of $x$ where the WKB solution almost agrees with the leading exact solution. 
Therefore, the WKB solution behaves for $x\to \infty$ as 
\begin{equation}\lb{Wlim}
\f{1}{\sqrt{p_i(r)}}\cos \l(\int^rdr'p_i(r')+\th_i  \r) \xrightarrow{x \to \infty} 
\sqrt{\f{\sqrt{2\s}}{r_0 \tilde \o }}e^{-\f{r_0}{4\s\eta}x}
\cos\l(\sqrt{2\s\eta^2}\tilde\o e^{\f{r_0}{2\s\eta}x}-\th_i\r).
\end{equation}
Comparing \Ref{Jlim} and \Ref{Wlim} in \Ref{k}, 
we obtain 
\begin{equation}
k=\sqrt{\f{\pi\s\eta}{r_0}},~~~\th_i=\f{\pi}{4}(1+2A).
\end{equation}

\section{Evaluation of the leading values of the regularized energy-momentum tensor \Ref{T0reg}}\lb{A:reg}
We first summarize the results about the leading solution:
\begin{align}\lb{Aphi}
\phi(x)&=\sum_{n,l,m}\int d^\e k (a_iu_i^{(0)}(x)+a_i^\dagger u_i^{(0)\ast}(x)),\\
\lb{Au}
u_i^{(0)}(t,r,\th,\phi,y^a)&= 
\sqrt{\f{\hbar\eta}{2r_0}}\sqrt{\f{\p \o_{nl}}{\p n}} e^{-i\o_{nl} t}e^{-\f{r^2}{8\s\eta}}J_A(\xi) Y_{lm}(\th,\phi) \f{e^{ik\cdot y}}{(2\pi)^{\e/2}},
\end{align}
where 
\begin{equation}\lb{AA}
A\equiv \sqrt{2\s\eta^2 (\tilde L +k^2) +\f{1}{4}},~~\xi \equiv \sqrt{2\s\eta^2}\tilde \o e^{\f{r_0}{2\s\eta}(r_0-r)},~~
\tilde \o \equiv \f{r_0}{\sqrt{2\s}}e^{-\f{r_0^2}{4\s\eta}}\o,~~\tilde L\equiv \f{l(l+1)}{r_0^2}.
\end{equation}
Here, according to the discussion at the beginning of subsection \ref{s:T0}, 
we include only the bound modes. 
$a_i$ and $a_j^\dagger$ satisfy 
\begin{equation}\lb{Aaa}
[a_i,a_j^\dagger]=\d_{ij},
\end{equation}
and the vacuum state $|0\ket$ satisfies 
\begin{equation}\lb{Avac}
a_i|0\ket=0.
\end{equation}
The metric is 
\begin{equation}\lb{Amet}
ds^2=-\f{2\s}{r^2} e^{\f{r^2}{2\s \eta}} d t^2 + \f{r^2}{2\s} d r^2 + r^2 d \Omega^2+\sum_{a=1}^\e (dy^a)^2.
\end{equation}

From these, we can make easily the formulae we are using below. 
We have at $r=r_0$
\begin{equation}\lb{u2}
|u_i^{(0)}(r_0)|^2 = \f{\hbar}{2r_0^2(2\pi)^\e}\f{\p \Omega_i}{\p n}J_A(\Omega_i)^2 |Y_{lm}|^2,
\end{equation}
where we have used 
\begin{equation}\lb{Omega}
\xi \xrightarrow{r \to r_0} \sqrt{2\s \eta^2} \tilde \o_i = \eta r_0 e^{-\f{r_0}{4\s\eta}}\o_i\equiv \Omega_i.
\end{equation}
Next, we take the derivative of \Ref{Au} with respect to $r$: 
\begin{align}
\p_r u_i^{(0)}(r) &=-\f{r}{4\s\eta}u_i^{(0)}+ \f{\p\xi}{\p r}\f{\p J_A(\xi)}{\p \xi} \f{1}{J_A(\xi)}u_i^{(0)} \nn \\
 &= -\f{r}{4\s\eta}\l[1 +\f{2r_0}{r} \l(\xi \f{J_{A-1}(\xi)}{J_A(\xi)}-A\r)\r]u_i^{(0)},
\end{align}
where we have used the second one of \Ref{AA} and a formula of Bessel function:
\begin{equation}
\f{\p}{\p \xi} J_A(\xi) =J_{A-1}(\xi)-\f{A}{\xi}J_{A}(\xi).
\end{equation}
Therefore, we have 
\begin{align}\lb{ur2}
|\p_r u_i^{(0)}(r_0) |^2 &=4 \l(\f{r_0}{4\s\eta}\r)^2 \l[\l(\f{1}{2}-A\r) +\Omega_i \f{J_{A-1}(\Omega_i)}{J_A(\Omega_i)}\r]^2 |u_i^{(0)}(r_0)|^2    \nn \\
 &= \f{\hbar}{8\s^2 \eta^2 (2\pi)^\e}\f{\p \Omega_i}{\p n} |Y_{lm}|^2\l[\l(\f{1}{2}-A\r)J_A(\Omega_i) +\Omega_i J_{A-1}(\Omega_i)\r]^2.
\end{align}

\subsection{Evaluation of $\bra0|g^{tt}(\p_t \phi)^2|0\ket^{(0)}_{reg}$}
Now, we can have 
\begin{align}
\bra0|g^{tt}(\p_t \phi)^2|0\ket^{(0)}_{reg}|_{r=r_0} 
 &=g^{tt}(r_0)\bra 0 | \sum_i(-i\o_i)(a_iu_i^{(0)}-a_i^\dagger u_i^{(0)\ast}) \sum_j(-i\o_j)(a_ju_j^{(0)}-a_j^\dagger u_j^{(0)\ast}) |0\ket   \nn\\
 &=-\f{r_0^2}{2\s}e^{-\f{r_0^2}{2\s\eta}} \sum_i\o_i^2 |u_i^{(0)}(r_0)|^2 \nonumber \\
 &=-\f{\hbar}{4\s\eta^2r_0^2(2\pi)^\e}\int d^\e k \sum_n \f{\p \Omega_i}{\p n}\Omega_i^2 \sum_{lm}|Y_{lm}|^2 J_{A}(\Omega_i)^2,
\end{align}
where in the second line we have used \Ref{Aaa} and \Ref{Avac} and in the third line \Ref{u2} and \Ref{Omega}. 
In evaluating the leading contribution, we can neglect the correction from the Euler-Maclaurin formula in $\sum_{n,l,m}$
\footnote{This correction and the subleading solution $\vp_i^{(1)}$ contribute to the subleading value $\bra0|g^{tt}(\p_t \phi)^2|0\ket^{(1)}_{reg}$.}
and use
\begin{equation}
\sum_n \f{\p \Omega_i}{\p n}\approx \int d\Omega,~~
\sum_{lm}|Y_{lm}|^2 = \sum_l \f{2l+1}{4\pi}\approx \f{1}{4\pi}\int d(l^2) = \f{r_0^2}{4\pi}\int d\tilde L.
\end{equation}
Then, we have 
\begin{align}
\bra0|g^{tt}(\p_t \phi)^2|0\ket^{(0)}_{reg}|_{r=r_0} 
 &=-\f{\hbar}{16\pi\s\eta^2(2\pi)^\e}\int d^\e k \int d\Omega \Omega^2 \int d\tilde L  J_{A}(\Omega)^2\nn\\
 &=-\f{\hbar}{32\pi\s^2\eta^4(2\s\eta^2)^{\f{\e}{2}}(2\pi)^\e}\int d^\e K \int d\Omega \Omega^2 \int dY  J_{A}(\Omega)^2,
\end{align}
where we have introduced
\begin{equation}
Y\equiv 2\s\eta^2\tilde L,~~~K^a\equiv \sqrt{2\s\eta^2}k^a.
\end{equation}
Instead of considering directly the index $A$, which is the first of \Ref{AA}, we can insert the identity 
\begin{equation}
1= \int dA \d\l(\sqrt{Y+K^2 + \f{1}{4}}-A\r)
\end{equation}
and consider 
\begin{align}\lb{teq1}
\bra0|g^{tt}(\p_t \phi)^2|0\ket^{(0)}_{reg}|_{r=r_0}&=
-\f{\hbar}{32\pi\s^2\eta^4(2\s\eta^2)^{\f{\e}{2}}(2\pi)^\e}\nn\\
&~~~~\int dA \int d^\e K \int d\Omega \Omega^2 \int dY  J_{A}(\Omega)^2\d\l(\sqrt{Y+K^2 + \f{1}{4}}-A\r),
\end{align}
This integration is performed to the extent that the square root is not negative.

\textbf{$K$-integration}: 
We first perform the $K$-integration as 
\begin{align}\lb{Kint}
\int\f{d^\e K}{(2\pi)^\e} \d\l(\sqrt{Y+K^2 + \f{1}{4}}-A\r)
 &=\f{S^\e}{(2\pi)^\e}\int^\infty_0 dK K^{\e-1} \d\l(\sqrt{Y+K^2 + \f{1}{4}}-A\r)\nn \\
 &=\f{2}{(4\pi)^{\e/2}\G(\f{\e}{2})}\int^\infty_0 dK K^{\e-1} \f{A}{K_0} \d(K-K_0) \nonumber \\
 &=\f{2}{(4\pi)^{\e/2}\G(\f{\e}{2})}A\l(A^2-\f{1}{4}-Y\r)^{\f{\e-2}{2}},
\end{align}
where $S^\e=\f{2\pi^{\f{\e}{2}}}{\Gamma(\f{\e}{2})}$ is the area of a $\e$-dimensional unit sphere, 
and $K_0=\sqrt{A^2-\f{1}{4}-Y}$. 

\textbf{$Y$-integration}: 
We next integrate it over $Y$ as 
\begin{equation}\lb{Yint}
\int_0^{A^2-1/4} dY\l(A^2-\f{1}{4}-Y\r)^{\f{\e}{2}-1}=-\l.\f{1}{\f{\e}{2}}\l(A^2-\f{1}{4}-Y\r)^{\f{\e}{2}}\r|^{A^2-\f{1}{4}}_0
=\f{1}{\f{\e}{2}}\l(A^2-\f{1}{4}\r)^{\f{\e}{2}}.
\end{equation}
We have discarded the contribution from the top of the integral 
because in the dimensional regularization we drop $0^{\e}$ and $\infty^\e$ by choosing $\e$ properly. 
Then, \Ref{teq1} becomes 
\begin{align}\lb{teq2}
\bra0|g^{tt}(\p_t \phi)^2|0\ket^{(0)}_{reg}|_{r=r_0}&
=-\f{\hbar}{16\pi\s^2\eta^4(8\pi\s\eta^2)^{\f{\e}{2}}\G(1+\f{\e}{2})}\int dA d\Omega h_t(\Omega,A),\\
\lb{ht}
h_t(\Omega,A)&\equiv A\l(A^2-\f{1}{4}\r)^{\f{\e}{2}}\Omega^2 J_A(\Omega)^2.
\end{align}

\textbf{$\Om$-integration}:
To evaluate the $\Omega$-integration of $h_t(\Omega,A)$, we introduce a damping factor $e^{-s\Omega}~(s>0)$, 
integrate it and then expand the result around $s=0$: 
\begin{align}\lb{teq3}
\int^\infty_0 d\Om e^{-s\Om} h_t(\Om,A) &= \f{A}{\pi s^2} \l(A^2-\f{1}{4}\r)^{\f{\e}{2}}+\f{1}{4\pi}\l(A^2-\f{1}{4}\r)^{\f{\e}{2}+1}\l[-1+\log \f{16}{s^2}\r] \nn \\
 &+\f{1}{4\pi}\l(A^2-\f{1}{4}\r)^{\f{\e}{2}+1}(H_{A-2}-H_{A+\f{1}{2}}-2H_{2A-3}) +\MO(s).
\end{align}
Here, $H_n$ is Harmonic number: 
\begin{equation}\lb{D_def}
H_n\equiv \sum_{k=1}^n\f{1}{k},
\end{equation}
which has useful formulae: 
\begin{align}\lb{H1}
H_n &=H_{n-l}+\sum_{k=n-l+1}^n\f{1}{k} \\
\lb{H2}
H_{2x} &=\f{1}{2}(H_x+H_{x-\f{1}{2}})+\log2. 
\end{align}
Using these properly, we can rewrite 
\begin{equation}\lb{H3}
H_{A-2}-H_{A+\f{1}{2}}-2H_{2A-3}=\f{1}{A^2-\f{1}{4}}\l(-2(A^2-\f{1}{4})H_{A-\f{1}{2}}+1-(A^2-\f{1}{4})\log4\r).
\end{equation}
Note here that any term of the form $A\times$(polynomial in $A^2$) can be dropped by choosing $\e$ properly in the $A$-integration.
Therefore, we can drop the first line of \Ref{teq3} and the second and third terms of the r.h.s. in \Ref{H3} 
and have in the limit $s\to 0$
\begin{equation}\lb{teq4}
\int^\infty_0 d\Om h_t(\Om,A) = -\f{1}{2\pi}A\l(A^2-\f{1}{4}\r)^{1+\f{\e}{2}}H_{A-\f{1}{2}}.
\end{equation}

\textbf{$A$-integration}:
Next, we consider the $A$-integration. 
Using the integral representation of Harmonic number
\begin{equation}\lb{H4}
H_A =\int^1_0 dx\f{1-x^A}{1-x},
\end{equation}
we have 
\begin{align}
\int dA d\Om h_t(\Om,A) &=-\f{1}{2\pi}\int^\infty_{1/2}dAA\l(A^2-\f{1}{4}\r)^{1+\f{\e}{2}}H_{A-\f{1}{2}}  \nn  \\
 &= -\f{1}{2\pi}\int^\infty_{1/2}dAA\l(A^2-\f{1}{4}\r)^{1+\f{\e}{2}}\int^1_0 dx\f{1-x^{A-\f{1}{2}}}{1-x}  \nn  \\
 &= \f{1}{2\pi} \int^1_0 dx\f{x^{-\f{1}{2}}}{1-x}\int^\infty_{\f{1}{2}}dAA\l(A^2-\f{1}{4}\r)^{1+\f{\e}{2}}x^A\nonumber \\
 &= \f{1}{2\pi} \int^1_0 dx\f{x^{-\f{1}{2}}}{1-x}\f{\G(2+\f{\e}{2})}{2\sqrt{\pi}}(-\log x)^{-\f{3}{2}-\f{\e}{2}} K_{\f{5+\e}{2}}\l(-\f{1}{2} \log x\r).
\end{align}
Here, in the third line we have discarded the term proportional to a polynomial in $A$ and 
$K_{s}(x)$ is the modified Bessel function of the second kind. 
Introducing $y=-\f{1}{2}\log x$, we have 
\begin{equation}\lb{teq5}
\int dA d\Om h_t(\Om,A)=\f{\G(2+\f{\e}{2})}{4\pi^{\f{3}{2}}}2^{-\f{3}{2}-\f{\e}{2}}\int^\infty_0 dy \f{y^{-\f{3}{2}-\f{\e}{2}}}{\sinh y}K_{\f{5+\e}{2}}(y),
\end{equation}
which is difficult to evaluate analytically. 

\textbf{Numerical evaluation}:
From \Ref{teq2} and \Ref{teq5}, thus we obtain the expression for an energy scale $\mu$ 
\begin{align}\lb{teq6}
&\mu^{-\e}\bra0|g^{tt}(\p_t \phi)^2|0\ket^{(0)}_{reg}|_{r=r_0}\nn\\
=&-\f{\hbar}{16\pi\s^2\eta^4(8\pi\s\eta^2\mu^2)^{\f{\e}{2}}\G(1+\e/2)}
\f{\G(2+\f{\e}{2})}{4\pi^{\f{3}{2}}}2^{-\f{3}{2}-\f{\e}{2}}\int^\infty_0 dy \f{y^{-\f{3}{2}-\f{\e}{2}}}{\sinh y}K_{\f{5+\e}{2}}(y)\nn\\
\equiv& C_\e \int^\infty_0 dy f_\e(y).
\end{align}
Let us  evaluate this numerically. 
Because the integration diverges around $y=0$, we divide it into the pole part and the finite one. 
We first expand $f_\e(y)$ around $y=0$: 
\begin{equation}
f_\e(y)= a_1 y^{-5-\e}+ a_2 y^{-3-\e}+ a_3 y^{-1-\e}+ \MO(y^{1-\e}),
\end{equation}
where 
\begin{equation}
a_1=3\sqrt{\f{\pi}{2}}+\MO(\e),~a_2=-\sqrt{\f{\pi}{2}}+\MO(\e),~
a_3=\f{2\sqrt{2\pi}}{15}-\f{\sqrt{\pi}}{180\sqrt{2}}(-29+24\g+24\log2)\e+\MO(\e^2).
\end{equation}
Here, we have
\begin{equation}
\int^1_0 dy y^{-n-\e}=\f{1}{1-n-\e},
\end{equation}
where we have discarded the contribution from the bottom of the integral. 
This means that $n=1$ provides a pole $-1/\e$. 
Therefore, we obtain the pole part: 
\begin{equation}\lb{ppart}
v_0\equiv C_\e a_3\f{-1}{\e}=\f{\hbar}{960\pi^2\s^2\eta^4\e}+\hbar\f{\f{53}{24}-(\g+\log(32\pi\s\eta^2\mu^2))}{1920\pi^2\s^2\eta^4}+\MO(\e),
\end{equation}
which also includes the finite contribution from $y^{-1}$ in the range $0\leq y \leq 1$.

Next, we consider the finite part in the range $0\leq y \leq 1$. 
The contributions from $y^{-5}$ and $y^{-3}$ are given by 
\begin{align}
v_1 &\equiv \l.C_\e a_1\f{1}{1-5-\e}\r|_{\e\to0}=\f{3\hbar}{1024\pi^2\s^2\eta^4} \\
v_2 &\equiv  \l.C_\e a_2\f{1}{1-3-\e}\r|_{\e\to0}=-\f{\hbar}{512\pi^2\s^2\eta^4}.
\end{align}
Subtracting the divergent part, we can evaluate numerically 
\begin{equation}
v_3\equiv C_{\e=0}\int^1_0 dy (f_{\e=0}(y)-a_1 y^{-5}- a_2 y^{-3}- a_3 y^{-1})=\f{0.000282257\hbar}{\pi^2 \s^2\eta^4}.
\end{equation}
These $v_0,v_1,v_2,v_3$ determine the value from $0\leq y \leq 1$ although $v_0$ diverges. 

On the other hand, all the contribution from the range $1\leq y \leq \infty$ is evaluated numerically as 
\begin{equation}
v_4\equiv C_{\e=0}\int^\infty_1 dy f_{\e=0}(y)=-\f{0.00174953\hbar}{\pi^2 \s^2\eta^4}.
\end{equation}

Thus, we obtain 
\begin{align}\lb{teq7}
\mu^{-\e}\bra0|g^{tt}(\p_t \phi)^2|0\ket^{(0)}_{reg}|_{r=r_0} &=\sum_{q=0}^4 v_q \nn\\
 &= \f{\hbar}{960\pi^2\s^2\eta^4}\l[\f{1}{\e}+\f{1}{2}(\g+\log(32\pi\s\eta^2\mu^2)^{-1})+c \r],
\end{align}
with 
\begin{equation}
c=0.055868.
\end{equation}
Note that  
$\mu^{-\e}\bra0|g^{tt}(\p_t \phi)^2|0\ket^{(0)}_{reg}|_{r=r_0}$ has become $\MO(1)$ as a result of the integration over $\o$ \textit{and} $l$.

\subsection{Evaluation of $\bra0|g^{rr}(\p_r \phi)^2|0\ket^{(0)}_{reg}$}
We next go to the evaluation of the $r$-derivative. 
\begin{align}\lb{req1}
&\bra0|g^{rr}(\p_r \phi)^2|0\ket^{(0)}_{reg}|_{r=r_0}\nn\\ 
 =&g^{rr}(r_0) \sum_i|\p_r u_i^{(0)}(r_0)|^2 \nonumber \\
 =& \f{\hbar}{4\s \eta^2 r_0^2 (2\pi)^\e}\int d^\e k \sum_n \f{\p \Omega_i}{\p n} \sum_{lm} |Y_{lm}|^2\l[\l(\f{1}{2}-A\r)J_A(\Omega_i) +\Omega_i J_{A-1}(\Omega_i)\r]^2\nn\\
 =& \f{\hbar}{32\pi \s^2 \eta^4(2\s\eta^2)^{\f{\e}{2}}(2\pi)^\e}
 \int d^\e K \int d\Om \int dY \int dA \d\l(\sqrt{Y+K^2 + \f{1}{4}}-A\r)\nn\\
 &~~~~~~~~~~~~~~~~~~~~~~~~~~~~~~~~~~~~~~~~~~~~~~~\l[\l(\f{1}{2}-A\r)J_A(\Omega) +\Omega J_{A-1}(\Omega)\r]^2.
\end{align}
Here, in the third line we have used \Ref{ur2} and in the forth line we have taken the same procedure as in the previous subsection. 
We can do the same calculation for $K$ and $Y$ as in \Ref{Kint} and \Ref{Yint} and obtain 
\begin{align}\lb{req2}
\bra0|g^{rr}(\p_r \phi)^2|0\ket^{(0)}_{reg}|_{r=r_0}&
=-\f{\hbar}{16\pi\s^2\eta^4(8\pi\s\eta^2)^{\f{\e}{2}}\G(1+\f{\e}{2})}\int dA d\Omega h_r(\Omega,A),\\
\lb{hr}
h_r(\Omega,A)&\equiv -A\l(A^2-\f{1}{4}\r)^{\e/2}\l[\l(\f{1}{2}-A\r)J_A(\Omega) +\Omega J_{A-1}(\Omega)\r]^2.
\end{align}

Using $e^{-s\Omega}$, we can perform the $\Om$-integration and expand it around $s=0$ again: 
\begin{align}
&\int^\infty_0 d\Om e^{-s\Om}h_r(\Om,A)\nn\\ 
=&-\f{A}{\pi s^2} \l(A^2-\f{1}{4}\r)^{\f{\e}{2}}-\f{A}{4\pi}\l(A^2-\f{1}{4}\r)^{1+\f{\e}{2}}\log\l(\f{s^2}{4}\r) \nn\\
 &-\f{2^{-4}}{\pi}A\l(A^2-\f{1}{4}\r)^{\f{\e}{2}}\f{1}{-3+2A}\l(-7-22A+44A^2-8A^3+2(-3+2A)(-1+4A^2)H_{A-\f{5}{2}}\r)  \nonumber \\
 =&-\f{2^{-4}}{\pi}A\l(A^2-\f{1}{4}\r)^{\f{\e}{2}}\l((-3-4A^2)+8(A^2-\f{1}{4})H_{A-\f{1}{2}}\r)  \nonumber\\
 =&-\f{A}{2\pi} \l(A^2-\f{1}{4}\r)^{1+\f{\e}{2}}H_{A-\f{1}{2}},
\end{align}
where we have dropped the two terms in the second line and the first one in the forth line, which are polynomials in $A$, 
and used \Ref{H1} to change from $H_{A-\f{5}{2}}$ to $H_{A-\f{1}{2}}$. 
Note that this is the same as \Ref{teq4}, 
which means 
\begin{equation}\lb{req3}
\bra0|g^{rr}(\p_r \phi)^2|0\ket^{(0)}_{reg}|_{r=r_0}=\bra0|g^{tt}(\p_t \phi)^2|0\ket^{(0)}_{reg}|_{r=r_0}.
\end{equation}

\subsection{Evaluation of $\f{1}{2}\bra0|g^{\th\th}(\p_\th \phi)^2+g^{\phi\phi}(\p_\phi \phi)^2|0\ket^{(0)}_{reg}$}
We do almost the same calculation again. 
\begin{align}\lb{theq1}
&\f{1}{2}\bra0|g^{\th\th}(\p_\th \phi)^2+g^{\phi\phi}(\p_\phi \phi)^2|0\ket^{(0)}_{reg}|_{r=r_0}\nn\\
=&\f{1}{2}\sum_i\l(g^{\th\th}|\p_\th u_i^{(0)}|^2 +g^{\phi\phi}|\p_\phi u_i^{(0)}|^2  \r)\nn\\
=&\f{1}{2}\sum_i\f{l(l+1)}{r_0^2}|u_i^{(0)}|^2 \nn\\
=&\f{\hbar}{4r_0^2(2\pi)^\e}\int d^\e k \sum_n \f{\p \Omega_i}{\p n}\sum_{lm}|Y_{lm}|^2\tilde L J_{A}(\Omega_i)^2\nn\\
=&\f{\hbar}{64\pi\s^2\eta^4(2\s\eta^2)^{\f{\e}{2}}(2\pi)^\e}
\int dA \int d^\e K \int d\Omega \int dY Y J_{A}(\Omega)^2\d\l(\sqrt{Y+K^2 + \f{1}{4}}-A\r)\nn\\
=&\f{\hbar}{32\pi\s^2\eta^4(8\pi\s\eta^2)^{\f{\e}{2}}\G(\f{\e}{2})}
\int dA A \int d\Omega  J_{A}(\Omega)^2 dY Y  \l(A^2-\f{1}{4}-Y\r)^{\f{\e}{2}-1}
\end{align}
where we have used \Ref{u2}, introduced the same variables and performed the $K$-integration \Ref{Kint}. 
Now, we integrate it over $Y$ as 
\begin{equation}
\int_0^{A^2-1/4} dYY\l(A^2-\f{1}{4}-Y\r)^{\f{\e}{2}-1}=B\l(2,\f{\e}{2}\r)\l(A^2-\f{1}{4}\r)^{1+\f{\e}{2}}=\f{1}{\f{\e}{2}\l(1+\f{\e}{2}\r)}\l(A^2-\f{1}{4}\r)^{1+\f{\e}{2}}.
\end{equation}
Thus, we have 
\begin{align}\lb{theq2}
\f{1}{2}\bra0|g^{\th\th}(\p_\th \phi)^2+g^{\phi\phi}(\p_\phi \phi)^2|0\ket^{(0)}_{reg}|_{r=r_0}&
=-\f{\hbar}{16\pi\s^2\eta^4(8\pi\s\eta^2)^{\f{\e}{2}}\G(1+\f{\e}{2})}\int dA d\Omega h_\th(\Omega,A),\\
\lb{hth}
h_\th(\Omega,A)&\equiv -\f{1}{2(1+\f{\e}{2})}A\l(A^2-\f{1}{4}\r)^{1+\f{\e}{2}}J_A(\Omega)^2.
\end{align}

We again perform the $\Om$-integration with $e^{-s\Om}$ and expand it around $s=0$: 
\begin{align}
\int^\infty_0 d\Om e^{-s\Om}h_\th(\Om,A) &= \f{A}{2\pi(1+\f{\e}{2})}\l(A^2-\f{1}{4}\r)^{1+\f{\e}{2}}\l(H_{A-\f{1}{2}}+\log\f{s}{2}\r)\nn\\
 &= \l(1-\f{\e}{2}\r)\f{A}{2\pi}\l(A^2-\f{1}{4}\r)^{1+\f{\e}{2}} H_{A-\f{1}{2}},
\end{align}
where we have dropped the second term, which is a polynomial in $A$, and considered $\e\ll1$. 
Comparing this with \Ref{teq4}, we conclude 
\begin{equation}\lb{theq3}
\f{1}{2}\bra0|g^{\th\th}(\p_\th \phi)^2+g^{\phi\phi}(\p_\phi \phi)^2|0\ket^{(0)}_{reg}|_{r=r_0}
=-\l(1-\f{\e}{2}\r)\bra0|g^{tt}(\p_t \phi)^2|0\ket^{(0)}_{reg}|_{r=r_0}.
\end{equation}
\subsection{Evaluation of $\sum_{a=1}^\e \bra0|(\p_{y^a}\phi)^2|0\ket^{(0)}_{reg}$}
In the dimensional regularization, $\sum_{a=1}^\e \bra0|(\p_{y^a}\phi)^2|0\ket^{(0)}_{reg}$ is also important. 
\begin{align}\lb{yeq1}
&\bra0|\sum_{a=1}^\e (\p_{y^a} \phi)^2|0\ket^{(0)}_{reg}|_{r=r_0}\nn\\
=&\sum_i\sum_{a=1}^\e k_ak^a |u_i^{(0)}|^2 \nn\\
=&\f{\hbar}{2r_0^2(2\pi)^\e}\int d^\e k k^2  \sum_n \f{\p \Omega_i}{\p n}\sum_{lm}|Y_{lm}|^2 J_{A}(\Omega_i)^2\nn\\
=&\f{\hbar}{32\pi\s^2\eta^4(2\s\eta^2)^{\f{\e}{2}}(2\pi)^\e}
\int dA \int d^\e K K^2 \int d\Omega \int dY J_{A}(\Omega)^2\d\l(\sqrt{Y+K^2 + \f{1}{4}}-A\r)\nn\\
=&\f{\hbar}{16\pi\s^2\eta^4(8\pi\s\eta^2)^{\f{\e}{2}}\G(\f{\e}{2})}
\int dA A \int d\Omega  J_{A}(\Omega)^2 \int dY \l(A^2-\f{1}{4}-Y\r)^{\f{\e}{2}},
\end{align}
where we have performed a similar integration to \Ref{Kint}. 
The $Y$-integration is done as 
\begin{equation}
\int_0^{A^2-1/4} dY \l(A^2-\f{1}{4}-Y\r)^{\f{\e}{2}}=-\l.\f{1}{1+\f{\e}{2}}\l(A^2-\f{1}{4}-Y\r)^{\f{\e}{2}+1}\r|^{A^2-1/4}_0=
\f{1}{1+\f{\e}{2}}\l(A^2-\f{1}{4}\r)^{\f{\e}{2}+1}.
\end{equation}
Then, we have 
\begin{align}\lb{yeq2}
\bra0|\sum_{a=1}^\e (\p_{y^a} \phi)^2|0\ket^{(0)}_{reg}|_{r=r_0}
&=-\f{\hbar}{16\pi\s^2\eta^4(8\pi\s\eta^2)^{\f{\e}{2}}\G(1+\f{\e}{2})}\int dA d\Omega h_y(\Omega,A),\\
\lb{hy}
h_y(\Omega,A)&\equiv -\f{\e}{2(1+\f{\e}{2})}A\l(A^2-\f{1}{4}\r)^{1+\f{\e}{2}}J_A(\Omega)^2.
\end{align}
Comparing this to \Ref{hth}, we find $h_y=\e h_{\th}$, which means through \Ref{theq3}
\begin{equation}\lb{yeq3}
\bra0|\sum_{a=1}^\e (\p_{y^a} \phi)^2|0\ket^{(0)}_{reg}|_{r=r_0}
=\e \f{1}{2}\bra0|g^{\th\th}(\p_\th \phi)^2+g^{\phi\phi}(\p_\phi \phi)^2|0\ket^{(0)}_{reg}|_{r=r_0}
= -\e \bra0|g^{tt}(\p_t \phi)^2|0\ket^{(0)}_{reg}|_{r=r_0}.
\end{equation}

\subsection{Evaluation of $\bra0|T^\mu{}_\nu|0\ket^{(0)}_{reg}$}
Combing the results we have obtained so far, we can evaluate each component of $\bra0|T^\mu{}_\nu|0\ket^{(0)}_{reg}$.
Because each free field $\phi_a(x)$ gives the same contribution, 
we can just multiply $N$ to obtain through \Ref{EMT} 
\begin{align}\lb{Att}
\bra0|T^t{}_t|0\ket^{(0)}_{reg} 
&= \f{N}{2} \bra0|g^{tt}(\p_t \phi)^2 - g^{rr}(\p_r \phi)^2-g^{\th\th}(\p_\th \phi)^2-g^{\phi\phi}(\p_\phi \phi)^2
-\sum_{a=1}^\e (\p_{y^a} \phi)^2|0\ket^{(0)}_{reg}\nn\\
&= \f{N}{2} \bra0|g^{tt}(\p_t \phi)^2 - g^{tt}(\p_t \phi)^2+2(1-\e/2)g^{tt}(\p_t \phi)^2+\e g^{tt}(\p_t \phi)^2|0\ket^{(0)}_{reg}\nn\\
 &= N \bra0|g^{tt}(\p_t \phi)^2|0\ket^{(0)}_{reg},
\end{align}
where we have used \Ref{req3}, \Ref{theq3} and \Ref{yeq3}. 
Similarly, we calculate 
\begin{align}\lb{Arr}
\bra0|T^r{}_r|0\ket^{(0)}_{reg} 
&= \f{N}{2} \bra0|g^{rr}(\p_r \phi)^2 - g^{tt}(\p_t \phi)^2-g^{\th\th}(\p_\th \phi)^2-g^{\phi\phi}(\p_\phi \phi)^2
-\sum_{a=1}^\e (\p_{y^a} \phi)^2|0\ket^{(0)}_{reg}\nn\\
&= \f{N}{2} \bra0|g^{tt}(\p_t \phi)^2 - g^{tt}(\p_t \phi)^2+2(1-\e/2)g^{tt}(\p_t \phi)^2+\e g^{tt}(\p_t \phi)^2|0\ket^{(0)}_{reg}\nn\\
 &= N \bra0|g^{tt}(\p_t \phi)^2|0\ket^{(0)}_{reg}.
\end{align}
Again, we have 
\begin{align}\lb{Ath}
\f{1}{2}\bra0|T^\th{}_\th+T^\phi{}_\phi|0\ket^{(0)}_{reg} 
&= \f{N}{2} \bra0|-g^{tt}(\p_t \phi)^2 - g^{rr}(\p_r \phi)^2-\sum_{a=1}^\e (\p_{y^a} \phi)^2|0\ket^{(0)}_{reg}\nn\\
&= \f{N}{2} \bra0|-g^{tt}(\p_t \phi)^2 - g^{tt}(\p_t \phi)^2+\e g^{tt}(\p_t \phi)^2|0\ket^{(0)}_{reg}\nn\\
 &=\l(-1+\f{\e}{2}\r) N \bra0|g^{tt}(\p_t \phi)^2|0\ket^{(0)}_{reg}.
\end{align}
This average value gives $\bra0|T^\th{}_\th|0\ket^{(0)}_{reg}=\bra 0| T^\phi{}_\phi|0\ket^{(0)}_{reg}$ 
because of the spherical symmetry. 
Note here that $\f{\e}{2}$ picks up the contribution from the pole $\f{1}{\e}$ of $N \bra0|g^{tt}(\p_t \phi)^2|0\ket^{(0)}_{reg}$, 
which is completely determined by the UV structure. 
In this sense, the term is anomalous. 

Thus, these and \Ref{teq7} provide \Ref{T0reg}.

\section{Derivation of $\tau_{vv}(u,v_s)$ \Ref{vv2}}\lb{A:vv}
First we derive \Ref{vp_vs}. 
For simplicity, we assume that for $u\lesssim u_\ast$ 
the shell falls approximately in the static Schwarzschild metric with the initial mass $\f{a_0}{2G}$. 
This is motivated by the fact that 
the time scale in which the shell approaches from, say, $r=2a$ to $r=a+\f{2\s}{a}$ 
is $\D u \sim a$ 
while the time scale in which the energy of the system changes significantly is $\D u\sim \f{a^3}{\s}$. 
Then, we can evaluate for $u\lesssim u_\ast$ 
\begin{align}\lb{vpvv1}
\vp(u,v_s) &=-\int^u_{-\infty}d\tu \f{a(\tu)}{2r_s(\tu)^2} \nn \\
 &\approx -a_0 \int^u_{-\infty}d\tu \f{1}{2r_s(\tu)^2}  \nonumber \\
 &= a_0 \int^{r_s(u)}_{\infty}dr_s  \f{1}{r_s(r_s-a_0)}  \nonumber\\
 &= \log \f{r_s(u)-a_0}{r_s(u)},
\end{align}
where 
in the third line we have changed the variable from $\tu$ to $r_s$ by using the second of \Ref{vpr} with $a_0$. 

Now, suppose that the black hole starts to evaporate from $u\sim u_*$, 
which means that $a_0\approx a(u_*)\equiv a_{\ast}$. 
Similarly, we can calculate for $u\gtrsim u_\ast$ 
\begin{align}\lb{vpvv2}
\vp(u,v_s) &=-\int^{u_*}_{-\infty}d\tu \f{a(\tu)}{2r_s(\tu)^2} -\int^{u}_{u_*}d\tu \f{a(\tu)}{2r_s(\tu)^2}\nn \\
 &\approx -a_0 \int^{u_*}_{-\infty}d\tu \f{1}{2r_s(\tu)^2}-\int^{u}_{u_*}d\tu \f{1}{2a(\tu)}  \nonumber \\
 &= a_0 \int^{r_s(u_*)}_{\infty}dr_s  \f{1}{r_s(r_s-a_0)}+\int^{a(u)}_{a(u_*)}da \f{a}{2\s}  \nonumber\\
 &\approx \log \f{2\s }{a_*^2}+\f{a(u)^2-a_*^2}{4\s},
\end{align}
where we have used $r_s(u)= a(u)+\f{2\s}{a(u)}$, and 
in the second term of the third line we have changed the variable from $\tu$ to $a$ 
by using \Ref{da}. 

Using these, we consider for $u\gtrsim u_*$
\begin{equation}
\int^u_{-\infty}d\tu \f{a(\tu)}{r(\tu,v_s)^4}e^{2\vp(\tu,v_s)}=
\int^{u_*}_{-\infty}d\tu \f{a(\tu)}{r_s(\tu)^4}e^{2\vp(\tu,v_s)}+ \int^u_{u_*}d\tu \f{a(\tu)}{r_s(\tu)^4}e^{2\vp(\tu,v_s)} 
\equiv I_1+I_2.
\end{equation}
$I_1$ can be evaluated from \Ref{vpvv1} as 
\begin{align}\lb{AI}
I_1 &\approx a_* \int^{u_*}_{-\infty}d\tu \f{1}{r_s(\tu)^4}\l(\f{r_s(\tu)-a_*}{r_s(\tu)}\r)^2 \nn\\
 &=-2a_* \int^{r_s(u_*)}_{\infty}dr_s \f{1}{r_s^4}\f{r_s-a_*}{r_s}  \nonumber \\
 &=\f{2a_*}{3r_s(u_*)^3}-\f{a_*^2}{2r_s(u_*)^4} \nonumber\\
 &\approx \f{1}{6a_*^2},
\end{align}
where we have used again the second of \Ref{vpr} and $r_s(u_*)\approx a_*$. 
From \Ref{vpvv2}, $I_2$ becomes 
\begin{align}
I_2 &\approx \int^u_{u_*}d\tu \f{1}{a(\tu)^3}\l(\f{2\s}{a_*^2}\r)^2e^{\f{a(\tu)^2-a_*^2}{2\s}} \nn\\
 &= \l(\f{2\s}{a_*^2}\r)^2 \int^{a(u)}_{a(u_*)}da \f{-1}{\s a}e^{\f{a^2-a_*^2}{2\s}}  \nonumber \\
 &\approx \f{-1}{\s a_*} \l(\f{2\s}{a_*^2}\r)^2 \int^{a(u)}_{a_*}da e^{\f{a_*}{\s}(a-a_*)}   \nonumber \\
 &\approx  \f{1}{a_*^2} \l(\f{2\s}{a_*^2}\r)^2=\MO\l(a^{-6}\r),
\end{align}
where we have used \Ref{da} again. This is negligible compared to $I_1$.

Thus, we reach the first term of \Ref{vv2}: 
\begin{equation}
\f{3}{2}\g \int^u_{-\infty}d\tu \f{a(\tu)}{r(\tu,v_s)^4}e^{2\vp(\tu,v_s)}\approx -\f{N\hbar}{192 \pi a(u)^2}.
\end{equation}
Here, we have replaced $\f{1}{a_*^2}$ with  $\f{1}{a(u)^2}$ 
because they are almost the same for a small mass of the outermost shell. 
\section{Classical and quantum contributions to Energy density $-T^{u}{}_{u}$}\lb{A:Bondi}
We derive \Ref{dBon}, \Ref{T_cl2} and \Ref{den_vac_A}. 
Before this, 
we note that in general, the Bondi mass is defined as the energy inside $r$ on a $u$-constant surface \ci{Poisson}, 
which is expressed in the $(u, r)$ coordinate of \Ref{Vaidya} as 
\begin{equation}\lb{Bondi}
M \equiv 4\pi \int^{r}_{0,~u={\rm const.}} dr' r'^2 \bra-T^{\bar u}{}_{\bar u}\ket.
\end{equation}
Here, the suffix $\bar u$ stands for the $u$-component in the $(u,r)$ coordinate of \Ref{Vaidya}. 
Then, we express $T^{\bar u}{}_{\bar u}$ in terms of the $(u,v)$ coordinate. 
From $d\bar u= du$ and $dr=\l(\f{\p r}{\p u}\r)_v du + \l(\f{\p r}{\p v}\r)_u dv$ in \Ref{Vaidya}, 
we have the coordinate transformation between $(\bar u,r)$ and $(u,v)$:
\begin{equation}
du=d\bar u,~~dv=2e^{-\vp}\l(dr+\f{1}{2}\f{r-a(\bar u)}{r}d\bar u\r),
\end{equation}
where we have used \Ref{vpr}. We calculate 
\begin{align}\lb{den_formula}
T^{\bar u}{}_{\bar u} &=\f{\p \bar u}{\p x^a}\f{\p x^b}{\p \bar u} T^a{}_b \nn\\
 &=T^u{}_u+e^{-\vp}\f{r-a}{r}T^u{}_v \nonumber \\
 &=-2 e^{-\vp}T_{vu}-2e^{-2\vp}\f{r-a}{r}T_{vv}.
\end{align}

Let us construct \Ref{dBon}. For the classical shell part \Ref{v_s}, 
we have only $T_{vv}^{(cl)}=\f{\tau_{vv}^{(0)}}{4\pi r^2}=\f{W}{4\pi r^2}\d(v-v_s)$ and obtain 
\begin{equation}\lb{T_cl1}
T^{\bar u(cl)}{}_{\bar u}=\f{1}{4\pi r^2}\l(-2We^{-2\vp}\f{r-a(u)}{r}\r)\d(v-v_s).
\end{equation}
From this, we evaluate , 
for $r_0\equiv r(u,v_0)>r_s\equiv r(u,v_s)$ where $v_0>v_s$,  
\begin{align}
\D M &\equiv 4\pi \int^{r_0}_{0,\bar u={\rm const.}}dr r^2 (-T^{\bar u(cl)}{}_{\bar u})\nn \\
 &= 4\pi \int^{v_0}_{-\infty}dv \l(\f{\p r}{\p v}\r)_u r^2 (-T^{\bar u(cl)}{}_{\bar u}) \nonumber \\
 &= \int^{v_0}_{-\infty}dv \f{1}{2}e^\vp  2We^{-2\vp}\f{r-a(u)}{r} \d(v-v_s) \nonumber \\
 &= \f{r_s-a(u)}{r_s} e^{-\vp(u,v_s)}   W \nonumber \\
 &\approx \f{2\s}{a(u)^2} \f{a_*^2}{2\s} e^{-\f{a(u)^2-a_*^2}{4\s}}W\nn\\
 &\approx e^{-\f{a(u)^2-a_*^2}{4\s}}W,
\end{align}
which is \Ref{dBon}. 
Here, in the fifth line we have used the part for $u\gtrsim u_*$ of \Ref{vp_vs} and $r_s\approx a+\f{2\s}{a}$; 
and in the last one we have made an approximation $a(u)\approx a_*$, 
since the difference does not contribute to the time evolution, compared to the exponential factor. 

Then, we check the energy density \Ref{T_cl2}. 
We can use \Ref{vp_vs} to evaluate the classical contribution \Ref{T_cl1} for $u\gtrsim u_*$ as
\begin{align}
-T^{(cl)\bar u}{}_{\bar u}&
\approx \f{2 W}{4\pi r^2} \l(\f{a^2}{2\s}\r)^2 e^{-\f{a^2-a_*^2}{2\s}} \f{2\s}{a^2} \d(v-v_s)\nn\\
&\approx \f{W}{4\pi \s} e^{-\f{r^2-a_*^2}{2\s}} \d(v-v_s),
\end{align}
where we have used $a\approx r$ approximately. 

Finally, we derive the energy density induced by the vacuum 
near the surface for $u\gtrsim u_*$, that is, \Ref{den_vac_A}.
We first evaluate 
\begin{align}\lb{T_uv_vac}
T_{uv}^{(vac)} &=\f{1}{4\pi r^2}\tau_{uv} \nn\\
 &= \f{1}{4\pi r^2} \f{-\hbar N }{24\pi} \p_u\p_v\vp \nonumber \\
 &\approx -\f{\hbar N}{192\pi a^4}e^\vp,
\end{align}
where we have used the third one of \Ref{tau_uv} and \Ref{vpr}. 
Putting this and the first term of \Ref{vv2} into \Ref{den_formula}, 
we have 
\begin{align}
-T^{(vac)\bar u}{}_{\bar u}
 &\approx-\f{\hbar N}{96\pi a^4} + 
 2 \l(\f{a^2}{2\s} \r)^2e^{-\f{a^2-a_*^2}{2\s}}\f{2\s}{a^2}\f{-\hbar N}{192\pi a^2}\f{1}{4\pi a^2}\nn\\
 &=-\f{\hbar N}{96\pi a^4} + 
 e^{-\f{a^2-a_*^2}{2\s}}\f{-\hbar N}{192\pi \s}\f{1}{4\pi a^2}\nn\\
 &\approx -\f{1}{8\pi G a^2}  e^{-\f{a^2-a_*^2}{2\s}},
\end{align}
where we have used $\s=\s_s$ \Ref{sigma_s} and kept only the second term as the leading. 
This gives \Ref{den_vac_A} because of $a\approx r$.


\end{document}